\documentclass[twocolumn,prmaterials,aps,10pt,bibnotes,amsfonts,amsmath,amssymb,floatfix,nolongbibliography]{revtex4-2}
\usepackage{color}
\usepackage{soul}
\usepackage{hhline}	
\usepackage{graphicx}
\usepackage{float}
\usepackage{color}
\usepackage{dcolumn}
\usepackage{multirow}
\usepackage{booktabs}
\usepackage{afterpage}
\usepackage{amsmath,cases}
\usepackage{braket}
\usepackage{array}
\usepackage{placeins}
\renewcommand{\bm}[1]{\boldsymbol{#1}}
\usepackage{newtxtext}
\usepackage[varg,upint,smallerops]{newtxmath}

\usepackage{hyperref}
\hypersetup{
 setpagesize=false,
 bookmarksnumbered=true,%
 bookmarksopen=true,%
 colorlinks=true,%
 linkcolor=blue,
 citecolor=blue,
 urlcolor=blue,
}

\makeatletter
\renewcommand*\email[1][]{\begingroup\sanitize@url\@email{#1}}%
\def\@email#1#2{%
 \endgroup
 \@AF@join{#1#2}%
}%
\makeatother

\begin{document}

\title{First-principles study on the electrical resistivity in zirconium dichalcogenides with multi-valley bands: mode-resolved analysis of electron-phonon scattering}

\author{Hitoshi Mori}
\author{Masayuki Ochi}
\author{Kazuhiko Kuroki}
\affiliation{Department of Physics, Osaka University, Machikaneyama-cho, Toyonaka, Osaka 560-0043, Japan}
\date{\today}
\begin{abstract}
Based on the first-principles calculations, we study the electron-phonon scattering effect on the resistivity in the zirconium dichalcogenides, $\text{Zr}_{}\text{S}_{2}$ and $\text{Zr}_{}\text{Se}_{2}$, whose electronic band structures possess multiple valleys at conduction band minimum. The computed resistivity exhibits non-linear temperature dependence, especially for $\text{Zr}_{}\text{S}_{2}$, which is also experimentally observed on some TMDCs such as $\text{Ti}_{}\text{S}_{2}$ and $\text{Zr}_{}\text{Se}_{2}$. By performing the decomposition of the contributions of scattering processes, we find that the intra-valley scattering by acoustic phonons mainly contributes to the resistivity around 50~K. Moreover, the contribution of the intra-valley scattering by optical phonons becomes dominant even above 80~K, which is a sufficiently low temperature compared with their frequencies. 
By contrast, the effect of the inter-valley scattering is found to be not significant. 
Our study identifies the characteristic scattering channels in the resistivity of the zirconium dichalcogenides, which provides critical knowledge to microscopically understand electron transport in systems with multi-valley band structure.
\end{abstract}

\maketitle

\section{Introduction}
Transition metal dichalcogenides (TMDCs)~\cite{Mattheiss1973-ua,Wilson1969-fe,Doni1986,Wang2012-iv} are layered materials of the form $MX_2$, where $M$ and $X$ are transition metal and chalcogen atoms, respectively. 
The TMDCs have been extensively investigated due to a wide range of physical properties and potential applications originating from the unique band structures, such as superconductivity~\cite{Taniguchi2012-wt,Ye2012-cw,Joe2014-kx,Liu2017-fd,De_la_Barrera2018-zr}, magnetic ordering~\cite{Ma2012-vv,Zhu2016-dh,Xiang2016-ln,Chiew2020-ce}, charge density waves (CDWs)~\cite{PhysRevB.14.4321,PhysRevB.17.1839,Suzuki1985-jc,Sugai1985-ip,PhysRevB.65.235101,PhysRevLett.88.226402,Clerc2007-aa,Porer2014-xb,Dolui2016-ap,Li2016-fj}, exciton dynamics~\cite{PhysRevB.89.205303,PhysRevB.90.041414,PhysRevB.92.235425,PhysRevLett.117.187401,Hao2016-ck,PhysRevLett.117.187401,Rivera2018-sk,RevModPhys.90.021001}, topological semimetal~\cite{Bruno2016-rs,Huang2016-qm,Huang2016-eo,Belopolski2016-gr,Wang2016-vn,Jiang2017-mk,Yan2017-sb,Zhang2017-xb,Li2017-jb,Wang2019-fa,Kar2020-gq}, and thermoelectricity~\cite{Koyano1986-av,Sasaki1987-md,PhysRevB.64.241104,Guilmeau2011-da,Wan2010-hj,Wickramaratne2014-rh,PhysRevB.90.174301,BOURGES20161183,Huang2016-na,Zhang2017-cf}. 
In some of typical TMDCs, the electronic band structure shows multiple valleys around the Fermi level, where a valley implies a local extremum of the band structure ; the manipulation of valley degree of freedom has attracted considerable attention for valleytronics applications~\cite{Schaibley2016-dn,Rycerz2007-bl,Xiao2007-zr,Yao2008-yr,Cao2012-nh,Mak2012-me,Zeng2012-bn,Wu2013-lu,Yuan2013-oa,Jones2013-uy,Mak2014-zs,Tong2017-bj,Liu2019-ia}. In addition, the effect of scattering between valleys, namely inter-valley scattering, have been investigated from various perspectives such as excitonic electron-hole exchange interaction~\cite{PhysRevB.89.205303,PhysRevB.90.041414,PhysRevB.92.235425,RevModPhys.90.021001}, CDW instability~\cite{PhysRevB.14.4321,PhysRevB.17.1839,Suzuki1985-jc,Sugai1985-ip,PhysRevB.65.235101,Dolui2016-ap}, and electrical transport~\cite{PhysRevB.10.1409,PhysRevB.64.241104,PhysRevLett.29.163,PhysRevLett.35.1786,Klipstein1981-hy,Onuki1982-dz,Koyano1986-av,Zheng1989-li,Patel1998-zd,Zandt2007-lm,Suri_2017}. 
\par
The effect of inter-valley scattering is also considered to be a key factor to the thermoelectric performance~\cite{Herring1955-nt,Popescu2012-xy,Pei2012-uj,Xin2018-zw}. 
Higher valley degeneracy leads to higher carrier density, but at the same time, results in higher possibility of electron scattering, which in turn suppresses the electrical conductivity. 
In particular, it has been argued that inter-valley scattering induced by acoustic phonons strongly affects the electrical transport in addition to intra-valley scattering by acoustic phonons, and it may cause the quadratic-like temperature dependence of electrical resistivity of $\text{Ti}_{}\text{S}_{2}$ observed over a wide temperature range~\cite{PhysRevB.64.241104,Klipstein1981-hy,Koyano1986-av,Suri_2017}. 
The quadratic-like temperature dependence has been observed not only in $\text{Ti}_{}\text{S}_{2}$~\cite{PhysRevB.10.1409,PhysRevB.64.241104,PhysRevLett.29.163,PhysRevLett.35.1786,Wilson1977-mt,Wilson1978-zs,PhysRevB.24.1691,Klipstein1981-hy,Onuki1982-dz,Koyano1986-av,Suri_2017} but also in some other TMDCs such as $\text{Zr}_{}\text{Se}_{2}$~\cite{Onuki1982-dz,Zheng1989-li,Patel1998-zd}, $\text{Hf}_{}\text{Se}_{2}$~\cite{Zheng1989-li}, and $\text{Mo}_{}\text{Te}_{2}$~\cite{Zandt2007-lm}, and the electron scattering mechanism is still controversial. 
Other than inter-valley scattering by acoustic phonons, some other scattering mechanisms such as optical phonon scattering~\cite{Wilson1977-mt,Wilson1978-zs,PhysRevB.24.1691,Onuki1982-dz,Zheng1989-li,PhysRev.163.743}, electron-electron scattering~\cite{PhysRevLett.35.1786,PhysRevB.19.6172}, electron-hole scattering~\cite{PhysRevLett.37.782,PhysRevB.19.6172,PhysRevB.24.1691}, and ionized impurity scattering~\cite{Patel1998-zd} have also been proposed as crucial mechanisms determining the transport properties. 
Although some first-principles studies on the inter-valley scattering effect in some TMDCs and other materials have recently been performed in terms of electron-phonon scattering~\cite{Kaasbjerg2012-za,Zhao2018-yq,doi:10.1063/1.5040752,PhysRevX.9.031019,Wu2021-db}, they have not pursued the cause of the quadratic temperature dependence of resistivity.
Elucidating the origin of the temperature dependence will lead to acquiring knowledge essential for controlling transport properties such as thermoelectric performance. 
\par
Given this background, in the present study, we theoretically investigate the electron-phonon scattering effect on the resistivity in the zirconium dichalcogenides, $\text{Zr}_{}\text{S}_{2}$ and $\text{Zr}_{}\text{Se}_{2}$, by means of first-principles calculations. 
The theoretical approach employed in the present study can take into account the electron-phonon scattering by all the phonon modes. 
We shall show the intra-valley scattering is a dominant contributor to the resistivity in a wide temperature range and that the contribution from the intra-valley scattering by optical phonons increases with increasing temperature. 
Our detailed analysis unambiguously identifies the phonon states which strongly scatter the electrons and thus dominantly contribute to the resistivity at both low and high temperatures, 50~K and 300~K. 
Based on these results, we reveal that the intra-valley scattering caused by optical phonons is a primary factor in making the temperature dependence of resistivity non-linear. 
We also conclude that the electrical resistivity of $\text{Zr}_{}\text{Se}_{2}$ exhibits temperature dependence closer to linear than that of $\text{Zr}_{}\text{S}_{2}$ because the frequencies of optical phonons of $\text{Zr}_{}\text{Se}_{2}$ are lower.\par
The paper is organized as follows: In Sec.~\ref{sec:method} we show the details of the calculation methods we used in the present study. Sections~\hyperref[subsec:ZrS2]{III A} and \hyperref[subsec:ZrSe2]{III B} present calculation results for $\text{Zr}_{}\text{S}_{2}$ and $\text{Zr}_{}\text{Se}_{2}$, respectively. We discuss the non-linear temperature dependence of the resistivity in $\text{Zr}_{}\text{S}_{2}$ in Sec.~\hyperref[subsec:ZrS2]{III A} and the difference in the temperature dependence between $\text{Zr}_{}\text{S}_{2}$ and $\text{Zr}_{}\text{Se}_{2}$ in Sec.~\hyperref[subsec:ZrSe2]{III B}. Finally, a summary of the present study is presented in Sec.~\ref{sec:sum}.
\section{Methods of Calculations\label{sec:method}}
All the first-principles calculations were performed using \textsc{quantum espresso} package~\cite{QE-2009,QE-2017,doi:10.1063/5.0005082} (ver.~6.3). Perdew-Burke-Ernzerhof parametrization adapted for solids of the generalized gradient approximation (PBEsol-GGA)~\cite{PhysRevB.79.075126,PhysRevLett.100.136406} was used. Spin-orbit coupling was not included in the calculations. First, we determined the crystal structures through structural optimization, and then we performed the band-structure calculations. We used the optimized norm-conserving Vanderbilt (ONCV) pseudopotentials~\cite{PhysRevB.88.085117} extracted from the PseudoDojo library~\cite{VANSETTEN201839} and an $18\times18\times12$ $\bm{k}$-mesh for both materials. The plane wave cutoff was set as 100~Ry. After the band-structure calculations, we performed phonon calculations within the density-functional perturbation theory (DFPT) on a $6\times6\times4$ $\bm{q}$-mesh sampling. Hereafter, the vector $\bm{k}$ implies the electronic wave number vector, and the vector $\bm{q}$ is the phonon wave number vector in this paper. In order to construct the effective models by using the maximally localized Wannier functions~\cite{wannier1} and obtain the electron-phonon matrix elements on ultra-fine $\bm{k}$- and $\bm{q}$-grids, we employed the \textsc{epw} code~\cite{PhysRevB.76.165108,NOFFSINGER20102140,PONCE2016116,PhysRevB.97.121201} (ver.~5.2) of the \textsc{quantum espresso} distribution, which is in conjunction with the \textsc{wannier90} library~\cite{wannier2,wannier3,wannier4,wannier5} (ver.~3.0). We chose Zr-$s$ and S(Se)-$p$ orbitals as initial guesses for the Wannier functions and constructed 11-orbital effective models using a $12\times12\times8$ $\bm{k}$-mesh sampling. 
We interpolated the electron-phonon matrix elements to $96\times96\times72$ $\bm{k}$- and $\bm{q}$-grids to calculate the resistivity. Indeed, as for $\bm{k}$-grid, only the 57109 $\bm{k}$-points in the irreducible wedge were used. In this study, we focused on the in-plane resistivity of both materials. 
For this purpose, we calculated the electrical conductivity tensor by the Boltzmann transport equation within the relaxation time approximation (RTA) as follows:
\begin{align}
  \bm{\sigma}=&2\times\frac{e^2}{\Omega}\sum_{\bm{k},n}\tau_{\bm{k}n} \bm{v}_{\bm{k}n}\otimes\bm{v}_{\bm{k}n}\left(-\frac{\partial f_{\bm{k}n}}{\partial \varepsilon}\right)\label{transportcoeff}\\
  \intertext{with the Fermi-Dirac distribution function}
  f_{\bm{k}n}=&\frac{1}{e^{\beta(\varepsilon_{\bm{k}n}-\mu)}+1}.
\end{align}
The factor ``2'' at the head of the right-hand side of Eq.~\eqref{transportcoeff} comes from spin degeneracy. 
$\Omega$ is the volume of crystal, $e$ $(>0)$ is the elementary charge, $\beta$ is the inverse temperature defined as $\beta=(k_{\text{B}}T)^{-1}$ with the Boltzmann constant $k_{\text{B}}$, $\mu$ is the Fermi level. $\tau_{\bm{k}n}$, $\varepsilon_{\bm{k}n}$, and $\bm{v}_{\bm{k}n}$ are the relaxation time, the energy level of electron, and the group velocity on the $n$th band at a certain $\bm{k}$-point, respectively. 
Here, the Fermi level $\mu$ was determined so as to provide a fixed electron carrier density of $1.0\times 10^{20}\text{~cm}^{-3}$ for each temperature. 
The in-plane electrical resistivity was obtained by taking the inverse of the corresponding diagonal component of the electrical conductivity tensor $\bm{\sigma}$. In this study, we computed the scattering rate, which is the inverse of the relaxation time, by using the following equation:
\begin{align}
  \frac{1}{\tau_{\bm{k}n}}=&\frac{2\pi}{N_{\text{p}} \hbar}\sum_{\bm{q},\nu}\sum_{n'}|g_{n'n\nu}(\bm{k},\bm{q})|^2\notag\\
  &\hspace{5em}\times\Bigl[W_{n'n\nu}^{(-)}(\bm{k},\bm{q})+W_{n'n\nu}^{(+)}(\bm{k},\bm{q})\Bigr]\label{im-sigma-el-ph}\\
  \intertext{with}
  &W_{n'n\nu}^{(\pm)}(\bm{k},\bm{q})\notag\\
  =&
    \left\{
      \begin{array}{c}
        f_{\bm{k}+\bm{q}n'}+n_{\bm{q}\nu} \\
        1-f_{\bm{k}+\bm{q}n'}+n_{\bm{q}\nu}
      \end{array}
    \right\}
    \delta(\varepsilon_{\bm{k}n}-\varepsilon_{\bm{k}+\bm{q}n'}\pm\hbar\omega_{\bm{q}\nu}),\label{W_kq}
\end{align}
where $N_{\text{p}}$ is the total number of $\bm{q}$-points in the 1st Brillouin zone, $n_{\bm{q}\nu}=(e^{\beta\hbar\omega_{\bm{q}\nu}}-1)^{-1}$ is the Bose-Einstein distribution function, $\omega_{\bm{q}\nu}$ is the frequency of phonon on the $\nu$th phonon-branch at a certain $\bm{q}$-point. $g_{n'n\nu}(\bm{k},\bm{q})$ is the electron-phonon matrix element, which represents the scattering from the initial electronic state $\ket{\bm{k}n}$ to the final state $\ket{\bm{k}+\bm{q}n'}$ via a phonon in the state $(\bm{q},\nu)$. Equation~\eqref{im-sigma-el-ph} coincides with the expression of the scattering rate given by the Fermi's golden rule~\cite{PhysRevB.13.768,wagner1978radio,grimvall1981electron}. 

\section{Result and Discussion\label{sec:RandD}}
\subsection{\texorpdfstring{$\text{Zr}_{}\text{S}_{2}$}{ZrS2}\label{subsec:ZrS2}}
\begin{figure*}
  \begin{center}
    \includegraphics[width=16cm]{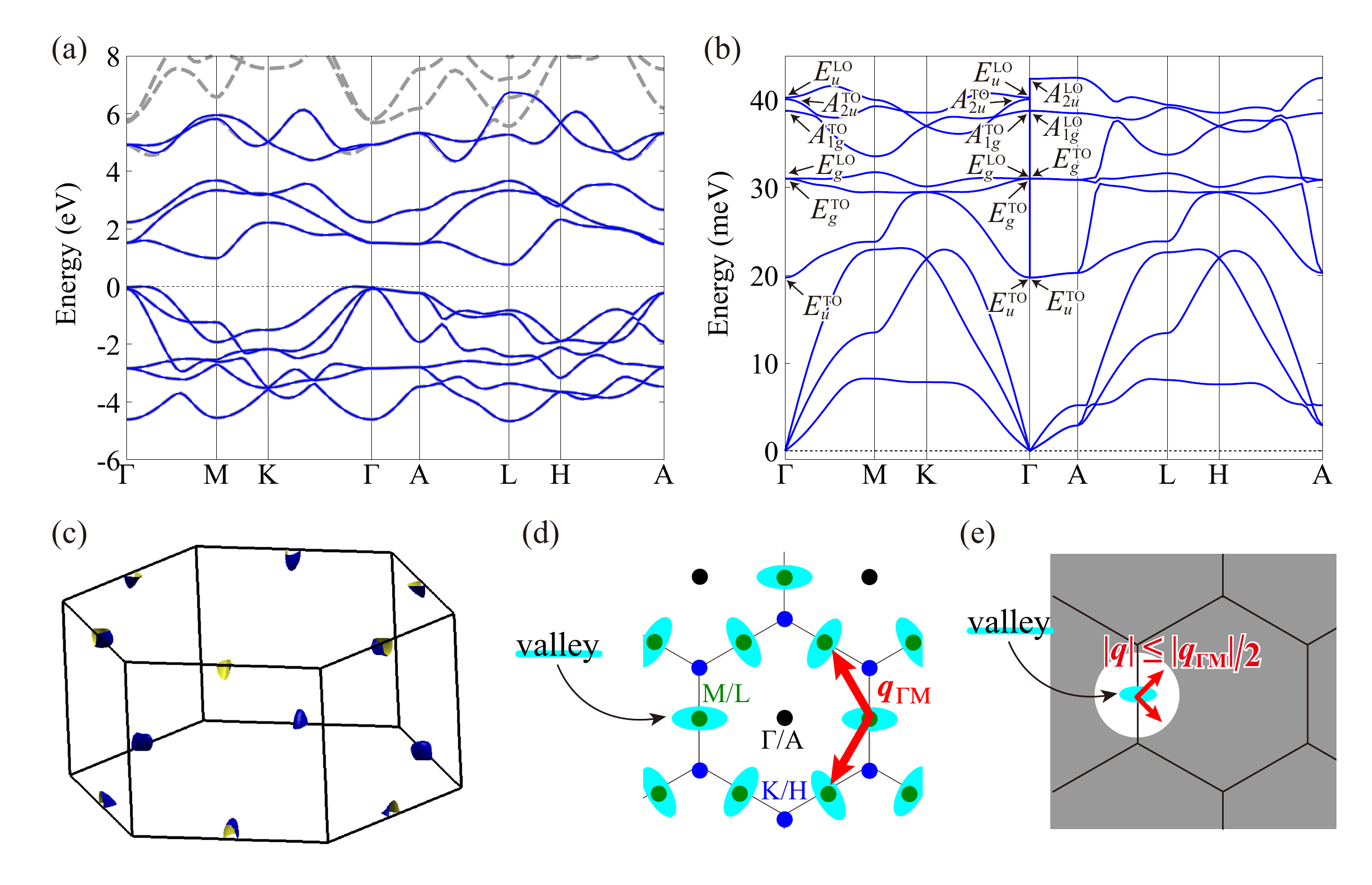}
    \caption{\label{fig:1}(a) Electronic band structures of $\text{Zr}_{}\text{S}_{2}$ obtained from first-principles calculation (dashed gray lines) and the effective model constructed by Wannier orbitals (solid blue lines). The phonon dispersion is shown in panel (b). The optical modes in the long-wavelength limit are specified by Mulliken symbols~\cite{Mulliken1955-rx}. (c) The Fermi surface for the electron carrier density of $1.0\times 10^{20}\text{~cm}^{-3}$ at $300\text{~K}$, which was plotted with \textsc{xcrysden} software~\cite{Kokalj1999-al}. (d) Schematic illustration of the Fermi surfaces in the Brillouin zone. The red arrows correspond to the scattering vectors. Panel (e) visually shows the condition $|\bm{q}|\le |\bm{q}_{\Gamma\text{M}}|/2$ which is used to sum over wave vectors $\bm{q}$ only corresponding to intra-valley scattering vectors.}
 \end{center}
\end{figure*}
Figure~\ref{fig:1} presents the calculated electron and phonon band structures of $\text{Zr}_{}\text{S}_{2}$. As shown in Fig.~\hyperref[fig:1]{1(a)}, the band structure of the effective model well reproduces the band structure obtained with first-principles calculations. 
Since n-type conductor behavior has been experimentally reported in related materials, $\text{Ti}_{}\text{S}_{2}$~\cite{PhysRevB.64.241104,Koyano1986-av,Klipstein1981-hy,Suri_2017} and $\text{Zr}_{}\text{Se}_{2}$~\cite{Onuki1982-dz,Zheng1989-li,Patel1998-zd}, we focused on the transport properties of electron-doped $\text{Zr}_{}\text{S}_{2}$ in this study. As shown in Fig.~\hyperref[fig:1]{1(c)}, there exist three Fermi pockets of electrons in the 1st Brillouin zone. The schematic illustration of Fermi surfaces is shown in Fig.~\hyperref[fig:1]{1(d)}. We note that the Fermi pockets are situated around the L points, and the scattering between the states on different pockets is the inter-valley scattering. \par
\begin{figure*}
 \begin{center}
  \includegraphics[width=16cm]{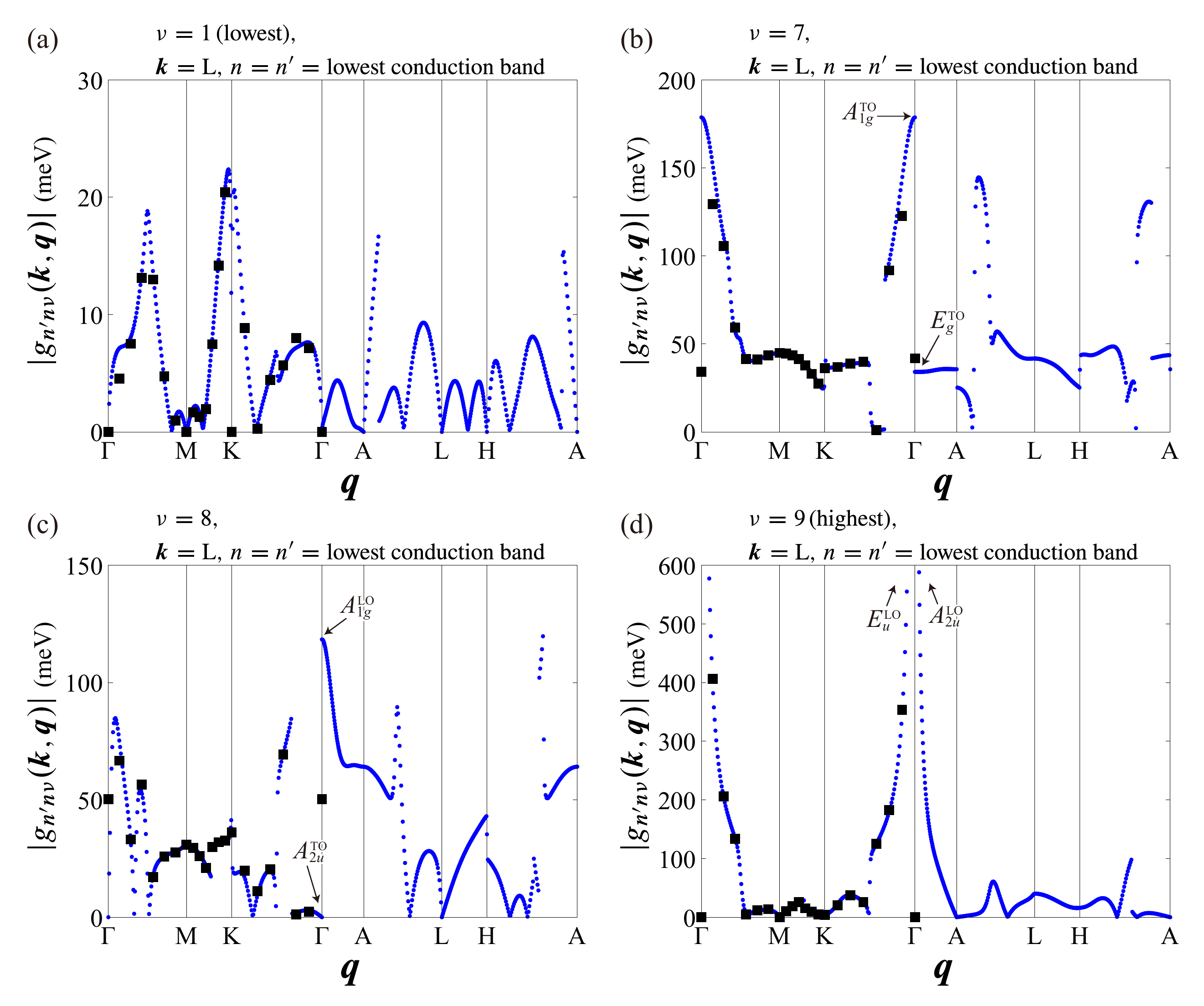}
  \caption{\label{fig:2}The $\bm{q}$-dependence of calculated electron-phonon matrix elements $g_{n'n\nu}(\bm{k},\bm{q})$ obtained with Wannier interpolations (blue dots) are compared with the elements obtained from direct DFPT calculations (black squares). In the calculation of $g_{n'n\nu}(\bm{k},\bm{q})$, the initial state $u_{\bm{k}n}$ is set to the bottom of the conduction band at L point, and the final states $u_{\bm{k}+\bm{q}n'}$ are set to the lowest conduction band. The phonon branch $\nu$ is set to the 1st (lowest) (a), the 7th (b), the 8th (c), and 9th (highest) (d). The electron-phonon matrix elements from direct DFPT calculations were obtained only along $\Gamma$-M-K-$\Gamma$. Mulliken symbols~\cite{Mulliken1955-rx} are shown for the $\bm{q}\rightarrow \bm{0}$ optical modes in panels (b)-(d).}
 \end{center}
\end{figure*}
Figure~\ref{fig:2} presents the calculated electron-phonon matrix elements, where the Wannier-interpolated ones and those explicitly obtained by DFPT calculations are shown by blue dots and black squares, respectively. 
It is seen that the $\bm{q}$-dependence of the interpolated electron-phonon matrix elements well reproduces the dependence obtained from DFPT calculations. We note that discontinuous jumps in Fig.~\ref{fig:2} come from switchings between phonon branches at their crossing points. 
As it can be seen in Fig.~\hyperref[fig:2]{2(d)}, the electron-phonon matrix elements diverge in the long-wavelength limit $\bm{q}\rightarrow \bm{0}$ because the corresponding coupling indicates that electrons couple with longitudinal optical (LO) phonons inducing macroscopic fields.

\subsubsection{Temperature dependence of the electrical resistivity}
From the interpolated electron-phonon matrix elements, we calculated the scattering rate and the electrical resistivity. 
To clarify which phonons mainly scatter electrons and whether intra-valley scattering or inter-valley scattering plays a significant role in the resistivity, we decomposed the scattering rate in terms of the $\bm{q}$-space region and the phonon branches as follows:
\begin{widetext}
\begin{align}
  \frac{1}{\tau_{\bm{k}n}^{\text{(intra-ac)}}}=&\frac{2\pi}{N_{\text{p}} \hbar}\sum_{|\bm{q}|\le |\bm{q}_{\Gamma\text{M}}|/2}\sum_{\nu=1}^{3}\sum_{n'}|g_{n'n\nu}(\bm{k},\bm{q})|^2\Bigl[W_{n'n\nu}^{(-)}(\bm{k},\bm{q})+W_{n'n\nu}^{(+)}(\bm{k},\bm{q})\Bigr],\label{intra-ac}\\
  \frac{1}{\tau_{\bm{k}n}^{\text{(intra-op)}}}=&\frac{2\pi}{N_{\text{p}} \hbar}
  \sum_{|\bm{q}|\le |\bm{q}_{\Gamma\text{M}}|/2}\sum_{\nu=4}^{9}
  \sum_{n'}|g_{n'n\nu}(\bm{k},\bm{q})|^2\Bigl[W_{n'n\nu}^{(-)}(\bm{k},\bm{q})+W_{n'n\nu}^{(+)}(\bm{k},\bm{q})\Bigr],\label{intra-op}\\
  \frac{1}{\tau_{\bm{k}n}^{\text{(inter-ac)}}}=&\frac{2\pi}{N_{\text{p}} \hbar}
  \sum_{|\bm{q}|> |\bm{q}_{\Gamma\text{M}}|/2}\sum_{\nu=1}^{3}  
  \sum_{n'}|g_{n'n\nu}(\bm{k},\bm{q})|^2\Bigl[W_{n'n\nu}^{(-)}(\bm{k},\bm{q})+W_{n'n\nu}^{(+)}(\bm{k},\bm{q})\Bigr],\label{inter-ac}\\
  \frac{1}{\tau_{\bm{k}n}^{\text{(inter-op)}}}=&\frac{2\pi}{N_{\text{p}} \hbar}\sum_{|\bm{q}|> |\bm{q}_{\Gamma\text{M}}|/2}\sum_{\nu=4}^{9}\sum_{n'}|g_{n'n\nu}(\bm{k},\bm{q})|^2\Bigl[W_{n'n\nu}^{(-)}(\bm{k},\bm{q})+W_{n'n\nu}^{(+)}(\bm{k},\bm{q})\Bigr],\label{inter-op}
\end{align}
\end{widetext}
where $|\bm{q}|=(q_x^2+q_y^2+q_z^2)^{1/2}$ and we regard the first three branches from the lowest frequency as acoustic-phonon branches. The summation with respect to $\bm{q}$ is taken over the 1st Brillouin zone. 
For example, $\tau_{\bm{k}n}^{\text{(intra-ac)}}$ can be regarded as the contribution of the intra-valley scattering by acoustic phonons to the relaxation time. Since the distance between the centers of the nearest Fermi surfaces is equal to the distance between $\Gamma$ and M points, i.e., $|\bm{q}_{\Gamma\text{M}}|$, the scattering processes with $\bm{q}\leq |\bm{q}_{\Gamma\text{M}}|/2$ can be regarded as intra-valley scattering as shown in Fig.~\hyperref[fig:1]{1(e)}.\par 
\begin{figure*}
\begin{center}
  \includegraphics[width=16cm]{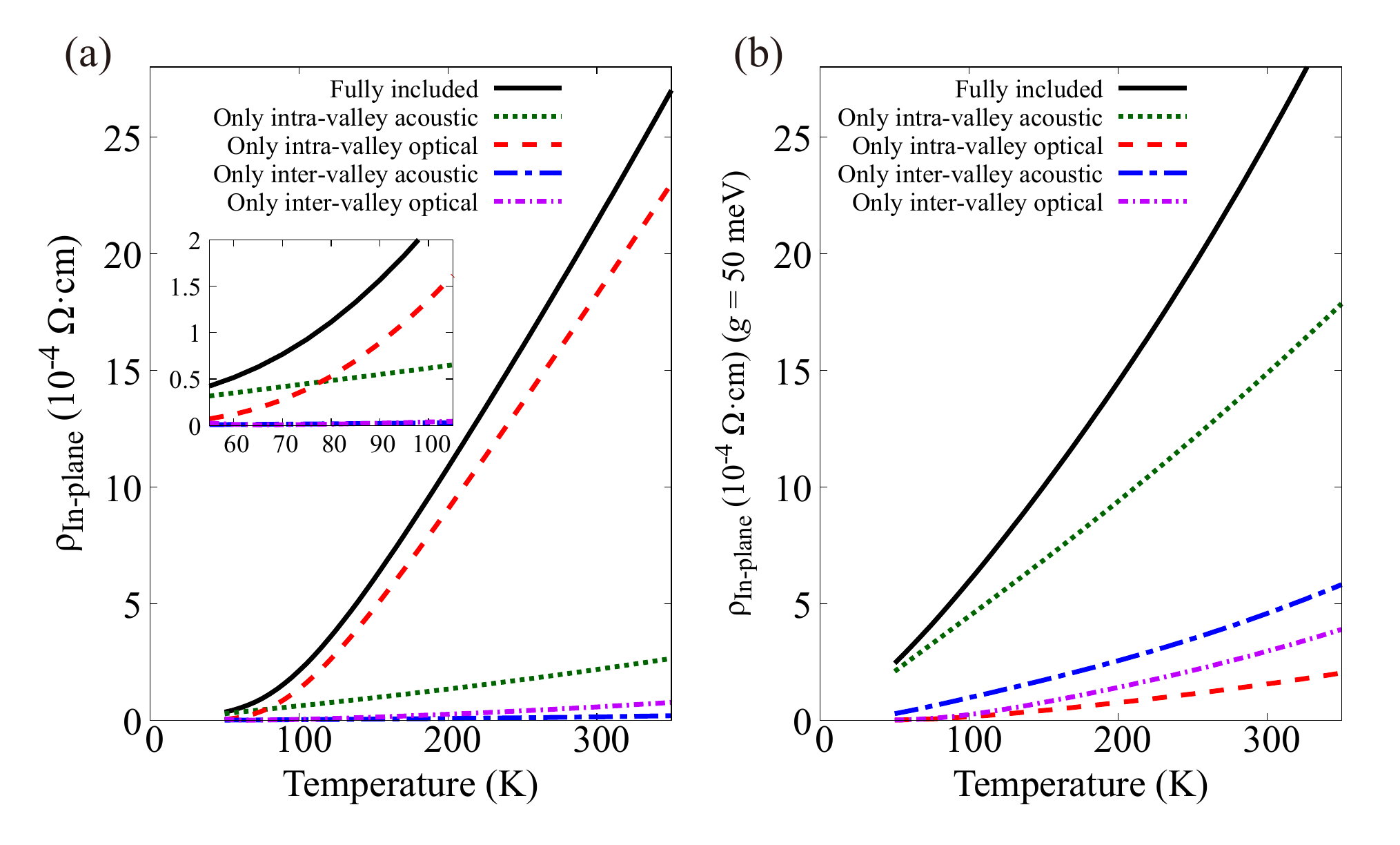}
  \caption{\label{fig:3}The temperature dependence of the in-plane electrical resistivity of $\text{Zr}_{}\text{S}_{2}$. The resistivity calculated with the electron-phonon matrix element obtained from first-principles calculation (a) and that calculated with the constant electron-phonon matrix element ($g=50\text{~meV}$) (b). We assumed the electron carrier density is $1\times 10^{20}\text{~cm}^{-3}$. The inset of panel (a) shows the closer view in the lower temperature region.}
 \end{center}
\end{figure*}
By using the relaxation times $\tau_{\bm{k}n}^{\text{(intra-ac)}}$, $\tau_{\bm{k}n}^{\text{(intra-op)}}$, $\tau_{\bm{k}n}^{\text{(inter-ac)}}$, and $\tau_{\bm{k}n}^{\text{(inter-op)}}$, instead of the total relaxation time $\tau_{\bm{k}n}$ in Eq.~\eqref{im-sigma-el-ph}, we calculated the decomposed resistivities respectively and compared them with the total resistivity which includes all the contributions as shown in Fig.~\hyperref[fig:3]{3(a)}. 
We can see that the intra-valley scattering processes by optical phonons play a significant role in the resistivity above 80~K. It is apparent that the temperature dependence of the resistivity, which includes all the contributions, is slightly curved, as shown in Fig.~\hyperref[fig:3]{3(a)}. 
(See also the logarithmic plots presented in Appendix~\ref{sec:log-plot}.) 
This behavior mainly comes from the intra-valley scattering processes by optical phonons. Similar behavior also has been experimentally observed on the resistivity of $\text{Ti}_{}\text{S}_{2}$~\cite{PhysRevLett.35.1786,PhysRevB.64.241104,Klipstein1981-hy,Koyano1986-av,Suri_2017} and that of $\text{Zr}_{}\text{Se}_{2}$~\cite{Onuki1982-dz,Zheng1989-li,Patel1998-zd}.\par
To investigate how the wave vector and frequency dependencies of the electron-phonon matrix elements $g_{n'n\nu}(\bm{k},\bm{q})$ contribute to the results, we also calculated the resistivity and the decomposed resistivities by substituting a constant value into the electron-phonon matrix elements ($g=50\text{~meV}$) as presented in Fig.~\hyperref[fig:3]{3(b)}. 
Comparing this with Fig.~\hyperref[fig:3]{3(a)}, it is clear that the contribution of the intra-valley scattering by optical phonons is vastly underestimated, compared with other kinds of scattering. 
Thus, we can conclude that the large intra-valley scattering by optical phonons shown in Fig.~\hyperref[fig:3]{3(a)} is due to a large electron-phonon coupling. 
In fact, we have seen a sharp increase of $|g_{n'n\nu}(\bm{k},\bm{q})|$ for some optical phonons near $\bm{q}=\bm{0}$ in Fig.~\ref{fig:2}. 
We can see that the temperature dependence of the total resistivity by regarding the electron-phonon matrix elements as a constant value is closer to a linear behavior than that obtained with explicit consideration of the matrix elements. Therefore, to understand the temperature dependence of the resistivity, the phonon wave vector and frequency dependence of the electron-phonon matrix elements cannot be ignored.

\subsubsection{The wavelength- and frequency- resolved resistivity}
Using the decomposition of the scattering rate as represented in Eqs.~\eqref{intra-ac}-\eqref{inter-op}, the temperature dependence of each contribution was obtained, and it was found that the intra-valley scattering makes a dominant contribution to resistivity in a wide temperature range. However, since we took partial summations over phonon branches and wavenumber vectors, it is difficult to identify the wavenumber vectors of phonons that scatter the electrons strongly just from the results shown above. 
We thus calculated the wavelength- and frequency-resolved resistivity as follows:
\begin{align}
    &\rho(q_{\text{cutoff}},\omega_{\text{cutoff}})\notag\\
    =&\left[\frac{e^2}{\Omega}\sum_{\bm{k},n} \tau_{\bm{k}n}(q_{\text{cutoff}},\omega_{\text{cutoff}})\, v^2_{\bm{k}n} \left(-\frac{\partial f_{\bm{k}n}}{\partial \varepsilon}\right)\right]^{-1},\label{rho-q-w}
\end{align}
using the following relaxation time where only the phonon modes satisfying $|\bm{q}|<q_{\text{cutoff}}$ and $\omega_{\bm{q}\nu}< \omega_{\text{cutoff}}$ are taken into account,
\begin{widetext}
\begin{align}
  \tau_{\bm{k}n}(q_{\text{cutoff}},\omega_{\text{cutoff}})
=&\biggl[\frac{2\pi}{N_{\text{p}}\hbar} \sum_{n',\bm{q},\nu}|g_{n'n\nu}(\bm{k},\bm{q})|^2
\Bigl[W_{n'n\nu}^{(-)}(\bm{k},\bm{q})+W_{n'n\nu}^{(+)}(\bm{k},\bm{q})\Bigr]
\theta (q_{\text{cutoff}}-|\bm{q}|)\theta (\omega_{\text{cutoff}}-\omega_{\bm{q}\nu})\biggl]^{-1} \label{tau-q-w}
\end{align}
\end{widetext}
\begin{figure*}
  \begin{center}
    \includegraphics[width=16cm]{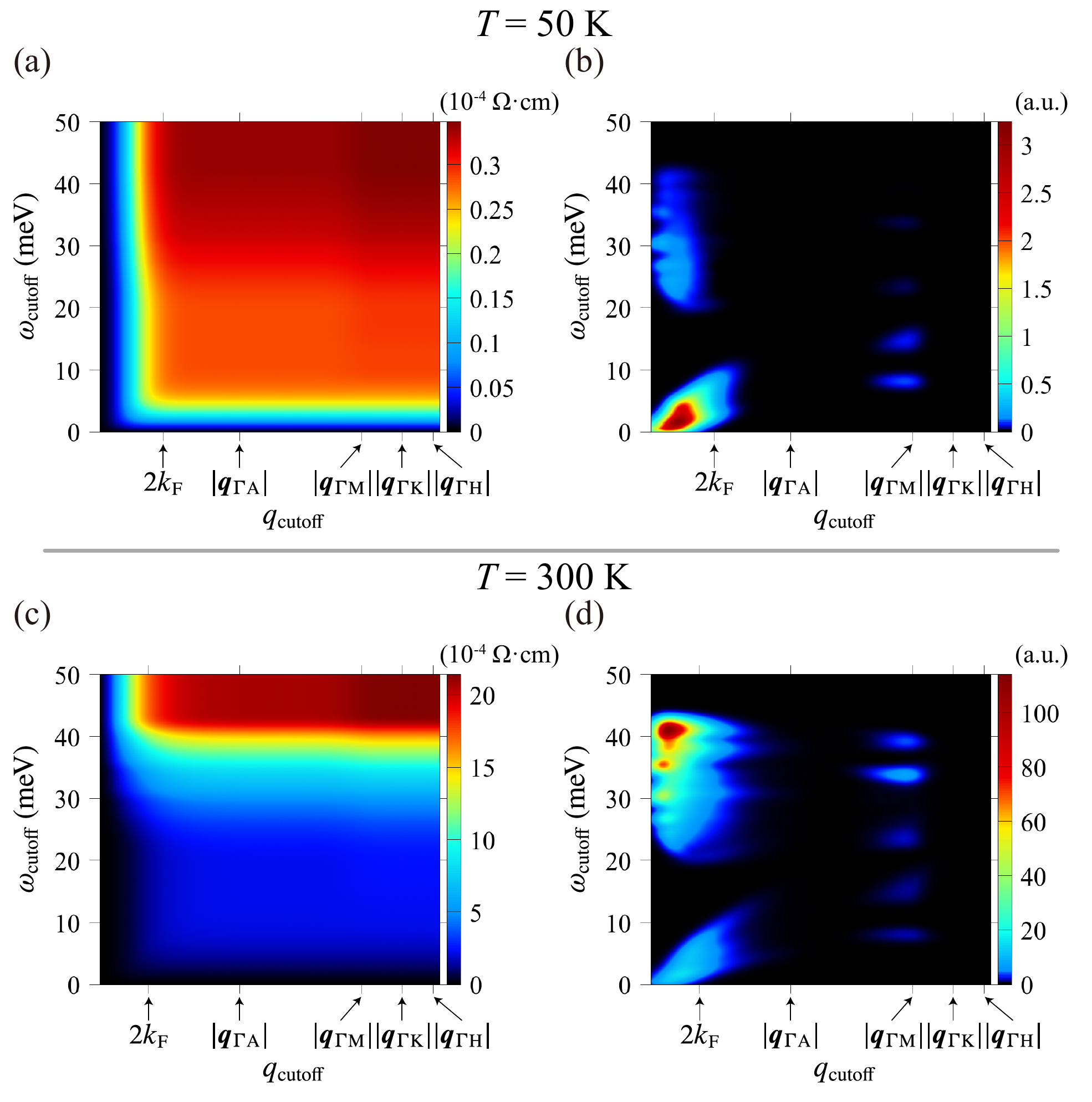}
    \caption{\label{fig:4}(a), (c) The cutoff dependences of the resolved electrical resistivities $\rho(q_{\text{cutoff}},\omega_{\text{cutoff}})$ of $\text{Zr}_{}\text{S}_{2}$ and (b), (d) those of the derivatives of them $\partial^2 \rho(q_{\text{cutoff}},\omega_{\text{cutoff}})/\partial q_{\text{cutoff}} \partial \omega_{\text{cutoff}}$ with the electron carrier density of $1\times 10^{20}\text{~cm}^{-3}$. Panels (a) and (b) show the cutoff dependences at 50~K, and other panels at 300~K. $k_{\text{F}}$ is defined in the main text.}
  \end{center}
\end{figure*}
where $\theta (x)$ is the Heaviside step function, whose value is $1$ if $x>0$ and $0$ otherwise~\footnote{In actual calculations, we replaced the step function $\theta (x)$ by $[1+\text{erf}(x)]/2$, where $\text{erf}(x)$ is the Gauss error function.}. 
We note that $\rho(q_{\text{cutoff}},\omega_{\text{cutoff}})$ is a monotonically increasing function for $q_{\text{cutoff}}$ and $\omega_{\text{cutoff}}$; it is equal to the original $\rho$ if $q_{\text{cutoff}}$ is greater than the maximum value of the norm of the phonon vector $|\bm{q}|$, which corresponds to the distance between $\Gamma$ and H points $|\bm{q}_{\Gamma\text{H}}|$ in this case, and $\omega_{\text{cutoff}}$ is greater than the maximum value of the phonon frequency $\omega_{\bm{q}\nu}$. One can regard that the phonon states corresponding to $(q_{\text{cutoff}},\omega_{\text{cutoff}})$ where $\rho(q_{\text{cutoff}},\omega_{\text{cutoff}})$ changes abruptly make a significant contribution to the electrical resistivity. We thus calculated $\rho(q_{\text{cutoff}},\omega_{\text{cutoff}})$ and the derivative of $\rho(q_{\text{cutoff}},\omega_{\text{cutoff}})$, namely $\partial^2 \rho(q_{\text{cutoff}},\omega_{\text{cutoff}})/\partial q_{\text{cutoff}} \partial \omega_{\text{cutoff}}$ for $T=50\text{~K}$ and $T=300\text{~K}$ as shown in Fig.~\ref{fig:4}. The derivative $\partial^2 \rho(q_{\text{cutoff}},\omega_{\text{cutoff}})/\partial q_{\text{cutoff}} \partial \omega_{\text{cutoff}}$ was calculated by using the finite-difference method. \par
\begin{figure*}
  \begin{center}
    \includegraphics[width=16cm]{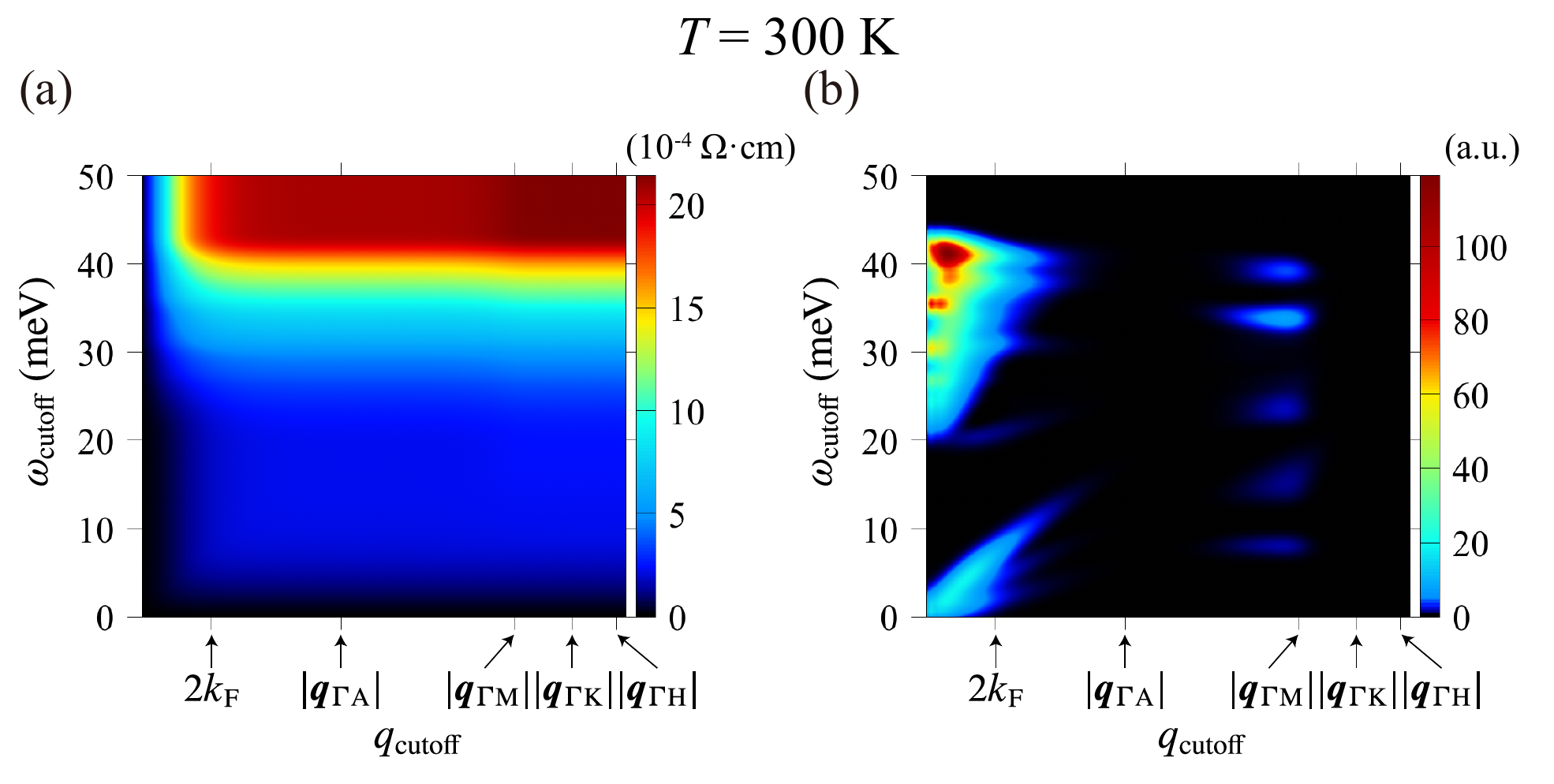}
    \caption{\label{fig:5}(a) The cutoff dependence of the resolved electrical resistivity $\rho(q_{\text{cutoff}},\omega_{\text{cutoff}})$ of $\text{Zr}_{}\text{S}_{2}$ and (b) that of the derivative of it $\partial^2 \rho(q_{\text{cutoff}},\omega_{\text{cutoff}})/\partial q_{\text{cutoff}} \partial \omega_{\text{cutoff}}$ at $300\text{~K}$ with the electron carrier density of $1\times 10^{20}\text{~cm}^{-3}$. The wavelength cutoff $q_{\text{cutoff}}$ is applied to $(q_x^2+q_y^2)^{1/2}$ instead of $|\bm{q}|=(q_x^2+q_y^2+q_z^2)^{1/2}$. $k_{\text{F}}$ is defined in the main text.}
  \end{center}
\end{figure*}
Figure~\ref{fig:4} reveals that the contributions to the resistivity are large within the radius 2$k_{\text{F}}$ from the $\Gamma$ point, where $k_{\text{F}}$ denotes the length of the semimajor axis of the Fermi surface along the A-L line. 
Thus, the strong intensity of $\partial^2 \rho(q_{\text{cutoff}},\omega_{\text{cutoff}})/\partial q_{\text{cutoff}} \partial \omega_{\text{cutoff}}$ for $q_{\text{cutoff}} < 2k_{\text{F}}$ in Figs.~\hyperref[fig:4]{4(b)} and \hyperref[fig:4]{4(d)} is consistent with the strong intra-valley scattering presented in Fig.~\hyperref[fig:3]{3(a)}. 
While the acoustic phonons have dominant contribution at 50~K as shown in Fig.~\hyperref[fig:4]{4(b)}, the contribution of optical phonons becomes dominant at 300~K as shown in Fig.~\hyperref[fig:4]{4(d)}. 
This switching can be naturally understood by the increased occupation of high-frequency phonons at high temperature. 
We note that, however, as shown in Fig.~\ref{fig:3}, this switching takes place at around 80~K, which is much lower than the optical phonon frequencies $\sim$30--40~meV. This is because some optical phonons bring about strong electron-phonon coupling as we shall discuss in the next paragraph. 
On the other hand, the inter-valley scattering is of less importance because less phonons occupy the states near the $\bm{q}_{\Gamma \text{M}}$ points than the states of acoustic phonons at $\bm{q}\sim \bm{0}$, and also the electron-phonon matrix elements coupled with phonons near the $\bm{q}_{\Gamma \text{M}}$ points are not significant.\par 

As mentioned above, the electron-phonon coupling $|g_{n'n\nu}(\bm{k},\bm{q})|$ with some optical phonons becomes quite large around $\bm{q}=\bm{0}$. 
In particular, the coupling with polar LO ($E_{u}$ and $A_{2u}$) modes diverges in the long wavelength limit ($\bm{q}\rightarrow\bm{0}$) as shown in Fig.~\hyperref[fig:2]{2(d)}. 
Several theoretical studies have also pointed out that the interaction between polar LO phonons and electrons, which was first investigated by Fr\"{o}hlich~\cite{Frohlich1954-zf}, plays an important role in the electronic transport also in other polar materials such as bulk $\text{Ga}_{}\text{As}_{}$~\cite{PhysRevB.94.201201,PhysRevB.95.075206,PhysRevB.97.045201}, mono- or multi-layer $\text{In}_{}\text{Se}_{}$~\cite{Chang2019-ii,Shi2020-sf}, and mono-layer $\text{Mo}_{}\text{S}_{2}$~\cite{Kaasbjerg2012-za}. 
Some previous studies also have indicated that the homopolar scattering induced by $A_{1g}$ phonons whose eigenmode corresponds to the vibration of chalcogen layers in counter-phase in the direction vertical to the layer plane (so-called ``Fivaz''-mode phonons~\cite{PhysRev.163.743}) is the primary scattering mechanism that determines the resistivity of $\text{Ti}_{}\text{S}_{2}$~\cite{Wilson1977-mt,Wilson1978-zs,PhysRevB.24.1691,Onuki1982-dz} or $\text{Zr}_{}\text{Se}_{2}$~\cite{Onuki1982-dz,Zheng1989-li}. 
Actually, in our calculations also, it can be considered that the $A_{1g}$ mode phonons contribute to the electrical resistivity to no small degree since the electron-phonon matrix elements coming from $A_{1g}$ phonons are quite large ($>100\text{~meV}$) around the $\Gamma$ point although they do not diverge, as shown in Figs.~\hyperref[fig:2]{2(b)} and \hyperref[fig:2]{2(c)}.\par
Here we discuss the non-linear temperature dependence of the resistivity shown in Fig.~\hyperref[fig:3]{3(a)}. 
The Bose distribution $n_{\bm{q}\nu}=(e^{\beta\hbar\omega_{\bm{q}\nu}}-1)^{-1}$ in Eqs. \eqref{im-sigma-el-ph}-\eqref{W_kq} can be approximated as $(k_{\text{B}}/\hbar\omega_{\bm{q}\nu})T$ if the temperature is higher than the phonon frequency ($\hbar\omega_{\bm{q}\nu}<k_{\text{B}}T$), and thus the scattering rate and the resistivity give a linear temperature dependence.
When acoustic phonons mainly contribute to the scattering rate, the resistivity shows a linear temperature dependence except in an extremely low-temperature range. The linear temperature dependence of the resistivity is often observed in materials whose scattering can be considered to come from acoustic phonons. On the other hand, when optical phonons have a significant contribution, 
since their frequencies are high, the linear approximation to the Bose factor is not valid even at moderate temperatures, and hence the resistivity can exhibit a non-linear temperature dependence. 
In fact, in the case of $\text{Zr}_{}\text{S}_{2}$, above 80~K (even though the temperature is still several times lower than the frequencies of the $E_{u}$ and $A_{2u}$ LO phonons and the $A_{1g}$ optical phonons), the optical phonons already play a significant role, and the non-linear temperature dependence is observed in a wider temperature range. 
The present analysis may provide (at least partial) explanation for the non-linear temperature dependence on the resistivity of $\text{Ti}_{}\text{S}_{2}$~\cite{PhysRevLett.35.1786,PhysRevB.64.241104,Klipstein1981-hy,Koyano1986-av,Suri_2017} or that of $\text{Zr}_{}\text{Se}_{2}$~\cite{Onuki1982-dz,Zheng1989-li,Patel1998-zd}.\par
\begin{figure*}
  \begin{center}
    \includegraphics[width=16cm]{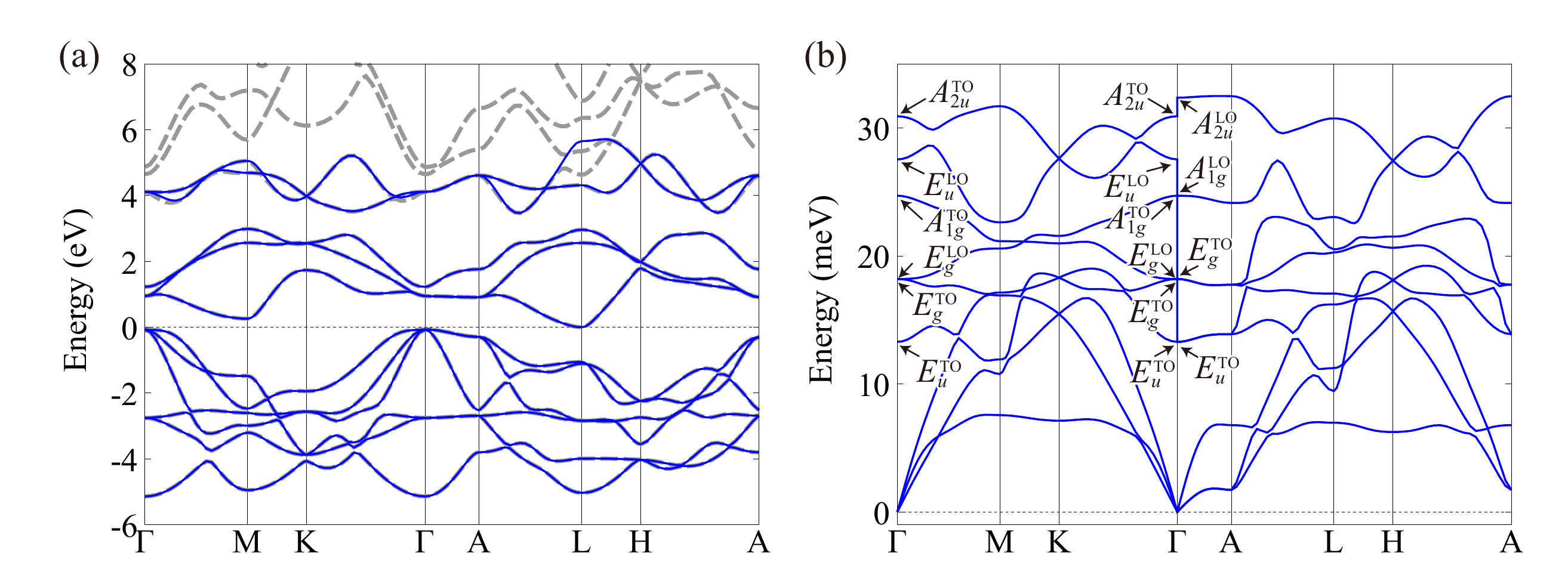}
    \caption{\label{fig:6}(a) Electronic band structures of $\text{Zr}_{}\text{Se}_{2}$ obtained from first-principles calculation (dashed gray lines) and the effective model constructed by Wannier orbitals (solid blue lines). The phonon dispersion is shown in panel (b). The optical modes at the $\Gamma$ point are distinguished by Mulliken symbols~\cite{Mulliken1955-rx}.}
 \end{center}
\end{figure*}
We note that the RTA, which is symbolized by Eq.~\eqref{im-sigma-el-ph} and adopted in this study, can overestimate the forward scattering effects by phonons at $\bm{q}\sim \bm{0}$ (and hence the electrical resistivity). This is because when the group velocity of the electrons changes little by the scattering process, it should not significantly affect the electrical current, while this effect is not taken into account within the RTA. This problem can become significant when the size of the Fermi surfaces, limiting the size of $\bm{q}$ required to change the direction of the electron group velocity, is much larger than wave vectors of phonons that mainly scatter the electrons. However, in our calculation, the scale of scattering vectors is the same length as the size of the Fermi surface ($\sim 2k_{\text{F}}$) as shown in Fig.~\ref{fig:4}. 
Therefore, we can regard that the velocity of the electrons changes its direction by intra-valley scattering.\par
Since we discuss the in-plane electrical resistivity in the layered compound, we also calculated the wavelength and frequency-resolved resistivity
$\tilde{\rho}(q_{\text{cutoff}},\omega_{\text{cutoff}})$ with the following relaxation time: 
\begin{widetext}
\begin{align}
  \tilde{\tau}_{\bm{k}n}(q_{\text{cutoff}},\omega_{\text{cutoff}})
=&\biggl[\frac{2\pi}{N_{\text{p}}\hbar} \sum_{n',\bm{q},\nu}|g_{n'n\nu}(\bm{k},\bm{q})|^2
\Bigl[W_{n'n\nu}^{(-)}(\bm{k},\bm{q})+W_{n'n\nu}^{(+)}(\bm{k},\bm{q})\Bigr]
\theta \left(q_{\text{cutoff}}-\sqrt{q_x^2+q_y^2}\right)\theta(\omega_{\text{cutoff}}-\omega_{\bm{q}\nu})\biggl]^{-1}
\end{align}
\end{widetext}
where all the $q_z$ are considered but the $q_{\text{cutoff}}$ is applied to $\sqrt{q_x^2+q_y^2}$. 
Comparing Figs.~\hyperref[fig:5]{5(a)} and \hyperref[fig:5]{5(b)} to Figs.~\hyperref[fig:4]{4(c)} and \hyperref[fig:4]{4(d)}, the results turned out to be almost the same as those calculated with applying $q_{\text{cutoff}}$ to $|\bm{q}|=\sqrt{q_x^2+q_y^2+q_z^2}$. 
Therefore, we can conclude that it is acceptable to compare $|\bm{q}|$ with $|\bm{q}_{\Gamma\text{M}}|$ and $k_{\text{F}}$ as done in the above discussion, while the $\bm{q}_{\Gamma\text{M}}$ and $k_{\text{F}}$ are defined in the $q_xq_y$($k_xk_y$)-plane.

\subsection{\texorpdfstring{$\text{Zr}_{}\text{Se}_{2}$}{ZrSe2}\label{subsec:ZrSe2}}
We also investigated the electrical resistivity of $\text{Zr}_{}\text{Se}_{2}$, which is one of the TMDCs whose electrical resistivity has been measured in previous studies~\cite{Onuki1982-dz,Zheng1989-li,Patel1998-zd}. 
Figure~\ref{fig:6} presents the calculated electron and phonon band structures of $\text{Zr}_{}\text{Se}_{2}$. Similar to $\text{Zr}_{}\text{S}_{2}$, there also exist three Fermi pockets of electrons in the 1st Brillouin zone. As shown in Fig.~\hyperref[fig:6]{6(b)}, the frequencies optical phonons of $\text{Zr}_{}\text{Se}_{2}$ are lower than those of $\text{Zr}_{}\text{S}_{2}$.
\subsubsection{Analysis of the electrical resistivity}
\begin{figure}
  \begin{center}
    \includegraphics[width=8.2cm]{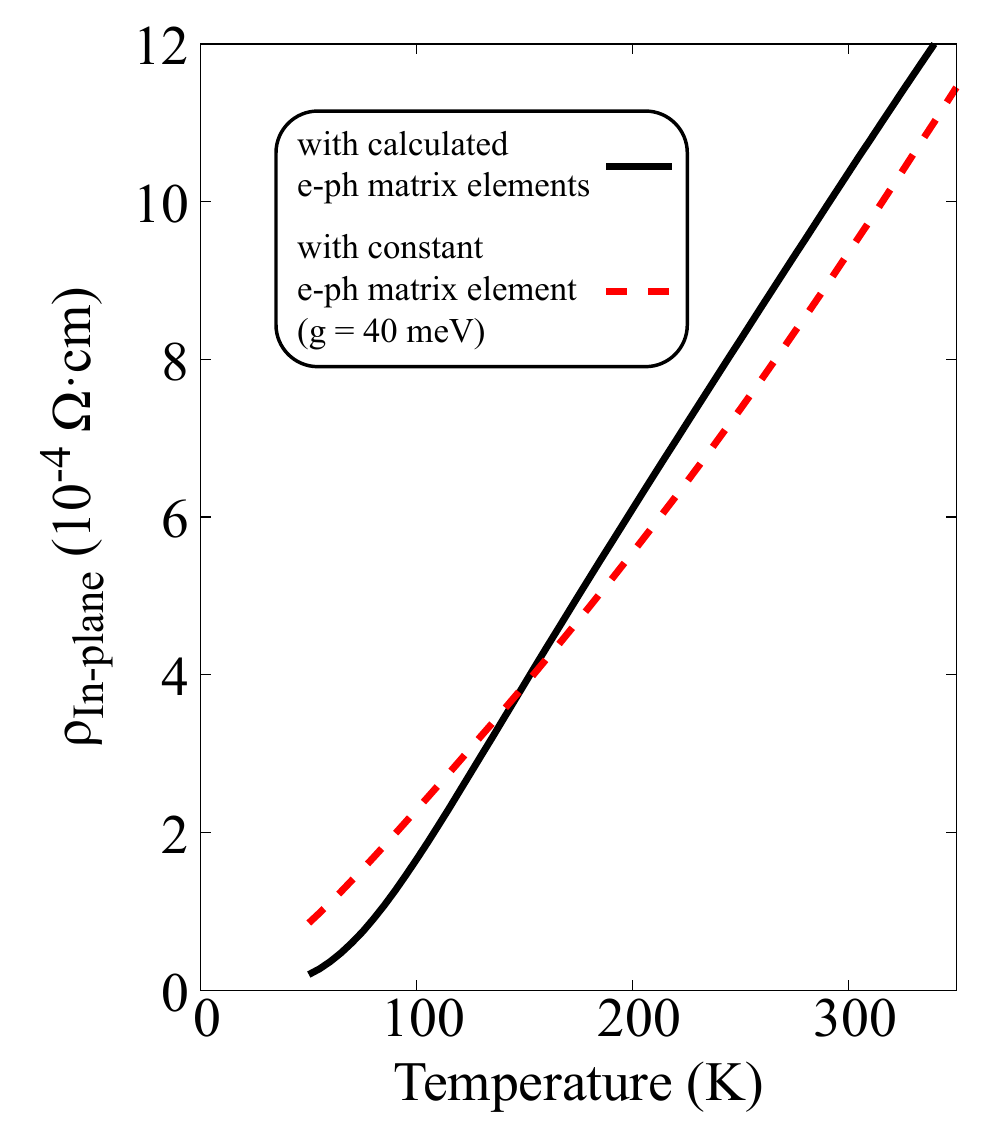}
    \caption{\label{fig:7}The temperature dependence of the in-plane electrical resistivity of $\text{Zr}_{}\text{Se}_{2}$. The resistivity calculated with the electron-phonon matrix element obtained from first-principles calculation (solid black line) and that calculated with the constant electron-phonon matrix element ($g=40\text{~meV}$) (dashed red line). We assumed the electron carrier density is $1\times 10^{20}\text{~cm}^{-3}$.}
  \end{center}
\end{figure}
To see the effect of the atomic substitution from sulfur atoms to selenium atoms on the resistivity, we calculated the in-plane resistivity of n-type $\text{Zr}_{}\text{Se}_{2}$, which is determined from the electron scattering by phonons. 
For comparison, we also calculated the electrical resistivity using a constant value for the electron-phonon matrix element ($g=40\text{~meV}$) instead of the calculated matrix elements $g_{n'n\nu}(\bm{k},\bm{q})$. 
Note that the result obtained from the calculated matrix elements is roughly consistent with the experimental value of resistivity, which is $1.25\times10^{-2}~\Omega\cdot\text{cm}$ for the Hall carrier concentration of $3.97\times 10^{19}\text{~cm}^{-3}$ at room temperature, reported by {\=O}nuki \textit{et al.}~\cite{Onuki1982-dz} 
As shown in Fig~\ref{fig:7}, the non-linear behavior on the temperature dependence of the resistivity is weakened compared with the dependence in $\text{Zr}_{}\text{S}_{2}$ (see also Appendix~\ref{sec:log-plot}); the temperature dependence appears to be almost linear in a wide temperature range. 

\begin{figure*}
  \begin{center}
    \includegraphics[width=16cm]{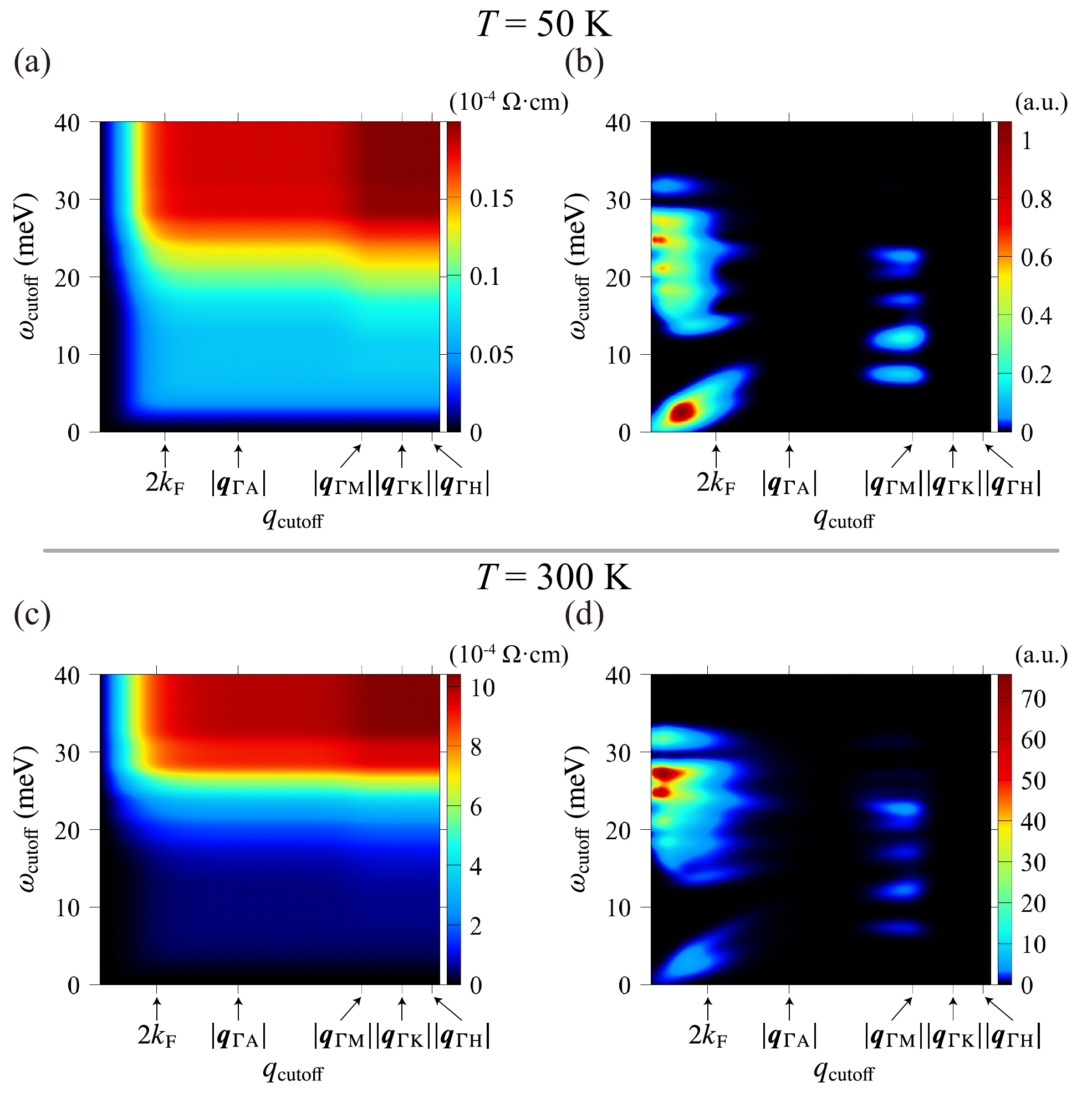}
    \caption{\label{fig:8}(a), (c) The cutoff dependences of the resolved electrical resistivities $\rho(q_{\text{cutoff}},\omega_{\text{cutoff}})$ of $\text{Zr}_{}\text{Se}_{2}$ and (b), (d) those of the derivatives of them $\partial^2 \rho(q_{\text{cutoff}},\omega_{\text{cutoff}})/\partial q_{\text{cutoff}} \partial \omega_{\text{cutoff}}$ with the electron carrier density of $1\times 10^{20}\text{~cm}^{-3}$. Panels (a) and (b) show the cutoff dependences at 50~K, and other panels at 300~K. The definition of $k_{\text{F}}$ is the same as that of $\text{Zr}_{}\text{S}_{2}$, but the value was determined from the size of the Fermi surface of $\text{Zr}_{}\text{Se}_{2}$.}
  \end{center}
\end{figure*}
One may expect that a similar analysis as in the case of $\text{Zr}_{}\text{S}_{2}$ can be performed, but the scattering rate of $\text{Zr}_{}\text{Se}_{2}$ cannot be decomposed into the contributions coming from acoustic and optical phonons as represented in Eqs.~\eqref{intra-ac}-\eqref{inter-op}, because the phonon branches of acoustic phonons are entangled with those of optical phonons, as shown in Fig.~\hyperref[fig:6]{6(b)}. 
To understand the difference in the temperature dependence of the resistivity between $\text{Zr}_{}\text{S}_{2}$ and $\text{Zr}_{}\text{Se}_{2}$, we calculated the wavelength- and frequency-resolved resistivity given by Eqs.~\eqref{rho-q-w} and \eqref{tau-q-w} for $50\text{~K}$ and $300\text{~K}$ shown by Fig~\ref{fig:8}. First, the difference in the temperature dependence can be understood as follows. The reason for the linear-like temperature dependence in $\text{Zr}_{}\text{Se}_{2}$ is that the frequencies of optical phonons decrease compared to those of $\text{Zr}_{}\text{S}_{2}$, and the Bose factor $(e^{\beta\hbar\omega_{\bm{q}\nu}}-1)^{-1}$ can be regarded as $(k_{\text{B}}/\hbar\omega_{\bm{q}\nu})T$ in a wider temperature range. 
Here, we should understand the reason why the value of resistivity is lower than that of $\text{Zr}_{}\text{S}_{2}$. Since the frequencies of the optical phonons become lower, and hence the states of them are more occupied, one would expect that the resistivity becomes higher. However, the calculated results in Figs.~\ref{fig:3} and \ref{fig:8} show the opposite trend. 
One of the possible reasons is that the electron-phonon matrix elements coupled with the polar LO phonons in $\text{Zr}_{}\text{Se}_{2}$ at $\bm{q}\rightarrow\bm{0}$ are smaller than those in $\text{Zr}_{}\text{S}_{2}$. The divergent behavior of the electron-phonon matrix elements for $\bm{q}\rightarrow\bm{0}$ depends on the electrostatic potential, which is screened by the electronic permittivity. The in-plane components of dielectric constants $\epsilon_{\infty}$ obtained from the DFPT calculations are $11.99$ for $\text{Zr}_{}\text{S}_{2}$ and $18.39$ for $\text{Zr}_{}\text{Se}_{2}$; the cross-plane components are $5.93$ and $9.29$, respectively. Therefore, it can be considered that the resistivity of $\text{Zr}_{}\text{Se}_{2}$ is lower than that of $\text{Zr}_{}\text{S}_{2}$ due to the smaller electron-phonon matrix elements coming from the stronger screening effect on the electrostatic potential.

\section{Conclusion\label{sec:sum}}
In summary, we have investigated the electron-phonon scattering effect on the resistivity of $\text{Zr}_{}\text{S}_{2}$ and $\text{Zr}_{}\text{Se}_{2}$. We have found that the calculated resistivity exhibits a non-linear temperature behavior, and the tendency is stronger in $\text{Zr}_{}\text{S}_{2}$ than in $\text{Zr}_{}\text{Se}_{2}$. According to our analysis of the mode-resolved electrical resistivity, the intra-valley scattering by optical phonons mainly contributes to the resistivity at around room temperature. 
Although optical phonons are less excited than acoustic phonons, the contributions of optical phonons become more extensive than the contributions of the acoustic phonons due to the larger electron-phonon matrix elements coupled with some optical phonons in the higher temperature range. 
The inter-valley scattering is less significant than the intra-valley scattering in both $\text{Zr}_{}\text{S}_{2}$ and $\text{Zr}_{}\text{Se}_{2}$. Conversely, the inter-valley scattering can contribute largely to the resistivity if the following conditions are satisfied: (1) the temperature is much lower than the energy scale of optical phonons but the low-frequency phonons whose vector connects between electronic valleys sufficiently occupy, (2) the electron-phonon matrix elements coupled with optical phonons are not large even at $\bm{q}\sim \bm{0}$. The difference in the temperature dependence of the resistivity between $\text{Zr}_{}\text{S}_{2}$ and $\text{Zr}_{}\text{Se}_{2}$ can be explained by the difference in the strength of the screening effect on the electron-phonon matrix elements. The present study deepens our understanding of the electron-phonon scattering effect on the transport properties not only for zirconium dichalcogenides but also for other materials possessing multiple valleys in the electronic band structure.

\acknowledgments
This study was supported by Grant-in-Aid for JSPS Fellows (Grant No. JP19J10443) and JST CREST (Grant No. JPMJCR20Q4). We appreciate fruitful discussion with T. Tadano and R. Mizuno.

\appendix
\section{Optimized crystal structures\label{sec:crystal}}
Tables~\ref{lat_ZS}, \ref{lat_ZSe} present the structural parameters of target materials obtained by our first-principles calculations. 
\begin{table}[t]
  \centering
  \caption{\label{lat_ZS} Optimized crystal structures of $\text{Zr}_{}\text{S}_{2}$.}
\begin{tabular}{|c|c|c|c|l|l|l|}\hline
  \multicolumn{7}{c}{Lattice constants: $a=3.627\text{ }\mathring{\text{A}}$, $c=5.887\text{ }\mathring{\text{A}}$}  \\ \hline
    \multicolumn{1}{|c|}{\multirow{2}{*}{Element}}&\multicolumn{1}{c|}{\multirow{2}{*}{Multiplicity}}&\multicolumn{1}{c|}{Wyckoff}&\multicolumn{1}{c|}{Site}&\multicolumn{3}{c|}{\multirow{2}{*}{Atomic positions}} \\
    &&\multicolumn{1}{c|}{letter}&\multicolumn{1}{c|}{Symmetry}&\multicolumn{3}{c|}{} \\ \hline
    $\text{Zr}$ & 1 & $a$ & $-3m.$ & 0 & 0 & 0 \\
    $\text{S}$ & 2 & $d$& $3m.$ & 1/3 & 2/3 & 0.2470 \\ \hline
\end{tabular}

  \centering
  \caption{\label{lat_ZSe} Optimized crystal structures of $\text{Zr}_{}\text{Se}_{2}$.}
\begin{tabular}{|c|c|c|c|l|l|l|}\hline
    \multicolumn{7}{c}{Lattice constants: $a=3.724\text{ }\mathring{\text{A}}$, $c=6.116\text{ }\mathring{\text{A}}$}  \\ \hline
    \multicolumn{1}{|c|}{\multirow{2}{*}{Element}}&\multicolumn{1}{c|}{\multirow{2}{*}{Multiplicity}}&\multicolumn{1}{c|}{Wyckoff}&\multicolumn{1}{c|}{Site}&\multicolumn{3}{c|}{\multirow{2}{*}{Atomic positions}} \\
    &&\multicolumn{1}{c|}{letter}&\multicolumn{1}{c|}{Symmetry}&\multicolumn{3}{c|}{} \\ \hline
    $\text{Zr}$ & 1 & $a$ & $-3m.$ & 0 & 0 & 0 \\
    $\text{Se}$ & 2 & $d$& $3m.$ & 1/3 & 2/3 & 0.2599 \\ \hline
\end{tabular}
\end{table}

\section{Normal modes of lattice vibrations\label{sec:vib-mode}}

\begin{figure}
  \begin{center}
    \includegraphics[width=8.2cm]{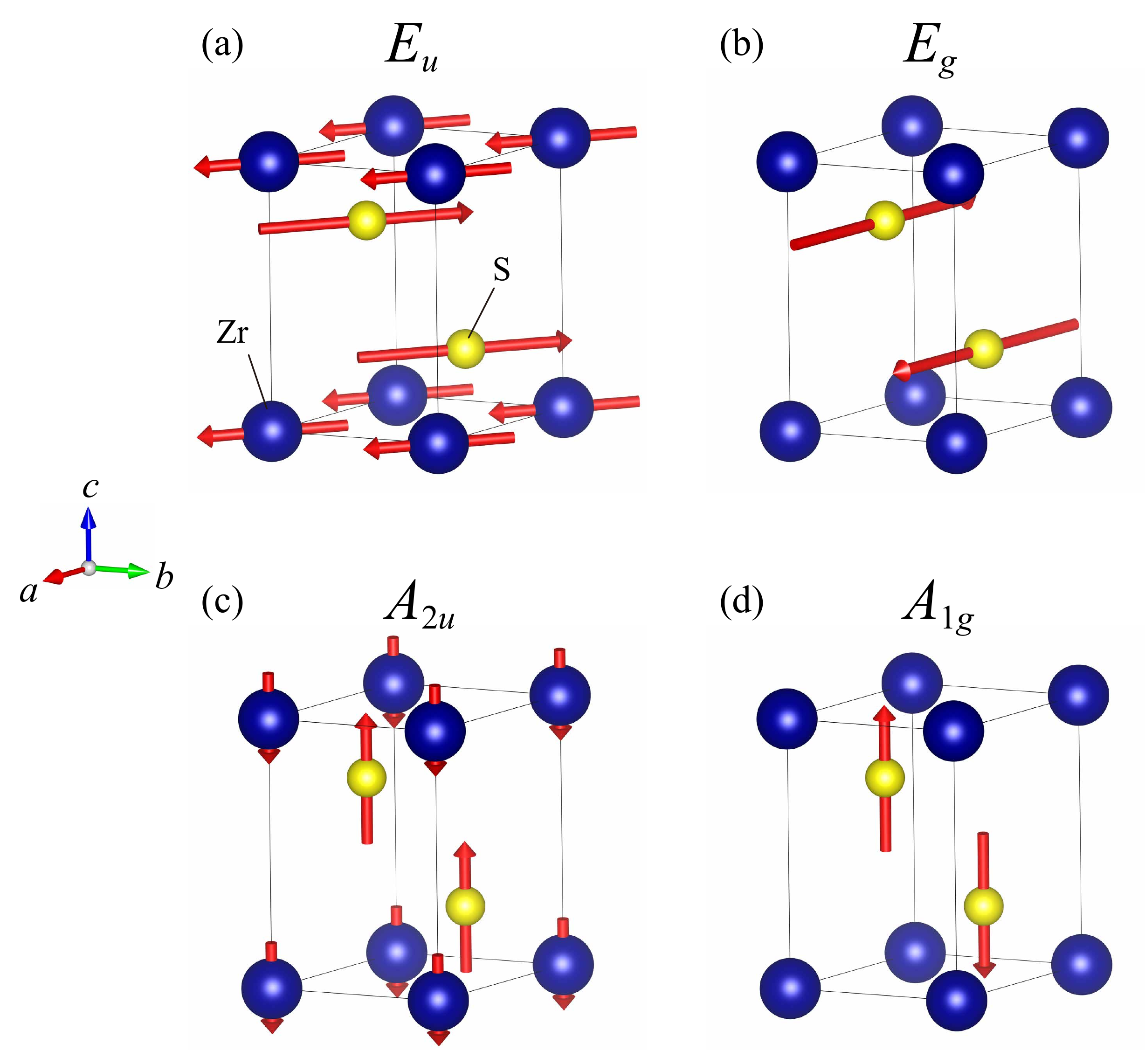}
    \caption{\label{fig:9}Atomic displacements in each optical vibration mode of $\text{Zr}_{}\text{S}_{2}$. This figure was depicted using \textsc{vesta} software\cite{Momma2011-fj}.}
  \end{center}
\end{figure}

In the long-wavelength limit $\bm{q}\rightarrow \bm{0}$, the nine normal vibration modes in $\text{Zr}_{}\text{S}_{2}$ and $\text{Zr}_{}\text{Se}_{2}$ can be expressed by using the irreducible representations of point group $D_{3d}$ as follows:
\begin{align}
\Gamma_{\text{vib}} = A_{1g} + E_{g} + 2A_{2u} + 2E_{u},
\end{align}
where $E$ modes are two-fold degenerate. The atomic displacements of each optical mode in the long-wavelength limit are presented in Fig.~\ref{fig:9}. The electron-phonon matrix elements coupled with the $E_{u}$ and $A_{2u}$ LO phonons are divergently large in the long wave-length limit, as shown in Fig.~\hyperref[fig:2]{2(d)} since they macroscopically induce an electric field.

\section{Double logarithmic plots of the electric resistivity against the temperature\label{sec:log-plot}}
\begin{figure}
  \begin{center}
    \includegraphics[width=8.2cm]{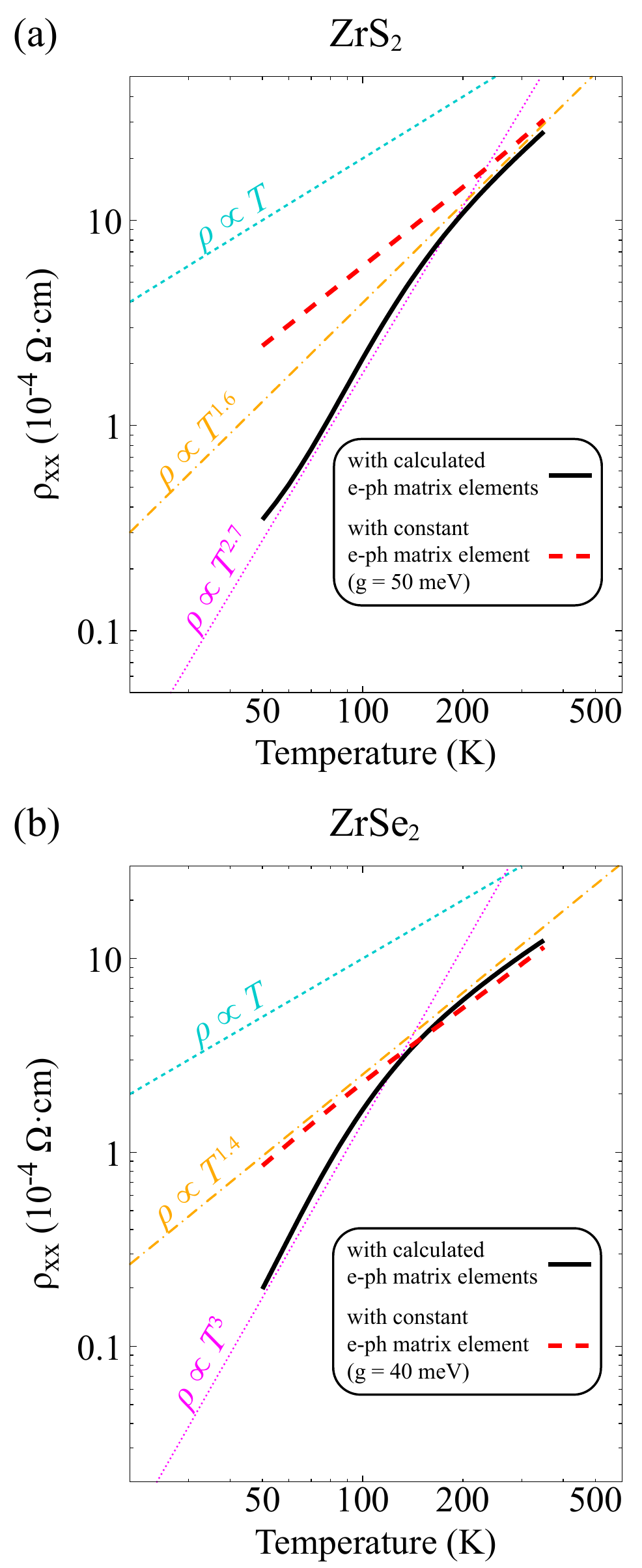}
    \caption{\label{fig:10}The temperature dependence of the in-plane electrical resistivity of (a) $\text{Zr}_{}\text{S}_{2}$ and (b) $\text{Zr}_{}\text{Se}_{2}$. The logarithmic scale is used for both the vertical and horizontal axes, and some power-law curves are also represented in each panel.  The resistivity calculated with the electron-phonon matrix element obtained from first-principles calculation (solid black line) and that calculated with the constant electron-phonon matrix element (dashed red line), which is set as 50~meV for $\text{Zr}_{}\text{S}_{2}$ and 40~meV for $\text{Zr}_{}\text{Se}_{2}$. We assumed the electron carrier density is $1\times 10^{20}\text{~cm}^{-3}$.}
  \end{center}
\end{figure}

This appendix includes additional figures to show the temperature dependence of the resistivity in both materials on a logarithmic scale (Fig.~\ref{fig:10}).

\FloatBarrier


\begin{thebibliography}{117}%
  \makeatletter
  \providecommand \@ifxundefined [1]{%
   \@ifx{#1\undefined}
  }%
  \providecommand \@ifnum [1]{%
   \ifnum #1\expandafter \@firstoftwo
   \else \expandafter \@secondoftwo
   \fi
  }%
  \providecommand \@ifx [1]{%
   \ifx #1\expandafter \@firstoftwo
   \else \expandafter \@secondoftwo
   \fi
  }%
  \providecommand \natexlab [1]{#1}%
  \providecommand \enquote  [1]{``#1''}%
  \providecommand \bibnamefont  [1]{#1}%
  \providecommand \bibfnamefont [1]{#1}%
  \providecommand \citenamefont [1]{#1}%
  \providecommand \href@noop [0]{\@secondoftwo}%
  \providecommand \href [0]{\begingroup \@sanitize@url \@href}%
  \providecommand \@href[1]{\@@startlink{#1}\@@href}%
  \providecommand \@@href[1]{\endgroup#1\@@endlink}%
  \providecommand \@sanitize@url [0]{\catcode `\\12\catcode `\$12\catcode
    `\&12\catcode `\#12\catcode `\^12\catcode `\_12\catcode `\%12\relax}%
  \providecommand \@@startlink[1]{}%
  \providecommand \@@endlink[0]{}%
  \providecommand \url  [0]{\begingroup\@sanitize@url \@url }%
  \providecommand \@url [1]{\endgroup\@href {#1}{\urlprefix }}%
  \providecommand \urlprefix  [0]{URL }%
  \providecommand \Eprint [0]{\href }%
  \providecommand \doibase [0]{https://doi.org/}%
  \providecommand \selectlanguage [0]{\@gobble}%
  \providecommand \bibinfo  [0]{\@secondoftwo}%
  \providecommand \bibfield  [0]{\@secondoftwo}%
  \providecommand \translation [1]{[#1]}%
  \providecommand \BibitemOpen [0]{}%
  \providecommand \bibitemStop [0]{}%
  \providecommand \bibitemNoStop [0]{.\EOS\space}%
  \providecommand \EOS [0]{\spacefactor3000\relax}%
  \providecommand \BibitemShut  [1]{\csname bibitem#1\endcsname}%
  \let\auto@bib@innerbib\@empty
  \bibitem [{\citenamefont {Mattheiss}(1973)}]{Mattheiss1973-ua}%
    \BibitemOpen
    \bibfield  {author} {\bibinfo {author} {\bibfnamefont {L.~F.}\ \bibnamefont
    {Mattheiss}},\ }\href {https://doi.org/10.1103/physrevb.8.3719} {\bibfield
    {journal} {\bibinfo  {journal} {Phys. Rev.}\ }\textbf {\bibinfo {volume}
    {8}},\ \bibinfo {pages} {3719} (\bibinfo {year} {1973})}\BibitemShut
    {NoStop}%
  \bibitem [{\citenamefont {Wilson}\ and\ \citenamefont
    {Yoffe}(1969)}]{Wilson1969-fe}%
    \BibitemOpen
    \bibfield  {author} {\bibinfo {author} {\bibfnamefont {J.~A.}\ \bibnamefont
    {Wilson}}\ and\ \bibinfo {author} {\bibfnamefont {A.~D.}\ \bibnamefont
    {Yoffe}},\ }\href {https://doi.org/10.1080/00018736900101307} {\bibfield
    {journal} {\bibinfo  {journal} {Adv. Phys.}\ }\textbf {\bibinfo {volume}
    {18}},\ \bibinfo {pages} {193} (\bibinfo {year} {1969})}\BibitemShut
    {NoStop}%
  \bibitem [{\citenamefont {Doni}\ and\ \citenamefont
    {Girlanda}(1986)}]{Doni1986}%
    \BibitemOpen
    \bibfield  {author} {\bibinfo {author} {\bibfnamefont {E.}~\bibnamefont
    {Doni}}\ and\ \bibinfo {author} {\bibfnamefont {R.}~\bibnamefont
    {Girlanda}},\ }in\ \href@noop {} {\emph {\bibinfo {booktitle} {Electronic
    Structure and Electronic Transitions in Layered Materials}}},\ \bibinfo
    {editor} {edited by\ \bibinfo {editor} {\bibfnamefont {V.}~\bibnamefont
    {Grasso}}}\ (\bibinfo  {publisher} {Reidel},\ \bibinfo {address}
    {Dordrecht},\ \bibinfo {year} {1986})\BibitemShut {NoStop}%
  \bibitem [{\citenamefont {Wang}\ \emph {et~al.}(2012)\citenamefont {Wang},
    \citenamefont {Kalantar-Zadeh}, \citenamefont {Kis}, \citenamefont
    {Coleman},\ and\ \citenamefont {Strano}}]{Wang2012-iv}%
    \BibitemOpen
    \bibfield  {author} {\bibinfo {author} {\bibfnamefont {Q.~H.}\ \bibnamefont
    {Wang}}, \bibinfo {author} {\bibfnamefont {K.}~\bibnamefont
    {Kalantar-Zadeh}}, \bibinfo {author} {\bibfnamefont {A.}~\bibnamefont {Kis}},
    \bibinfo {author} {\bibfnamefont {J.~N.}\ \bibnamefont {Coleman}},\ and\
    \bibinfo {author} {\bibfnamefont {M.~S.}\ \bibnamefont {Strano}},\ }\href
    {https://doi.org/10.1038/nnano.2012.193} {\bibfield  {journal} {\bibinfo
    {journal} {Nat. Nanotechnol.}\ }\textbf {\bibinfo {volume} {7}},\ \bibinfo
    {pages} {699} (\bibinfo {year} {2012})}\BibitemShut {NoStop}%
  \bibitem [{\citenamefont {Taniguchi}\ \emph {et~al.}(2012)\citenamefont
    {Taniguchi}, \citenamefont {Matsumoto}, \citenamefont {Shimotani},\ and\
    \citenamefont {Takagi}}]{Taniguchi2012-wt}%
    \BibitemOpen
    \bibfield  {author} {\bibinfo {author} {\bibfnamefont {K.}~\bibnamefont
    {Taniguchi}}, \bibinfo {author} {\bibfnamefont {A.}~\bibnamefont
    {Matsumoto}}, \bibinfo {author} {\bibfnamefont {H.}~\bibnamefont
    {Shimotani}},\ and\ \bibinfo {author} {\bibfnamefont {H.}~\bibnamefont
    {Takagi}},\ }\href {https://doi.org/10.1063/1.4740268} {\bibfield  {journal}
    {\bibinfo  {journal} {Appl. Phys. Lett.}\ }\textbf {\bibinfo {volume}
    {101}},\ \bibinfo {pages} {042603} (\bibinfo {year} {2012})}\BibitemShut
    {NoStop}%
  \bibitem [{\citenamefont {Ye}\ \emph {et~al.}(2012)\citenamefont {Ye},
    \citenamefont {Zhang}, \citenamefont {Akashi}, \citenamefont {Bahramy},
    \citenamefont {Arita},\ and\ \citenamefont {Iwasa}}]{Ye2012-cw}%
    \BibitemOpen
    \bibfield  {author} {\bibinfo {author} {\bibfnamefont {J.~T.}\ \bibnamefont
    {Ye}}, \bibinfo {author} {\bibfnamefont {Y.~J.}\ \bibnamefont {Zhang}},
    \bibinfo {author} {\bibfnamefont {R.}~\bibnamefont {Akashi}}, \bibinfo
    {author} {\bibfnamefont {M.~S.}\ \bibnamefont {Bahramy}}, \bibinfo {author}
    {\bibfnamefont {R.}~\bibnamefont {Arita}},\ and\ \bibinfo {author}
    {\bibfnamefont {Y.}~\bibnamefont {Iwasa}},\ }\href
    {https://doi.org/10.1126/science.1228006} {\bibfield  {journal} {\bibinfo
    {journal} {Science}\ }\textbf {\bibinfo {volume} {338}},\ \bibinfo {pages}
    {1193} (\bibinfo {year} {2012})}\BibitemShut {NoStop}%
  \bibitem [{\citenamefont {Joe}\ \emph {et~al.}(2014)\citenamefont {Joe},
    \citenamefont {Chen}, \citenamefont {Ghaemi}, \citenamefont {Finkelstein},
    \citenamefont {de~la Pe{\~n}a}, \citenamefont {Gan}, \citenamefont {Lee},
    \citenamefont {Yuan}, \citenamefont {Geck}, \citenamefont {MacDougall},
    \citenamefont {Chiang}, \citenamefont {Cooper}, \citenamefont {Fradkin},\
    and\ \citenamefont {Abbamonte}}]{Joe2014-kx}%
    \BibitemOpen
    \bibfield  {author} {\bibinfo {author} {\bibfnamefont {Y.~I.}\ \bibnamefont
    {Joe}}, \bibinfo {author} {\bibfnamefont {X.~M.}\ \bibnamefont {Chen}},
    \bibinfo {author} {\bibfnamefont {P.}~\bibnamefont {Ghaemi}}, \bibinfo
    {author} {\bibfnamefont {K.~D.}\ \bibnamefont {Finkelstein}}, \bibinfo
    {author} {\bibfnamefont {G.~A.}\ \bibnamefont {de~la Pe{\~n}a}}, \bibinfo
    {author} {\bibfnamefont {Y.}~\bibnamefont {Gan}}, \bibinfo {author}
    {\bibfnamefont {J.~C.~T.}\ \bibnamefont {Lee}}, \bibinfo {author}
    {\bibfnamefont {S.}~\bibnamefont {Yuan}}, \bibinfo {author} {\bibfnamefont
    {J.}~\bibnamefont {Geck}}, \bibinfo {author} {\bibfnamefont {G.~J.}\
    \bibnamefont {MacDougall}}, \bibinfo {author} {\bibfnamefont {T.~C.}\
    \bibnamefont {Chiang}}, \bibinfo {author} {\bibfnamefont {S.~L.}\
    \bibnamefont {Cooper}}, \bibinfo {author} {\bibfnamefont {E.}~\bibnamefont
    {Fradkin}},\ and\ \bibinfo {author} {\bibfnamefont {P.}~\bibnamefont
    {Abbamonte}},\ }\href {https://doi.org/10.1038/nphys2935} {\bibfield
    {journal} {\bibinfo  {journal} {Nat. Phys.}\ }\textbf {\bibinfo {volume}
    {10}},\ \bibinfo {pages} {421} (\bibinfo {year} {2014})}\BibitemShut
    {NoStop}%
  \bibitem [{\citenamefont {Liu}(2017)}]{Liu2017-fd}%
    \BibitemOpen
    \bibfield  {author} {\bibinfo {author} {\bibfnamefont {C.-X.}\ \bibnamefont
    {Liu}},\ }\href {https://doi.org/10.1103/PhysRevLett.118.087001} {\bibfield
    {journal} {\bibinfo  {journal} {Phys. Rev. Lett.}\ }\textbf {\bibinfo
    {volume} {118}},\ \bibinfo {pages} {087001} (\bibinfo {year}
    {2017})}\BibitemShut {NoStop}%
  \bibitem [{\citenamefont {de~la Barrera}\ \emph {et~al.}(2018)\citenamefont
    {de~la Barrera}, \citenamefont {Sinko}, \citenamefont {Gopalan},
    \citenamefont {Sivadas}, \citenamefont {Seyler}, \citenamefont {Watanabe},
    \citenamefont {Taniguchi}, \citenamefont {Tsen}, \citenamefont {Xu},
    \citenamefont {Xiao},\ and\ \citenamefont {Hunt}}]{De_la_Barrera2018-zr}%
    \BibitemOpen
    \bibfield  {author} {\bibinfo {author} {\bibfnamefont {S.~C.}\ \bibnamefont
    {de~la Barrera}}, \bibinfo {author} {\bibfnamefont {M.~R.}\ \bibnamefont
    {Sinko}}, \bibinfo {author} {\bibfnamefont {D.~P.}\ \bibnamefont {Gopalan}},
    \bibinfo {author} {\bibfnamefont {N.}~\bibnamefont {Sivadas}}, \bibinfo
    {author} {\bibfnamefont {K.~L.}\ \bibnamefont {Seyler}}, \bibinfo {author}
    {\bibfnamefont {K.}~\bibnamefont {Watanabe}}, \bibinfo {author}
    {\bibfnamefont {T.}~\bibnamefont {Taniguchi}}, \bibinfo {author}
    {\bibfnamefont {A.~W.}\ \bibnamefont {Tsen}}, \bibinfo {author}
    {\bibfnamefont {X.}~\bibnamefont {Xu}}, \bibinfo {author} {\bibfnamefont
    {D.}~\bibnamefont {Xiao}},\ and\ \bibinfo {author} {\bibfnamefont {B.~M.}\
    \bibnamefont {Hunt}},\ }\href {https://doi.org/10.1038/s41467-018-03888-4}
    {\bibfield  {journal} {\bibinfo  {journal} {Nat. Commun.}\ }\textbf {\bibinfo
    {volume} {9}},\ \bibinfo {pages} {1427} (\bibinfo {year} {2018})}\BibitemShut
    {NoStop}%
  \bibitem [{\citenamefont {Ma}\ \emph {et~al.}(2012)\citenamefont {Ma},
    \citenamefont {Dai}, \citenamefont {Guo}, \citenamefont {Niu}, \citenamefont
    {Zhu},\ and\ \citenamefont {Huang}}]{Ma2012-vv}%
    \BibitemOpen
    \bibfield  {author} {\bibinfo {author} {\bibfnamefont {Y.}~\bibnamefont
    {Ma}}, \bibinfo {author} {\bibfnamefont {Y.}~\bibnamefont {Dai}}, \bibinfo
    {author} {\bibfnamefont {M.}~\bibnamefont {Guo}}, \bibinfo {author}
    {\bibfnamefont {C.}~\bibnamefont {Niu}}, \bibinfo {author} {\bibfnamefont
    {Y.}~\bibnamefont {Zhu}},\ and\ \bibinfo {author} {\bibfnamefont
    {B.}~\bibnamefont {Huang}},\ }\href {https://doi.org/10.1021/nn204667z}
    {\bibfield  {journal} {\bibinfo  {journal} {ACS nano}\ }\textbf {\bibinfo
    {volume} {6}},\ \bibinfo {pages} {1695} (\bibinfo {year} {2012})}\BibitemShut
    {NoStop}%
  \bibitem [{\citenamefont {Zhu}\ \emph {et~al.}(2016)\citenamefont {Zhu},
    \citenamefont {Guo}, \citenamefont {Cheng}, \citenamefont {Dai},
    \citenamefont {An}, \citenamefont {Zhao}, \citenamefont {Tian}, \citenamefont
    {Wei}, \citenamefont {Cheng~Zeng}, \citenamefont {Wu},\ and\ \citenamefont
    {Xie}}]{Zhu2016-dh}%
    \BibitemOpen
    \bibfield  {author} {\bibinfo {author} {\bibfnamefont {X.}~\bibnamefont
    {Zhu}}, \bibinfo {author} {\bibfnamefont {Y.}~\bibnamefont {Guo}}, \bibinfo
    {author} {\bibfnamefont {H.}~\bibnamefont {Cheng}}, \bibinfo {author}
    {\bibfnamefont {J.}~\bibnamefont {Dai}}, \bibinfo {author} {\bibfnamefont
    {X.}~\bibnamefont {An}}, \bibinfo {author} {\bibfnamefont {J.}~\bibnamefont
    {Zhao}}, \bibinfo {author} {\bibfnamefont {K.}~\bibnamefont {Tian}}, \bibinfo
    {author} {\bibfnamefont {S.}~\bibnamefont {Wei}}, \bibinfo {author}
    {\bibfnamefont {X.}~\bibnamefont {Cheng~Zeng}}, \bibinfo {author}
    {\bibfnamefont {C.}~\bibnamefont {Wu}},\ and\ \bibinfo {author}
    {\bibfnamefont {Y.}~\bibnamefont {Xie}},\ }\href
    {https://doi.org/10.1038/ncomms11210} {\bibfield  {journal} {\bibinfo
    {journal} {Nat. Commun.}\ }\textbf {\bibinfo {volume} {7}},\ \bibinfo {pages}
    {11210} (\bibinfo {year} {2016})}\BibitemShut {NoStop}%
  \bibitem [{\citenamefont {Xiang}\ \emph {et~al.}(2016)\citenamefont {Xiang},
    \citenamefont {Xu}, \citenamefont {Xia}, \citenamefont {Yin},\ and\
    \citenamefont {Liu}}]{Xiang2016-ln}%
    \BibitemOpen
    \bibfield  {author} {\bibinfo {author} {\bibfnamefont {H.}~\bibnamefont
    {Xiang}}, \bibinfo {author} {\bibfnamefont {B.}~\bibnamefont {Xu}}, \bibinfo
    {author} {\bibfnamefont {Y.}~\bibnamefont {Xia}}, \bibinfo {author}
    {\bibfnamefont {J.}~\bibnamefont {Yin}},\ and\ \bibinfo {author}
    {\bibfnamefont {Z.}~\bibnamefont {Liu}},\ }\href
    {https://doi.org/10.1038/srep39218} {\bibfield  {journal} {\bibinfo
    {journal} {Sci. Rep.}\ }\textbf {\bibinfo {volume} {6}},\ \bibinfo {pages}
    {39218} (\bibinfo {year} {2016})}\BibitemShut {NoStop}%
  \bibitem [{\citenamefont {Chiew}\ \emph {et~al.}(2020)\citenamefont {Chiew},
    \citenamefont {Miyata}, \citenamefont {Koyano},\ and\ \citenamefont
    {Oshima}}]{Chiew2020-ce}%
    \BibitemOpen
    \bibfield  {author} {\bibinfo {author} {\bibfnamefont {Y.~L.}\ \bibnamefont
    {Chiew}}, \bibinfo {author} {\bibfnamefont {M.}~\bibnamefont {Miyata}},
    \bibinfo {author} {\bibfnamefont {M.}~\bibnamefont {Koyano}},\ and\ \bibinfo
    {author} {\bibfnamefont {Y.}~\bibnamefont {Oshima}},\ }\href
    {https://doi.org/10.7566/JPSJ.89.074601} {\bibfield  {journal} {\bibinfo
    {journal} {J. Phys. Soc. Jpn.}\ }\textbf {\bibinfo {volume} {89}},\ \bibinfo
    {pages} {074601} (\bibinfo {year} {2020})}\BibitemShut {NoStop}%
  \bibitem [{\citenamefont {Di~Salvo}\ \emph {et~al.}(1976)\citenamefont
    {Di~Salvo}, \citenamefont {Moncton},\ and\ \citenamefont
    {Waszczak}}]{PhysRevB.14.4321}%
    \BibitemOpen
    \bibfield  {author} {\bibinfo {author} {\bibfnamefont {F.~J.}\ \bibnamefont
    {Di~Salvo}}, \bibinfo {author} {\bibfnamefont {D.~E.}\ \bibnamefont
    {Moncton}},\ and\ \bibinfo {author} {\bibfnamefont {J.~V.}\ \bibnamefont
    {Waszczak}},\ }\href {https://doi.org/10.1103/PhysRevB.14.4321} {\bibfield
    {journal} {\bibinfo  {journal} {Phys. Rev. B}\ }\textbf {\bibinfo {volume}
    {14}},\ \bibinfo {pages} {4321} (\bibinfo {year} {1976})}\BibitemShut
    {NoStop}%
  \bibitem [{\citenamefont {Zunger}\ and\ \citenamefont
    {Freeman}(1978)}]{PhysRevB.17.1839}%
    \BibitemOpen
    \bibfield  {author} {\bibinfo {author} {\bibfnamefont {A.}~\bibnamefont
    {Zunger}}\ and\ \bibinfo {author} {\bibfnamefont {A.~J.}\ \bibnamefont
    {Freeman}},\ }\href {https://doi.org/10.1103/PhysRevB.17.1839} {\bibfield
    {journal} {\bibinfo  {journal} {Phys. Rev. B}\ }\textbf {\bibinfo {volume}
    {17}},\ \bibinfo {pages} {1839} (\bibinfo {year} {1978})}\BibitemShut
    {NoStop}%
  \bibitem [{\citenamefont {Suzuki}\ \emph {et~al.}(1985)\citenamefont {Suzuki},
    \citenamefont {Yamamoto},\ and\ \citenamefont {Motizuki}}]{Suzuki1985-jc}%
    \BibitemOpen
    \bibfield  {author} {\bibinfo {author} {\bibfnamefont {N.}~\bibnamefont
    {Suzuki}}, \bibinfo {author} {\bibfnamefont {A.}~\bibnamefont {Yamamoto}},\
    and\ \bibinfo {author} {\bibfnamefont {K.}~\bibnamefont {Motizuki}},\ }\href
    {https://doi.org/10.1143/JPSJ.54.4668} {\bibfield  {journal} {\bibinfo
    {journal} {J. Phys. Soc. Jpn.}\ }\textbf {\bibinfo {volume} {54}},\ \bibinfo
    {pages} {4668} (\bibinfo {year} {1985})}\BibitemShut {NoStop}%
  \bibitem [{\citenamefont {Sugai}(1985)}]{Sugai1985-ip}%
    \BibitemOpen
    \bibfield  {author} {\bibinfo {author} {\bibfnamefont {S.}~\bibnamefont
    {Sugai}},\ }\href {https://doi.org/10.1002/pssb.2221290103} {\bibfield
    {journal} {\bibinfo  {journal} {Phys. Status Solidi B Basic Res.}\ }\textbf
    {\bibinfo {volume} {129}},\ \bibinfo {pages} {13} (\bibinfo {year}
    {1985})}\BibitemShut {NoStop}%
  \bibitem [{\citenamefont {Rossnagel}\ \emph {et~al.}(2002)\citenamefont
    {Rossnagel}, \citenamefont {Kipp},\ and\ \citenamefont
    {Skibowski}}]{PhysRevB.65.235101}%
    \BibitemOpen
    \bibfield  {author} {\bibinfo {author} {\bibfnamefont {K.}~\bibnamefont
    {Rossnagel}}, \bibinfo {author} {\bibfnamefont {L.}~\bibnamefont {Kipp}},\
    and\ \bibinfo {author} {\bibfnamefont {M.}~\bibnamefont {Skibowski}},\ }\href
    {https://doi.org/10.1103/PhysRevB.65.235101} {\bibfield  {journal} {\bibinfo
    {journal} {Phys. Rev. B}\ }\textbf {\bibinfo {volume} {65}},\ \bibinfo
    {pages} {235101} (\bibinfo {year} {2002})}\BibitemShut {NoStop}%
  \bibitem [{\citenamefont {Kidd}\ \emph {et~al.}(2002)\citenamefont {Kidd},
    \citenamefont {Miller}, \citenamefont {Chou},\ and\ \citenamefont
    {Chiang}}]{PhysRevLett.88.226402}%
    \BibitemOpen
    \bibfield  {author} {\bibinfo {author} {\bibfnamefont {T.~E.}\ \bibnamefont
    {Kidd}}, \bibinfo {author} {\bibfnamefont {T.}~\bibnamefont {Miller}},
    \bibinfo {author} {\bibfnamefont {M.~Y.}\ \bibnamefont {Chou}},\ and\
    \bibinfo {author} {\bibfnamefont {T.-C.}\ \bibnamefont {Chiang}},\ }\href
    {https://doi.org/10.1103/PhysRevLett.88.226402} {\bibfield  {journal}
    {\bibinfo  {journal} {Phys. Rev. Lett.}\ }\textbf {\bibinfo {volume} {88}},\
    \bibinfo {pages} {226402} (\bibinfo {year} {2002})}\BibitemShut {NoStop}%
  \bibitem [{\citenamefont {Clerc}\ \emph {et~al.}(2007)\citenamefont {Clerc},
    \citenamefont {Battaglia}, \citenamefont {Cercellier}, \citenamefont
    {Monney}, \citenamefont {Berger}, \citenamefont {Despont}, \citenamefont
    {Garnier},\ and\ \citenamefont {Aebi}}]{Clerc2007-aa}%
    \BibitemOpen
    \bibfield  {author} {\bibinfo {author} {\bibfnamefont {F.}~\bibnamefont
    {Clerc}}, \bibinfo {author} {\bibfnamefont {C.}~\bibnamefont {Battaglia}},
    \bibinfo {author} {\bibfnamefont {H.}~\bibnamefont {Cercellier}}, \bibinfo
    {author} {\bibfnamefont {C.}~\bibnamefont {Monney}}, \bibinfo {author}
    {\bibfnamefont {H.}~\bibnamefont {Berger}}, \bibinfo {author} {\bibfnamefont
    {L.}~\bibnamefont {Despont}}, \bibinfo {author} {\bibfnamefont {M.~G.}\
    \bibnamefont {Garnier}},\ and\ \bibinfo {author} {\bibfnamefont
    {P.}~\bibnamefont {Aebi}},\ }\href
    {https://doi.org/10.1088/0953-8984/19/35/355002} {\bibfield  {journal}
    {\bibinfo  {journal} {J. Phys. Condens. Matter}\ }\textbf {\bibinfo {volume}
    {19}},\ \bibinfo {pages} {355002} (\bibinfo {year} {2007})}\BibitemShut
    {NoStop}%
  \bibitem [{\citenamefont {Porer}\ \emph {et~al.}(2014)\citenamefont {Porer},
    \citenamefont {Leierseder}, \citenamefont {M{\'e}nard}, \citenamefont
    {Dachraoui}, \citenamefont {Mouchliadis}, \citenamefont {Perakis},
    \citenamefont {Heinzmann}, \citenamefont {Demsar}, \citenamefont
    {Rossnagel},\ and\ \citenamefont {Huber}}]{Porer2014-xb}%
    \BibitemOpen
    \bibfield  {author} {\bibinfo {author} {\bibfnamefont {M.}~\bibnamefont
    {Porer}}, \bibinfo {author} {\bibfnamefont {U.}~\bibnamefont {Leierseder}},
    \bibinfo {author} {\bibfnamefont {J.-M.}\ \bibnamefont {M{\'e}nard}},
    \bibinfo {author} {\bibfnamefont {H.}~\bibnamefont {Dachraoui}}, \bibinfo
    {author} {\bibfnamefont {L.}~\bibnamefont {Mouchliadis}}, \bibinfo {author}
    {\bibfnamefont {I.~E.}\ \bibnamefont {Perakis}}, \bibinfo {author}
    {\bibfnamefont {U.}~\bibnamefont {Heinzmann}}, \bibinfo {author}
    {\bibfnamefont {J.}~\bibnamefont {Demsar}}, \bibinfo {author} {\bibfnamefont
    {K.}~\bibnamefont {Rossnagel}},\ and\ \bibinfo {author} {\bibfnamefont
    {R.}~\bibnamefont {Huber}},\ }\href {https://doi.org/10.1038/nmat4042}
    {\bibfield  {journal} {\bibinfo  {journal} {Nat. Mater.}\ }\textbf {\bibinfo
    {volume} {13}},\ \bibinfo {pages} {857} (\bibinfo {year} {2014})}\BibitemShut
    {NoStop}%
  \bibitem [{\citenamefont {Dolui}\ and\ \citenamefont
    {Sanvito}(2016)}]{Dolui2016-ap}%
    \BibitemOpen
    \bibfield  {author} {\bibinfo {author} {\bibfnamefont {K.}~\bibnamefont
    {Dolui}}\ and\ \bibinfo {author} {\bibfnamefont {S.}~\bibnamefont
    {Sanvito}},\ }\href {https://doi.org/10.1209/0295-5075/115/47001} {\bibfield
    {journal} {\bibinfo  {journal} {Europhys. Lett.}\ }\textbf {\bibinfo {volume}
    {115}},\ \bibinfo {pages} {47001} (\bibinfo {year} {2016})}\BibitemShut
    {NoStop}%
  \bibitem [{\citenamefont {Li}\ \emph {et~al.}(2016)\citenamefont {Li},
    \citenamefont {Zhao}, \citenamefont {Liu}, \citenamefont {Ren}, \citenamefont
    {Eda},\ and\ \citenamefont {Loh}}]{Li2016-fj}%
    \BibitemOpen
    \bibfield  {author} {\bibinfo {author} {\bibfnamefont {L.~J.}\ \bibnamefont
    {Li}}, \bibinfo {author} {\bibfnamefont {W.~J.}\ \bibnamefont {Zhao}},
    \bibinfo {author} {\bibfnamefont {B.}~\bibnamefont {Liu}}, \bibinfo {author}
    {\bibfnamefont {T.~H.}\ \bibnamefont {Ren}}, \bibinfo {author} {\bibfnamefont
    {G.}~\bibnamefont {Eda}},\ and\ \bibinfo {author} {\bibfnamefont {K.~P.}\
    \bibnamefont {Loh}},\ }\href {https://doi.org/10.1063/1.4963885} {\bibfield
    {journal} {\bibinfo  {journal} {Appl. Phys. Lett.}\ }\textbf {\bibinfo
    {volume} {109}},\ \bibinfo {pages} {141902} (\bibinfo {year}
    {2016})}\BibitemShut {NoStop}%
  \bibitem [{\citenamefont {Yu}\ and\ \citenamefont
    {Wu}(2014)}]{PhysRevB.89.205303}%
    \BibitemOpen
    \bibfield  {author} {\bibinfo {author} {\bibfnamefont {T.}~\bibnamefont
    {Yu}}\ and\ \bibinfo {author} {\bibfnamefont {M.~W.}\ \bibnamefont {Wu}},\
    }\href {https://doi.org/10.1103/PhysRevB.89.205303} {\bibfield  {journal}
    {\bibinfo  {journal} {Phys. Rev. B}\ }\textbf {\bibinfo {volume} {89}},\
    \bibinfo {pages} {205303} (\bibinfo {year} {2014})}\BibitemShut {NoStop}%
  \bibitem [{\citenamefont {Mai}\ \emph {et~al.}(2014)\citenamefont {Mai},
    \citenamefont {Semenov}, \citenamefont {Barrette}, \citenamefont {Yu},
    \citenamefont {Jin}, \citenamefont {Cao}, \citenamefont {Kim},\ and\
    \citenamefont {Gundogdu}}]{PhysRevB.90.041414}%
    \BibitemOpen
    \bibfield  {author} {\bibinfo {author} {\bibfnamefont {C.}~\bibnamefont
    {Mai}}, \bibinfo {author} {\bibfnamefont {Y.~G.}\ \bibnamefont {Semenov}},
    \bibinfo {author} {\bibfnamefont {A.}~\bibnamefont {Barrette}}, \bibinfo
    {author} {\bibfnamefont {Y.}~\bibnamefont {Yu}}, \bibinfo {author}
    {\bibfnamefont {Z.}~\bibnamefont {Jin}}, \bibinfo {author} {\bibfnamefont
    {L.}~\bibnamefont {Cao}}, \bibinfo {author} {\bibfnamefont {K.~W.}\
    \bibnamefont {Kim}},\ and\ \bibinfo {author} {\bibfnamefont {K.}~\bibnamefont
    {Gundogdu}},\ }\href {https://doi.org/10.1103/PhysRevB.90.041414} {\bibfield
    {journal} {\bibinfo  {journal} {Phys. Rev. B}\ }\textbf {\bibinfo {volume}
    {90}},\ \bibinfo {pages} {041414(R)} (\bibinfo {year} {2014})}\BibitemShut
    {NoStop}%
  \bibitem [{\citenamefont {Dal~Conte}\ \emph {et~al.}(2015)\citenamefont
    {Dal~Conte}, \citenamefont {Bottegoni}, \citenamefont {Pogna}, \citenamefont
    {De~Fazio}, \citenamefont {Ambrogio}, \citenamefont {Bargigia}, \citenamefont
    {D'Andrea}, \citenamefont {Lombardo}, \citenamefont {Bruna}, \citenamefont
    {Ciccacci}, \citenamefont {Ferrari}, \citenamefont {Cerullo},\ and\
    \citenamefont {Finazzi}}]{PhysRevB.92.235425}%
    \BibitemOpen
    \bibfield  {author} {\bibinfo {author} {\bibfnamefont {S.}~\bibnamefont
    {Dal~Conte}}, \bibinfo {author} {\bibfnamefont {F.}~\bibnamefont
    {Bottegoni}}, \bibinfo {author} {\bibfnamefont {E.~A.~A.}\ \bibnamefont
    {Pogna}}, \bibinfo {author} {\bibfnamefont {D.}~\bibnamefont {De~Fazio}},
    \bibinfo {author} {\bibfnamefont {S.}~\bibnamefont {Ambrogio}}, \bibinfo
    {author} {\bibfnamefont {I.}~\bibnamefont {Bargigia}}, \bibinfo {author}
    {\bibfnamefont {C.}~\bibnamefont {D'Andrea}}, \bibinfo {author}
    {\bibfnamefont {A.}~\bibnamefont {Lombardo}}, \bibinfo {author}
    {\bibfnamefont {M.}~\bibnamefont {Bruna}}, \bibinfo {author} {\bibfnamefont
    {F.}~\bibnamefont {Ciccacci}}, \bibinfo {author} {\bibfnamefont {A.~C.}\
    \bibnamefont {Ferrari}}, \bibinfo {author} {\bibfnamefont {G.}~\bibnamefont
    {Cerullo}},\ and\ \bibinfo {author} {\bibfnamefont {M.}~\bibnamefont
    {Finazzi}},\ }\href {https://doi.org/10.1103/PhysRevB.92.235425} {\bibfield
    {journal} {\bibinfo  {journal} {Phys. Rev. B}\ }\textbf {\bibinfo {volume}
    {92}},\ \bibinfo {pages} {235425} (\bibinfo {year} {2015})}\BibitemShut
    {NoStop}%
  \bibitem [{\citenamefont {Wang}\ \emph
    {et~al.}(2016{\natexlab{a}})\citenamefont {Wang}, \citenamefont {Marie},
    \citenamefont {Liu}, \citenamefont {Amand}, \citenamefont {Robert},
    \citenamefont {Cadiz}, \citenamefont {Renucci},\ and\ \citenamefont
    {Urbaszek}}]{PhysRevLett.117.187401}%
    \BibitemOpen
    \bibfield  {author} {\bibinfo {author} {\bibfnamefont {G.}~\bibnamefont
    {Wang}}, \bibinfo {author} {\bibfnamefont {X.}~\bibnamefont {Marie}},
    \bibinfo {author} {\bibfnamefont {B.~L.}\ \bibnamefont {Liu}}, \bibinfo
    {author} {\bibfnamefont {T.}~\bibnamefont {Amand}}, \bibinfo {author}
    {\bibfnamefont {C.}~\bibnamefont {Robert}}, \bibinfo {author} {\bibfnamefont
    {F.}~\bibnamefont {Cadiz}}, \bibinfo {author} {\bibfnamefont
    {P.}~\bibnamefont {Renucci}},\ and\ \bibinfo {author} {\bibfnamefont
    {B.}~\bibnamefont {Urbaszek}},\ }\href
    {https://doi.org/10.1103/PhysRevLett.117.187401} {\bibfield  {journal}
    {\bibinfo  {journal} {Phys. Rev. Lett.}\ }\textbf {\bibinfo {volume} {117}},\
    \bibinfo {pages} {187401} (\bibinfo {year} {2016}{\natexlab{a}})}\BibitemShut
    {NoStop}%
  \bibitem [{\citenamefont {Hao}\ \emph {et~al.}(2016)\citenamefont {Hao},
    \citenamefont {Moody}, \citenamefont {Wu}, \citenamefont {Dass},
    \citenamefont {Xu}, \citenamefont {Chen}, \citenamefont {Sun}, \citenamefont
    {Li}, \citenamefont {Li}, \citenamefont {MacDonald},\ and\ \citenamefont
    {Li}}]{Hao2016-ck}%
    \BibitemOpen
    \bibfield  {author} {\bibinfo {author} {\bibfnamefont {K.}~\bibnamefont
    {Hao}}, \bibinfo {author} {\bibfnamefont {G.}~\bibnamefont {Moody}}, \bibinfo
    {author} {\bibfnamefont {F.}~\bibnamefont {Wu}}, \bibinfo {author}
    {\bibfnamefont {C.~K.}\ \bibnamefont {Dass}}, \bibinfo {author}
    {\bibfnamefont {L.}~\bibnamefont {Xu}}, \bibinfo {author} {\bibfnamefont
    {C.-H.}\ \bibnamefont {Chen}}, \bibinfo {author} {\bibfnamefont
    {L.}~\bibnamefont {Sun}}, \bibinfo {author} {\bibfnamefont {M.-Y.}\
    \bibnamefont {Li}}, \bibinfo {author} {\bibfnamefont {L.-J.}\ \bibnamefont
    {Li}}, \bibinfo {author} {\bibfnamefont {A.~H.}\ \bibnamefont {MacDonald}},\
    and\ \bibinfo {author} {\bibfnamefont {X.}~\bibnamefont {Li}},\ }\href
    {https://doi.org/10.1038/nphys3674} {\bibfield  {journal} {\bibinfo
    {journal} {Nat. Phys.}\ }\textbf {\bibinfo {volume} {12}},\ \bibinfo {pages}
    {677} (\bibinfo {year} {2016})}\BibitemShut {NoStop}%
  \bibitem [{\citenamefont {Rivera}\ \emph {et~al.}(2018)\citenamefont {Rivera},
    \citenamefont {Yu}, \citenamefont {Seyler}, \citenamefont {Wilson},
    \citenamefont {Yao},\ and\ \citenamefont {Xu}}]{Rivera2018-sk}%
    \BibitemOpen
    \bibfield  {author} {\bibinfo {author} {\bibfnamefont {P.}~\bibnamefont
    {Rivera}}, \bibinfo {author} {\bibfnamefont {H.}~\bibnamefont {Yu}}, \bibinfo
    {author} {\bibfnamefont {K.~L.}\ \bibnamefont {Seyler}}, \bibinfo {author}
    {\bibfnamefont {N.~P.}\ \bibnamefont {Wilson}}, \bibinfo {author}
    {\bibfnamefont {W.}~\bibnamefont {Yao}},\ and\ \bibinfo {author}
    {\bibfnamefont {X.}~\bibnamefont {Xu}},\ }\href
    {https://doi.org/10.1038/s41565-018-0193-0} {\bibfield  {journal} {\bibinfo
    {journal} {Nat. Nanotechnol.}\ }\textbf {\bibinfo {volume} {13}},\ \bibinfo
    {pages} {1004} (\bibinfo {year} {2018})}\BibitemShut {NoStop}%
  \bibitem [{\citenamefont {Wang}\ \emph {et~al.}(2018)\citenamefont {Wang},
    \citenamefont {Chernikov}, \citenamefont {Glazov}, \citenamefont {Heinz},
    \citenamefont {Marie}, \citenamefont {Amand},\ and\ \citenamefont
    {Urbaszek}}]{RevModPhys.90.021001}%
    \BibitemOpen
    \bibfield  {author} {\bibinfo {author} {\bibfnamefont {G.}~\bibnamefont
    {Wang}}, \bibinfo {author} {\bibfnamefont {A.}~\bibnamefont {Chernikov}},
    \bibinfo {author} {\bibfnamefont {M.~M.}\ \bibnamefont {Glazov}}, \bibinfo
    {author} {\bibfnamefont {T.~F.}\ \bibnamefont {Heinz}}, \bibinfo {author}
    {\bibfnamefont {X.}~\bibnamefont {Marie}}, \bibinfo {author} {\bibfnamefont
    {T.}~\bibnamefont {Amand}},\ and\ \bibinfo {author} {\bibfnamefont
    {B.}~\bibnamefont {Urbaszek}},\ }\href
    {https://doi.org/10.1103/RevModPhys.90.021001} {\bibfield  {journal}
    {\bibinfo  {journal} {Rev. Mod. Phys.}\ }\textbf {\bibinfo {volume} {90}},\
    \bibinfo {pages} {021001} (\bibinfo {year} {2018})}\BibitemShut {NoStop}%
  \bibitem [{\citenamefont {Bruno}\ \emph {et~al.}(2016)\citenamefont {Bruno},
    \citenamefont {Tamai}, \citenamefont {Wu}, \citenamefont {Cucchi},
    \citenamefont {Barreteau}, \citenamefont {de~la Torre}, \citenamefont
    {McKeown~Walker}, \citenamefont {Ricc\`o}, \citenamefont {Wang},
    \citenamefont {Kim}, \citenamefont {Hoesch}, \citenamefont {Shi},
    \citenamefont {Plumb}, \citenamefont {Giannini}, \citenamefont {Soluyanov},\
    and\ \citenamefont {Baumberger}}]{Bruno2016-rs}%
    \BibitemOpen
    \bibfield  {author} {\bibinfo {author} {\bibfnamefont {F.~Y.}\ \bibnamefont
    {Bruno}}, \bibinfo {author} {\bibfnamefont {A.}~\bibnamefont {Tamai}},
    \bibinfo {author} {\bibfnamefont {Q.~S.}\ \bibnamefont {Wu}}, \bibinfo
    {author} {\bibfnamefont {I.}~\bibnamefont {Cucchi}}, \bibinfo {author}
    {\bibfnamefont {C.}~\bibnamefont {Barreteau}}, \bibinfo {author}
    {\bibfnamefont {A.}~\bibnamefont {de~la Torre}}, \bibinfo {author}
    {\bibfnamefont {S.}~\bibnamefont {McKeown~Walker}}, \bibinfo {author}
    {\bibfnamefont {S.}~\bibnamefont {Ricc\`o}}, \bibinfo {author} {\bibfnamefont
    {Z.}~\bibnamefont {Wang}}, \bibinfo {author} {\bibfnamefont {T.~K.}\
    \bibnamefont {Kim}}, \bibinfo {author} {\bibfnamefont {M.}~\bibnamefont
    {Hoesch}}, \bibinfo {author} {\bibfnamefont {M.}~\bibnamefont {Shi}},
    \bibinfo {author} {\bibfnamefont {N.~C.}\ \bibnamefont {Plumb}}, \bibinfo
    {author} {\bibfnamefont {E.}~\bibnamefont {Giannini}}, \bibinfo {author}
    {\bibfnamefont {A.~A.}\ \bibnamefont {Soluyanov}},\ and\ \bibinfo {author}
    {\bibfnamefont {F.}~\bibnamefont {Baumberger}},\ }\href
    {https://doi.org/10.1103/PhysRevB.94.121112} {\bibfield  {journal} {\bibinfo
    {journal} {Phys. Rev. B}\ }\textbf {\bibinfo {volume} {94}},\ \bibinfo
    {pages} {121112(R)} (\bibinfo {year} {2016})}\BibitemShut {NoStop}%
  \bibitem [{\citenamefont {Huang}\ \emph
    {et~al.}(2016{\natexlab{a}})\citenamefont {Huang}, \citenamefont {Zhou},\
    and\ \citenamefont {Duan}}]{Huang2016-qm}%
    \BibitemOpen
    \bibfield  {author} {\bibinfo {author} {\bibfnamefont {H.}~\bibnamefont
    {Huang}}, \bibinfo {author} {\bibfnamefont {S.}~\bibnamefont {Zhou}},\ and\
    \bibinfo {author} {\bibfnamefont {W.}~\bibnamefont {Duan}},\ }\href
    {https://doi.org/10.1103/PhysRevB.94.121117} {\bibfield  {journal} {\bibinfo
    {journal} {Phys. Rev. B}\ }\textbf {\bibinfo {volume} {94}},\ \bibinfo
    {pages} {121117(R)} (\bibinfo {year} {2016}{\natexlab{a}})}\BibitemShut
    {NoStop}%
  \bibitem [{\citenamefont {Huang}\ \emph
    {et~al.}(2016{\natexlab{b}})\citenamefont {Huang}, \citenamefont {McCormick},
    \citenamefont {Ochi}, \citenamefont {Zhao}, \citenamefont {Suzuki},
    \citenamefont {Arita}, \citenamefont {Wu}, \citenamefont {Mou}, \citenamefont
    {Cao}, \citenamefont {Yan}, \citenamefont {Trivedi},\ and\ \citenamefont
    {Kaminski}}]{Huang2016-eo}%
    \BibitemOpen
    \bibfield  {author} {\bibinfo {author} {\bibfnamefont {L.}~\bibnamefont
    {Huang}}, \bibinfo {author} {\bibfnamefont {T.~M.}\ \bibnamefont
    {McCormick}}, \bibinfo {author} {\bibfnamefont {M.}~\bibnamefont {Ochi}},
    \bibinfo {author} {\bibfnamefont {Z.}~\bibnamefont {Zhao}}, \bibinfo {author}
    {\bibfnamefont {M.-T.}\ \bibnamefont {Suzuki}}, \bibinfo {author}
    {\bibfnamefont {R.}~\bibnamefont {Arita}}, \bibinfo {author} {\bibfnamefont
    {Y.}~\bibnamefont {Wu}}, \bibinfo {author} {\bibfnamefont {D.}~\bibnamefont
    {Mou}}, \bibinfo {author} {\bibfnamefont {H.}~\bibnamefont {Cao}}, \bibinfo
    {author} {\bibfnamefont {J.}~\bibnamefont {Yan}}, \bibinfo {author}
    {\bibfnamefont {N.}~\bibnamefont {Trivedi}},\ and\ \bibinfo {author}
    {\bibfnamefont {A.}~\bibnamefont {Kaminski}},\ }\href
    {https://doi.org/10.1038/nmat4685} {\bibfield  {journal} {\bibinfo  {journal}
    {Nat. Mater.}\ }\textbf {\bibinfo {volume} {15}},\ \bibinfo {pages} {1155}
    (\bibinfo {year} {2016}{\natexlab{b}})}\BibitemShut {NoStop}%
  \bibitem [{\citenamefont {Belopolski}\ \emph {et~al.}(2016)\citenamefont
    {Belopolski}, \citenamefont {Sanchez}, \citenamefont {Ishida}, \citenamefont
    {Pan}, \citenamefont {Yu}, \citenamefont {Xu}, \citenamefont {Chang},
    \citenamefont {Chang}, \citenamefont {Zheng}, \citenamefont {Alidoust},
    \citenamefont {Bian}, \citenamefont {Neupane}, \citenamefont {Huang},
    \citenamefont {Lee}, \citenamefont {Song}, \citenamefont {Bu}, \citenamefont
    {Wang}, \citenamefont {Li}, \citenamefont {Eda}, \citenamefont {Jeng},
    \citenamefont {Kondo}, \citenamefont {Lin}, \citenamefont {Liu},
    \citenamefont {Song}, \citenamefont {Shin},\ and\ \citenamefont
    {Hasan}}]{Belopolski2016-gr}%
    \BibitemOpen
    \bibfield  {author} {\bibinfo {author} {\bibfnamefont {I.}~\bibnamefont
    {Belopolski}}, \bibinfo {author} {\bibfnamefont {D.~S.}\ \bibnamefont
    {Sanchez}}, \bibinfo {author} {\bibfnamefont {Y.}~\bibnamefont {Ishida}},
    \bibinfo {author} {\bibfnamefont {X.}~\bibnamefont {Pan}}, \bibinfo {author}
    {\bibfnamefont {P.}~\bibnamefont {Yu}}, \bibinfo {author} {\bibfnamefont
    {S.-Y.}\ \bibnamefont {Xu}}, \bibinfo {author} {\bibfnamefont
    {G.}~\bibnamefont {Chang}}, \bibinfo {author} {\bibfnamefont {T.-R.}\
    \bibnamefont {Chang}}, \bibinfo {author} {\bibfnamefont {H.}~\bibnamefont
    {Zheng}}, \bibinfo {author} {\bibfnamefont {N.}~\bibnamefont {Alidoust}},
    \bibinfo {author} {\bibfnamefont {G.}~\bibnamefont {Bian}}, \bibinfo {author}
    {\bibfnamefont {M.}~\bibnamefont {Neupane}}, \bibinfo {author} {\bibfnamefont
    {S.-M.}\ \bibnamefont {Huang}}, \bibinfo {author} {\bibfnamefont {C.-C.}\
    \bibnamefont {Lee}}, \bibinfo {author} {\bibfnamefont {Y.}~\bibnamefont
    {Song}}, \bibinfo {author} {\bibfnamefont {H.}~\bibnamefont {Bu}}, \bibinfo
    {author} {\bibfnamefont {G.}~\bibnamefont {Wang}}, \bibinfo {author}
    {\bibfnamefont {S.}~\bibnamefont {Li}}, \bibinfo {author} {\bibfnamefont
    {G.}~\bibnamefont {Eda}}, \bibinfo {author} {\bibfnamefont {H.-T.}\
    \bibnamefont {Jeng}}, \bibinfo {author} {\bibfnamefont {T.}~\bibnamefont
    {Kondo}}, \bibinfo {author} {\bibfnamefont {H.}~\bibnamefont {Lin}}, \bibinfo
    {author} {\bibfnamefont {Z.}~\bibnamefont {Liu}}, \bibinfo {author}
    {\bibfnamefont {F.}~\bibnamefont {Song}}, \bibinfo {author} {\bibfnamefont
    {S.}~\bibnamefont {Shin}},\ and\ \bibinfo {author} {\bibfnamefont {M.~Z.}\
    \bibnamefont {Hasan}},\ }\href {https://doi.org/10.1038/ncomms13643}
    {\bibfield  {journal} {\bibinfo  {journal} {Nat. Commun.}\ }\textbf {\bibinfo
    {volume} {7}},\ \bibinfo {pages} {13643} (\bibinfo {year}
    {2016})}\BibitemShut {NoStop}%
  \bibitem [{\citenamefont {Wang}\ \emph
    {et~al.}(2016{\natexlab{b}})\citenamefont {Wang}, \citenamefont {Zhang},
    \citenamefont {Huang}, \citenamefont {Nie}, \citenamefont {Liu},
    \citenamefont {Liang}, \citenamefont {Zhang}, \citenamefont {Shen},
    \citenamefont {Liu}, \citenamefont {Hu}, \citenamefont {Ding}, \citenamefont
    {Liu}, \citenamefont {Hu}, \citenamefont {He}, \citenamefont {Zhao},
    \citenamefont {Yu}, \citenamefont {Hu}, \citenamefont {Wei}, \citenamefont
    {Mao}, \citenamefont {Shi}, \citenamefont {Jia}, \citenamefont {Zhang},
    \citenamefont {Zhang}, \citenamefont {Yang}, \citenamefont {Wang},
    \citenamefont {Peng}, \citenamefont {Weng}, \citenamefont {Dai},
    \citenamefont {Fang}, \citenamefont {Xu}, \citenamefont {Chen},\ and\
    \citenamefont {Zhou}}]{Wang2016-vn}%
    \BibitemOpen
    \bibfield  {author} {\bibinfo {author} {\bibfnamefont {C.}~\bibnamefont
    {Wang}}, \bibinfo {author} {\bibfnamefont {Y.}~\bibnamefont {Zhang}},
    \bibinfo {author} {\bibfnamefont {J.}~\bibnamefont {Huang}}, \bibinfo
    {author} {\bibfnamefont {S.}~\bibnamefont {Nie}}, \bibinfo {author}
    {\bibfnamefont {G.}~\bibnamefont {Liu}}, \bibinfo {author} {\bibfnamefont
    {A.}~\bibnamefont {Liang}}, \bibinfo {author} {\bibfnamefont
    {Y.}~\bibnamefont {Zhang}}, \bibinfo {author} {\bibfnamefont
    {B.}~\bibnamefont {Shen}}, \bibinfo {author} {\bibfnamefont {J.}~\bibnamefont
    {Liu}}, \bibinfo {author} {\bibfnamefont {C.}~\bibnamefont {Hu}}, \bibinfo
    {author} {\bibfnamefont {Y.}~\bibnamefont {Ding}}, \bibinfo {author}
    {\bibfnamefont {D.}~\bibnamefont {Liu}}, \bibinfo {author} {\bibfnamefont
    {Y.}~\bibnamefont {Hu}}, \bibinfo {author} {\bibfnamefont {S.}~\bibnamefont
    {He}}, \bibinfo {author} {\bibfnamefont {L.}~\bibnamefont {Zhao}}, \bibinfo
    {author} {\bibfnamefont {L.}~\bibnamefont {Yu}}, \bibinfo {author}
    {\bibfnamefont {J.}~\bibnamefont {Hu}}, \bibinfo {author} {\bibfnamefont
    {J.}~\bibnamefont {Wei}}, \bibinfo {author} {\bibfnamefont {Z.}~\bibnamefont
    {Mao}}, \bibinfo {author} {\bibfnamefont {Y.}~\bibnamefont {Shi}}, \bibinfo
    {author} {\bibfnamefont {X.}~\bibnamefont {Jia}}, \bibinfo {author}
    {\bibfnamefont {F.}~\bibnamefont {Zhang}}, \bibinfo {author} {\bibfnamefont
    {S.}~\bibnamefont {Zhang}}, \bibinfo {author} {\bibfnamefont
    {F.}~\bibnamefont {Yang}}, \bibinfo {author} {\bibfnamefont {Z.}~\bibnamefont
    {Wang}}, \bibinfo {author} {\bibfnamefont {Q.}~\bibnamefont {Peng}}, \bibinfo
    {author} {\bibfnamefont {H.}~\bibnamefont {Weng}}, \bibinfo {author}
    {\bibfnamefont {X.}~\bibnamefont {Dai}}, \bibinfo {author} {\bibfnamefont
    {Z.}~\bibnamefont {Fang}}, \bibinfo {author} {\bibfnamefont {Z.}~\bibnamefont
    {Xu}}, \bibinfo {author} {\bibfnamefont {C.}~\bibnamefont {Chen}},\ and\
    \bibinfo {author} {\bibfnamefont {X.~J.}\ \bibnamefont {Zhou}},\ }\href
    {https://doi.org/10.1103/PhysRevB.94.241119} {\bibfield  {journal} {\bibinfo
    {journal} {Phys. Rev. B}\ }\textbf {\bibinfo {volume} {94}},\ \bibinfo
    {pages} {241119(R)} (\bibinfo {year} {2016}{\natexlab{b}})}\BibitemShut
    {NoStop}%
  \bibitem [{\citenamefont {Jiang}\ \emph {et~al.}(2017)\citenamefont {Jiang},
    \citenamefont {Liu}, \citenamefont {Sun}, \citenamefont {Yang}, \citenamefont
    {Rajamathi}, \citenamefont {Qi}, \citenamefont {Yang}, \citenamefont {Chen},
    \citenamefont {Peng}, \citenamefont {Hwang}, \citenamefont {Sun},
    \citenamefont {Mo}, \citenamefont {Vobornik}, \citenamefont {Fujii},
    \citenamefont {Parkin}, \citenamefont {Felser}, \citenamefont {Yan},\ and\
    \citenamefont {Chen}}]{Jiang2017-mk}%
    \BibitemOpen
    \bibfield  {author} {\bibinfo {author} {\bibfnamefont {J.}~\bibnamefont
    {Jiang}}, \bibinfo {author} {\bibfnamefont {Z.~K.}\ \bibnamefont {Liu}},
    \bibinfo {author} {\bibfnamefont {Y.}~\bibnamefont {Sun}}, \bibinfo {author}
    {\bibfnamefont {H.~F.}\ \bibnamefont {Yang}}, \bibinfo {author}
    {\bibfnamefont {C.~R.}\ \bibnamefont {Rajamathi}}, \bibinfo {author}
    {\bibfnamefont {Y.~P.}\ \bibnamefont {Qi}}, \bibinfo {author} {\bibfnamefont
    {L.~X.}\ \bibnamefont {Yang}}, \bibinfo {author} {\bibfnamefont
    {C.}~\bibnamefont {Chen}}, \bibinfo {author} {\bibfnamefont {H.}~\bibnamefont
    {Peng}}, \bibinfo {author} {\bibfnamefont {C.-C.}\ \bibnamefont {Hwang}},
    \bibinfo {author} {\bibfnamefont {S.~Z.}\ \bibnamefont {Sun}}, \bibinfo
    {author} {\bibfnamefont {S.-K.}\ \bibnamefont {Mo}}, \bibinfo {author}
    {\bibfnamefont {I.}~\bibnamefont {Vobornik}}, \bibinfo {author}
    {\bibfnamefont {J.}~\bibnamefont {Fujii}}, \bibinfo {author} {\bibfnamefont
    {S.~S.~P.}\ \bibnamefont {Parkin}}, \bibinfo {author} {\bibfnamefont
    {C.}~\bibnamefont {Felser}}, \bibinfo {author} {\bibfnamefont {B.~H.}\
    \bibnamefont {Yan}},\ and\ \bibinfo {author} {\bibfnamefont {Y.~L.}\
    \bibnamefont {Chen}},\ }\href {https://doi.org/10.1038/ncomms13973}
    {\bibfield  {journal} {\bibinfo  {journal} {Nat. Commun.}\ }\textbf {\bibinfo
    {volume} {8}},\ \bibinfo {pages} {13973} (\bibinfo {year}
    {2017})}\BibitemShut {NoStop}%
  \bibitem [{\citenamefont {Yan}\ \emph {et~al.}(2017)\citenamefont {Yan},
    \citenamefont {Huang}, \citenamefont {Zhang}, \citenamefont {Wang},
    \citenamefont {Yao}, \citenamefont {Deng}, \citenamefont {Wan}, \citenamefont
    {Zhang}, \citenamefont {Arita}, \citenamefont {Yang}, \citenamefont {Sun},
    \citenamefont {Yao}, \citenamefont {Wu}, \citenamefont {Fan}, \citenamefont
    {Duan},\ and\ \citenamefont {Zhou}}]{Yan2017-sb}%
    \BibitemOpen
    \bibfield  {author} {\bibinfo {author} {\bibfnamefont {M.}~\bibnamefont
    {Yan}}, \bibinfo {author} {\bibfnamefont {H.}~\bibnamefont {Huang}}, \bibinfo
    {author} {\bibfnamefont {K.}~\bibnamefont {Zhang}}, \bibinfo {author}
    {\bibfnamefont {E.}~\bibnamefont {Wang}}, \bibinfo {author} {\bibfnamefont
    {W.}~\bibnamefont {Yao}}, \bibinfo {author} {\bibfnamefont {K.}~\bibnamefont
    {Deng}}, \bibinfo {author} {\bibfnamefont {G.}~\bibnamefont {Wan}}, \bibinfo
    {author} {\bibfnamefont {H.}~\bibnamefont {Zhang}}, \bibinfo {author}
    {\bibfnamefont {M.}~\bibnamefont {Arita}}, \bibinfo {author} {\bibfnamefont
    {H.}~\bibnamefont {Yang}}, \bibinfo {author} {\bibfnamefont {Z.}~\bibnamefont
    {Sun}}, \bibinfo {author} {\bibfnamefont {H.}~\bibnamefont {Yao}}, \bibinfo
    {author} {\bibfnamefont {Y.}~\bibnamefont {Wu}}, \bibinfo {author}
    {\bibfnamefont {S.}~\bibnamefont {Fan}}, \bibinfo {author} {\bibfnamefont
    {W.}~\bibnamefont {Duan}},\ and\ \bibinfo {author} {\bibfnamefont
    {S.}~\bibnamefont {Zhou}},\ }\href
    {https://doi.org/10.1038/s41467-017-00280-6} {\bibfield  {journal} {\bibinfo
    {journal} {Nat. Commun.}\ }\textbf {\bibinfo {volume} {8}},\ \bibinfo {pages}
    {257} (\bibinfo {year} {2017})}\BibitemShut {NoStop}%
  \bibitem [{\citenamefont {Zhang}\ \emph {et~al.}(2017)\citenamefont {Zhang},
    \citenamefont {Yan}, \citenamefont {Zhang}, \citenamefont {Huang},
    \citenamefont {Arita}, \citenamefont {Sun}, \citenamefont {Duan},
    \citenamefont {Wu},\ and\ \citenamefont {Zhou}}]{Zhang2017-xb}%
    \BibitemOpen
    \bibfield  {author} {\bibinfo {author} {\bibfnamefont {K.}~\bibnamefont
    {Zhang}}, \bibinfo {author} {\bibfnamefont {M.}~\bibnamefont {Yan}}, \bibinfo
    {author} {\bibfnamefont {H.}~\bibnamefont {Zhang}}, \bibinfo {author}
    {\bibfnamefont {H.}~\bibnamefont {Huang}}, \bibinfo {author} {\bibfnamefont
    {M.}~\bibnamefont {Arita}}, \bibinfo {author} {\bibfnamefont
    {Z.}~\bibnamefont {Sun}}, \bibinfo {author} {\bibfnamefont {W.}~\bibnamefont
    {Duan}}, \bibinfo {author} {\bibfnamefont {Y.}~\bibnamefont {Wu}},\ and\
    \bibinfo {author} {\bibfnamefont {S.}~\bibnamefont {Zhou}},\ }\href
    {https://doi.org/10.1103/PhysRevB.96.125102} {\bibfield  {journal} {\bibinfo
    {journal} {Phys. Rev. B}\ }\textbf {\bibinfo {volume} {96}},\ \bibinfo
    {pages} {125102} (\bibinfo {year} {2017})}\BibitemShut {NoStop}%
  \bibitem [{\citenamefont {Li}\ \emph {et~al.}(2017)\citenamefont {Li},
    \citenamefont {Wen}, \citenamefont {He}, \citenamefont {Zhang}, \citenamefont
    {Xia}, \citenamefont {Yu}, \citenamefont {Yang}, \citenamefont {Zhu},
    \citenamefont {Alshareef},\ and\ \citenamefont {Zhang}}]{Li2017-jb}%
    \BibitemOpen
    \bibfield  {author} {\bibinfo {author} {\bibfnamefont {P.}~\bibnamefont
    {Li}}, \bibinfo {author} {\bibfnamefont {Y.}~\bibnamefont {Wen}}, \bibinfo
    {author} {\bibfnamefont {X.}~\bibnamefont {He}}, \bibinfo {author}
    {\bibfnamefont {Q.}~\bibnamefont {Zhang}}, \bibinfo {author} {\bibfnamefont
    {C.}~\bibnamefont {Xia}}, \bibinfo {author} {\bibfnamefont {Z.-M.}\
    \bibnamefont {Yu}}, \bibinfo {author} {\bibfnamefont {S.~A.}\ \bibnamefont
    {Yang}}, \bibinfo {author} {\bibfnamefont {Z.}~\bibnamefont {Zhu}}, \bibinfo
    {author} {\bibfnamefont {H.~N.}\ \bibnamefont {Alshareef}},\ and\ \bibinfo
    {author} {\bibfnamefont {X.-X.}\ \bibnamefont {Zhang}},\ }\href
    {https://doi.org/10.1038/s41467-017-02237-1} {\bibfield  {journal} {\bibinfo
    {journal} {Nat. Commun.}\ }\textbf {\bibinfo {volume} {8}},\ \bibinfo {pages}
    {2150} (\bibinfo {year} {2017})}\BibitemShut {NoStop}%
  \bibitem [{\citenamefont {Wang}\ \emph {et~al.}(2019)\citenamefont {Wang},
    \citenamefont {Wieder}, \citenamefont {Li}, \citenamefont {Yan},\ and\
    \citenamefont {Bernevig}}]{Wang2019-fa}%
    \BibitemOpen
    \bibfield  {author} {\bibinfo {author} {\bibfnamefont {Z.}~\bibnamefont
    {Wang}}, \bibinfo {author} {\bibfnamefont {B.~J.}\ \bibnamefont {Wieder}},
    \bibinfo {author} {\bibfnamefont {J.}~\bibnamefont {Li}}, \bibinfo {author}
    {\bibfnamefont {B.}~\bibnamefont {Yan}},\ and\ \bibinfo {author}
    {\bibfnamefont {B.~A.}\ \bibnamefont {Bernevig}},\ }\href
    {https://doi.org/10.1103/PhysRevLett.123.186401} {\bibfield  {journal}
    {\bibinfo  {journal} {Phys. Rev. Lett.}\ }\textbf {\bibinfo {volume} {123}},\
    \bibinfo {pages} {186401} (\bibinfo {year} {2019})}\BibitemShut {NoStop}%
  \bibitem [{\citenamefont {Kar}\ \emph {et~al.}(2020)\citenamefont {Kar},
    \citenamefont {Chatterjee}, \citenamefont {Harnagea}, \citenamefont
    {Kushnirenko}, \citenamefont {Fedorov}, \citenamefont {Shrivastava},
    \citenamefont {B{\"u}chner}, \citenamefont {Mahadevan},\ and\ \citenamefont
    {Thirupathaiah}}]{Kar2020-gq}%
    \BibitemOpen
    \bibfield  {author} {\bibinfo {author} {\bibfnamefont {I.}~\bibnamefont
    {Kar}}, \bibinfo {author} {\bibfnamefont {J.}~\bibnamefont {Chatterjee}},
    \bibinfo {author} {\bibfnamefont {L.}~\bibnamefont {Harnagea}}, \bibinfo
    {author} {\bibfnamefont {Y.}~\bibnamefont {Kushnirenko}}, \bibinfo {author}
    {\bibfnamefont {A.~V.}\ \bibnamefont {Fedorov}}, \bibinfo {author}
    {\bibfnamefont {D.}~\bibnamefont {Shrivastava}}, \bibinfo {author}
    {\bibfnamefont {B.}~\bibnamefont {B{\"u}chner}}, \bibinfo {author}
    {\bibfnamefont {P.}~\bibnamefont {Mahadevan}},\ and\ \bibinfo {author}
    {\bibfnamefont {S.}~\bibnamefont {Thirupathaiah}},\ }\href
    {https://doi.org/10.1103/PhysRevB.101.165122} {\bibfield  {journal} {\bibinfo
     {journal} {Phys. Rev. B}\ }\textbf {\bibinfo {volume} {101}},\ \bibinfo
    {pages} {165122} (\bibinfo {year} {2020})}\BibitemShut {NoStop}%
  \bibitem [{\citenamefont {Koyano}\ \emph {et~al.}(1986)\citenamefont {Koyano},
    \citenamefont {Negishi}, \citenamefont {Ueda}, \citenamefont {Sasaki},\ and\
    \citenamefont {Inoue}}]{Koyano1986-av}%
    \BibitemOpen
    \bibfield  {author} {\bibinfo {author} {\bibfnamefont {M.}~\bibnamefont
    {Koyano}}, \bibinfo {author} {\bibfnamefont {H.}~\bibnamefont {Negishi}},
    \bibinfo {author} {\bibfnamefont {Y.}~\bibnamefont {Ueda}}, \bibinfo {author}
    {\bibfnamefont {M.}~\bibnamefont {Sasaki}},\ and\ \bibinfo {author}
    {\bibfnamefont {M.}~\bibnamefont {Inoue}},\ }\href
    {https://doi.org/10.1002/pssb.2221380137} {\bibfield  {journal} {\bibinfo
    {journal} {Phys. Status Solidi B}\ }\textbf {\bibinfo {volume} {138}},\
    \bibinfo {pages} {357} (\bibinfo {year} {1986})}\BibitemShut {NoStop}%
  \bibitem [{\citenamefont {Sasaki}\ \emph {et~al.}(1987)\citenamefont {Sasaki},
    \citenamefont {Koyano},\ and\ \citenamefont {Inoue}}]{Sasaki1987-md}%
    \BibitemOpen
    \bibfield  {author} {\bibinfo {author} {\bibfnamefont {M.}~\bibnamefont
    {Sasaki}}, \bibinfo {author} {\bibfnamefont {M.}~\bibnamefont {Koyano}},\
    and\ \bibinfo {author} {\bibfnamefont {M.}~\bibnamefont {Inoue}},\ }\href
    {https://doi.org/10.1063/1.337988} {\bibfield  {journal} {\bibinfo  {journal}
    {J. Appl. Phys.}\ }\textbf {\bibinfo {volume} {61}},\ \bibinfo {pages} {2267}
    (\bibinfo {year} {1987})}\BibitemShut {NoStop}%
  \bibitem [{\citenamefont {Imai}\ \emph {et~al.}(2001)\citenamefont {Imai},
    \citenamefont {Shimakawa},\ and\ \citenamefont {Kubo}}]{PhysRevB.64.241104}%
    \BibitemOpen
    \bibfield  {author} {\bibinfo {author} {\bibfnamefont {H.}~\bibnamefont
    {Imai}}, \bibinfo {author} {\bibfnamefont {Y.}~\bibnamefont {Shimakawa}},\
    and\ \bibinfo {author} {\bibfnamefont {Y.}~\bibnamefont {Kubo}},\ }\href
    {https://doi.org/10.1103/PhysRevB.64.241104} {\bibfield  {journal} {\bibinfo
    {journal} {Phys. Rev. B}\ }\textbf {\bibinfo {volume} {64}},\ \bibinfo
    {pages} {241104(R)} (\bibinfo {year} {2001})}\BibitemShut {NoStop}%
  \bibitem [{\citenamefont {Guilmeau}\ \emph {et~al.}(2011)\citenamefont
    {Guilmeau}, \citenamefont {Br{\'e}ard},\ and\ \citenamefont
    {Maignan}}]{Guilmeau2011-da}%
    \BibitemOpen
    \bibfield  {author} {\bibinfo {author} {\bibfnamefont {E.}~\bibnamefont
    {Guilmeau}}, \bibinfo {author} {\bibfnamefont {Y.}~\bibnamefont
    {Br{\'e}ard}},\ and\ \bibinfo {author} {\bibfnamefont {A.}~\bibnamefont
    {Maignan}},\ }\href {https://doi.org/10.1063/1.3621834} {\bibfield  {journal}
    {\bibinfo  {journal} {Appl. Phys. Lett.}\ }\textbf {\bibinfo {volume} {99}},\
    \bibinfo {pages} {052107} (\bibinfo {year} {2011})}\BibitemShut {NoStop}%
  \bibitem [{\citenamefont {Wan}\ \emph {et~al.}(2010)\citenamefont {Wan},
    \citenamefont {Wang}, \citenamefont {Wang},\ and\ \citenamefont
    {Koumoto}}]{Wan2010-hj}%
    \BibitemOpen
    \bibfield  {author} {\bibinfo {author} {\bibfnamefont {C.}~\bibnamefont
    {Wan}}, \bibinfo {author} {\bibfnamefont {Y.}~\bibnamefont {Wang}}, \bibinfo
    {author} {\bibfnamefont {N.}~\bibnamefont {Wang}},\ and\ \bibinfo {author}
    {\bibfnamefont {K.}~\bibnamefont {Koumoto}},\ }\href
    {https://doi.org/10.3390/ma3042606} {\bibfield  {journal} {\bibinfo
    {journal} {Materials}\ }\textbf {\bibinfo {volume} {3}},\ \bibinfo {pages}
    {2606} (\bibinfo {year} {2010})}\BibitemShut {NoStop}%
  \bibitem [{\citenamefont {Wickramaratne}\ \emph {et~al.}(2014)\citenamefont
    {Wickramaratne}, \citenamefont {Zahid},\ and\ \citenamefont
    {Lake}}]{Wickramaratne2014-rh}%
    \BibitemOpen
    \bibfield  {author} {\bibinfo {author} {\bibfnamefont {D.}~\bibnamefont
    {Wickramaratne}}, \bibinfo {author} {\bibfnamefont {F.}~\bibnamefont
    {Zahid}},\ and\ \bibinfo {author} {\bibfnamefont {R.~K.}\ \bibnamefont
    {Lake}},\ }\href {https://doi.org/10.1063/1.4869142} {\bibfield  {journal}
    {\bibinfo  {journal} {J. Chem. Phys.}\ }\textbf {\bibinfo {volume} {140}},\
    \bibinfo {pages} {124710} (\bibinfo {year} {2014})}\BibitemShut {NoStop}%
  \bibitem [{\citenamefont {Samanta}\ \emph {et~al.}(2014)\citenamefont
    {Samanta}, \citenamefont {Pandey},\ and\ \citenamefont
    {Singh}}]{PhysRevB.90.174301}%
    \BibitemOpen
    \bibfield  {author} {\bibinfo {author} {\bibfnamefont {A.}~\bibnamefont
    {Samanta}}, \bibinfo {author} {\bibfnamefont {T.}~\bibnamefont {Pandey}},\
    and\ \bibinfo {author} {\bibfnamefont {A.~K.}\ \bibnamefont {Singh}},\ }\href
    {https://doi.org/10.1103/PhysRevB.90.174301} {\bibfield  {journal} {\bibinfo
    {journal} {Phys. Rev. B}\ }\textbf {\bibinfo {volume} {90}},\ \bibinfo
    {pages} {174301} (\bibinfo {year} {2014})}\BibitemShut {NoStop}%
  \bibitem [{\citenamefont {Bourg^^c3^^a8s}\ \emph {et~al.}(2016)\citenamefont
    {Bourg^^c3^^a8s}, \citenamefont {Barbier}, \citenamefont {Gu^^c3^^a9lou},
    \citenamefont {Vaqueiro}, \citenamefont {Powell}, \citenamefont {Lebedev},
    \citenamefont {Barrier}, \citenamefont {Kinemuchi},\ and\ \citenamefont
    {Guilmeau}}]{BOURGES20161183}%
    \BibitemOpen
    \bibfield  {author} {\bibinfo {author} {\bibfnamefont {C.}~\bibnamefont
    {Bourg^^c3^^a8s}}, \bibinfo {author} {\bibfnamefont {T.}~\bibnamefont
    {Barbier}}, \bibinfo {author} {\bibfnamefont {G.}~\bibnamefont
    {Gu^^c3^^a9lou}}, \bibinfo {author} {\bibfnamefont {P.}~\bibnamefont
    {Vaqueiro}}, \bibinfo {author} {\bibfnamefont {A.~V.}\ \bibnamefont
    {Powell}}, \bibinfo {author} {\bibfnamefont {O.~I.}\ \bibnamefont {Lebedev}},
    \bibinfo {author} {\bibfnamefont {N.}~\bibnamefont {Barrier}}, \bibinfo
    {author} {\bibfnamefont {Y.}~\bibnamefont {Kinemuchi}},\ and\ \bibinfo
    {author} {\bibfnamefont {E.}~\bibnamefont {Guilmeau}},\ }\href
    {https://doi.org/https://doi.org/10.1016/j.jeurceramsoc.2015.11.025}
    {\bibfield  {journal} {\bibinfo  {journal} {J. Eur. Ceram. Soc.}\ }\textbf
    {\bibinfo {volume} {36}},\ \bibinfo {pages} {1183} (\bibinfo {year}
    {2016})}\BibitemShut {NoStop}%
  \bibitem [{\citenamefont {Huang}\ \emph
    {et~al.}(2016{\natexlab{c}})\citenamefont {Huang}, \citenamefont {Wu},
    \citenamefont {Kong}, \citenamefont {Meng}, \citenamefont {Zhuang},
    \citenamefont {Jiang},\ and\ \citenamefont {Bao}}]{Huang2016-na}%
    \BibitemOpen
    \bibfield  {author} {\bibinfo {author} {\bibfnamefont {Z.}~\bibnamefont
    {Huang}}, \bibinfo {author} {\bibfnamefont {T.}~\bibnamefont {Wu}}, \bibinfo
    {author} {\bibfnamefont {S.}~\bibnamefont {Kong}}, \bibinfo {author}
    {\bibfnamefont {Q.-L.}\ \bibnamefont {Meng}}, \bibinfo {author}
    {\bibfnamefont {W.}~\bibnamefont {Zhuang}}, \bibinfo {author} {\bibfnamefont
    {P.}~\bibnamefont {Jiang}},\ and\ \bibinfo {author} {\bibfnamefont
    {X.}~\bibnamefont {Bao}},\ }\href {https://doi.org/10.1039/C6TA03122F}
    {\bibfield  {journal} {\bibinfo  {journal} {J. Mater. Chem. A Mater. Energy
    Sustain.}\ }\textbf {\bibinfo {volume} {4}},\ \bibinfo {pages} {10159}
    (\bibinfo {year} {2016}{\natexlab{c}})}\BibitemShut {NoStop}%
  \bibitem [{\citenamefont {Zhang}\ and\ \citenamefont
    {Zhang}(2017)}]{Zhang2017-cf}%
    \BibitemOpen
    \bibfield  {author} {\bibinfo {author} {\bibfnamefont {G.}~\bibnamefont
    {Zhang}}\ and\ \bibinfo {author} {\bibfnamefont {Y.-W.}\ \bibnamefont
    {Zhang}},\ }\href {https://doi.org/10.1039/C7TC01088E} {\bibfield  {journal}
    {\bibinfo  {journal} {J. Mater. Chem.}\ }\textbf {\bibinfo {volume} {5}},\
    \bibinfo {pages} {7684} (\bibinfo {year} {2017})}\BibitemShut {NoStop}%
  \bibitem [{\citenamefont {Schaibley}\ \emph {et~al.}(2016)\citenamefont
    {Schaibley}, \citenamefont {Yu}, \citenamefont {Clark}, \citenamefont
    {Rivera}, \citenamefont {Ross}, \citenamefont {Seyler}, \citenamefont {Yao},\
    and\ \citenamefont {Xu}}]{Schaibley2016-dn}%
    \BibitemOpen
    \bibfield  {author} {\bibinfo {author} {\bibfnamefont {J.~R.}\ \bibnamefont
    {Schaibley}}, \bibinfo {author} {\bibfnamefont {H.}~\bibnamefont {Yu}},
    \bibinfo {author} {\bibfnamefont {G.}~\bibnamefont {Clark}}, \bibinfo
    {author} {\bibfnamefont {P.}~\bibnamefont {Rivera}}, \bibinfo {author}
    {\bibfnamefont {J.~S.}\ \bibnamefont {Ross}}, \bibinfo {author}
    {\bibfnamefont {K.~L.}\ \bibnamefont {Seyler}}, \bibinfo {author}
    {\bibfnamefont {W.}~\bibnamefont {Yao}},\ and\ \bibinfo {author}
    {\bibfnamefont {X.}~\bibnamefont {Xu}},\ }\href
    {https://doi.org/10.1038/natrevmats.2016.55} {\bibfield  {journal} {\bibinfo
    {journal} {Nat. Rev. Mater.}\ }\textbf {\bibinfo {volume} {1}},\ \bibinfo
    {pages} {1} (\bibinfo {year} {2016})}\BibitemShut {NoStop}%
  \bibitem [{\citenamefont {Rycerz}\ \emph {et~al.}(2007)\citenamefont {Rycerz},
    \citenamefont {Tworzyd{\l}o},\ and\ \citenamefont
    {Beenakker}}]{Rycerz2007-bl}%
    \BibitemOpen
    \bibfield  {author} {\bibinfo {author} {\bibfnamefont {A.}~\bibnamefont
    {Rycerz}}, \bibinfo {author} {\bibfnamefont {J.}~\bibnamefont
    {Tworzyd{\l}o}},\ and\ \bibinfo {author} {\bibfnamefont {C.~W.~J.}\
    \bibnamefont {Beenakker}},\ }\href {https://doi.org/10.1038/nphys547}
    {\bibfield  {journal} {\bibinfo  {journal} {Nat. Phys.}\ }\textbf {\bibinfo
    {volume} {3}},\ \bibinfo {pages} {172} (\bibinfo {year} {2007})}\BibitemShut
    {NoStop}%
  \bibitem [{\citenamefont {Xiao}\ \emph {et~al.}(2007)\citenamefont {Xiao},
    \citenamefont {Yao},\ and\ \citenamefont {Niu}}]{Xiao2007-zr}%
    \BibitemOpen
    \bibfield  {author} {\bibinfo {author} {\bibfnamefont {D.}~\bibnamefont
    {Xiao}}, \bibinfo {author} {\bibfnamefont {W.}~\bibnamefont {Yao}},\ and\
    \bibinfo {author} {\bibfnamefont {Q.}~\bibnamefont {Niu}},\ }\href
    {https://doi.org/10.1103/PhysRevLett.99.236809} {\bibfield  {journal}
    {\bibinfo  {journal} {Phys. Rev. Lett.}\ }\textbf {\bibinfo {volume} {99}},\
    \bibinfo {pages} {236809} (\bibinfo {year} {2007})}\BibitemShut {NoStop}%
  \bibitem [{\citenamefont {Yao}\ \emph {et~al.}(2008)\citenamefont {Yao},
    \citenamefont {Xiao},\ and\ \citenamefont {Niu}}]{Yao2008-yr}%
    \BibitemOpen
    \bibfield  {author} {\bibinfo {author} {\bibfnamefont {W.}~\bibnamefont
    {Yao}}, \bibinfo {author} {\bibfnamefont {D.}~\bibnamefont {Xiao}},\ and\
    \bibinfo {author} {\bibfnamefont {Q.}~\bibnamefont {Niu}},\ }\href
    {https://doi.org/10.1103/PhysRevB.77.235406} {\bibfield  {journal} {\bibinfo
    {journal} {Phys. Rev. B}\ }\textbf {\bibinfo {volume} {77}},\ \bibinfo
    {pages} {235406} (\bibinfo {year} {2008})}\BibitemShut {NoStop}%
  \bibitem [{\citenamefont {Cao}\ \emph {et~al.}(2012)\citenamefont {Cao},
    \citenamefont {Wang}, \citenamefont {Han}, \citenamefont {Ye}, \citenamefont
    {Zhu}, \citenamefont {Shi}, \citenamefont {Niu}, \citenamefont {Tan},
    \citenamefont {Wang}, \citenamefont {Liu},\ and\ \citenamefont
    {Feng}}]{Cao2012-nh}%
    \BibitemOpen
    \bibfield  {author} {\bibinfo {author} {\bibfnamefont {T.}~\bibnamefont
    {Cao}}, \bibinfo {author} {\bibfnamefont {G.}~\bibnamefont {Wang}}, \bibinfo
    {author} {\bibfnamefont {W.}~\bibnamefont {Han}}, \bibinfo {author}
    {\bibfnamefont {H.}~\bibnamefont {Ye}}, \bibinfo {author} {\bibfnamefont
    {C.}~\bibnamefont {Zhu}}, \bibinfo {author} {\bibfnamefont {J.}~\bibnamefont
    {Shi}}, \bibinfo {author} {\bibfnamefont {Q.}~\bibnamefont {Niu}}, \bibinfo
    {author} {\bibfnamefont {P.}~\bibnamefont {Tan}}, \bibinfo {author}
    {\bibfnamefont {E.}~\bibnamefont {Wang}}, \bibinfo {author} {\bibfnamefont
    {B.}~\bibnamefont {Liu}},\ and\ \bibinfo {author} {\bibfnamefont
    {J.}~\bibnamefont {Feng}},\ }\href {https://doi.org/10.1038/ncomms1882}
    {\bibfield  {journal} {\bibinfo  {journal} {Nat. Commun.}\ }\textbf {\bibinfo
    {volume} {3}},\ \bibinfo {pages} {887} (\bibinfo {year} {2012})}\BibitemShut
    {NoStop}%
  \bibitem [{\citenamefont {Mak}\ \emph {et~al.}(2012)\citenamefont {Mak},
    \citenamefont {He}, \citenamefont {Shan},\ and\ \citenamefont
    {Heinz}}]{Mak2012-me}%
    \BibitemOpen
    \bibfield  {author} {\bibinfo {author} {\bibfnamefont {K.~F.}\ \bibnamefont
    {Mak}}, \bibinfo {author} {\bibfnamefont {K.}~\bibnamefont {He}}, \bibinfo
    {author} {\bibfnamefont {J.}~\bibnamefont {Shan}},\ and\ \bibinfo {author}
    {\bibfnamefont {T.~F.}\ \bibnamefont {Heinz}},\ }\href
    {https://doi.org/10.1038/nnano.2012.96} {\bibfield  {journal} {\bibinfo
    {journal} {Nat. Nanotechnol.}\ }\textbf {\bibinfo {volume} {7}},\ \bibinfo
    {pages} {494} (\bibinfo {year} {2012})}\BibitemShut {NoStop}%
  \bibitem [{\citenamefont {Zeng}\ \emph {et~al.}(2012)\citenamefont {Zeng},
    \citenamefont {Dai}, \citenamefont {Yao}, \citenamefont {Xiao},\ and\
    \citenamefont {Cui}}]{Zeng2012-bn}%
    \BibitemOpen
    \bibfield  {author} {\bibinfo {author} {\bibfnamefont {H.}~\bibnamefont
    {Zeng}}, \bibinfo {author} {\bibfnamefont {J.}~\bibnamefont {Dai}}, \bibinfo
    {author} {\bibfnamefont {W.}~\bibnamefont {Yao}}, \bibinfo {author}
    {\bibfnamefont {D.}~\bibnamefont {Xiao}},\ and\ \bibinfo {author}
    {\bibfnamefont {X.}~\bibnamefont {Cui}},\ }\href
    {https://doi.org/10.1038/nnano.2012.95} {\bibfield  {journal} {\bibinfo
    {journal} {Nat. Nanotechnol.}\ }\textbf {\bibinfo {volume} {7}},\ \bibinfo
    {pages} {490} (\bibinfo {year} {2012})}\BibitemShut {NoStop}%
  \bibitem [{\citenamefont {Wu}\ \emph {et~al.}(2013)\citenamefont {Wu},
    \citenamefont {Ross}, \citenamefont {Liu}, \citenamefont {Aivazian},
    \citenamefont {Jones}, \citenamefont {Fei}, \citenamefont {Zhu},
    \citenamefont {Xiao}, \citenamefont {Yao}, \citenamefont {Cobden},\ and\
    \citenamefont {Xu}}]{Wu2013-lu}%
    \BibitemOpen
    \bibfield  {author} {\bibinfo {author} {\bibfnamefont {S.}~\bibnamefont
    {Wu}}, \bibinfo {author} {\bibfnamefont {J.~S.}\ \bibnamefont {Ross}},
    \bibinfo {author} {\bibfnamefont {G.-B.}\ \bibnamefont {Liu}}, \bibinfo
    {author} {\bibfnamefont {G.}~\bibnamefont {Aivazian}}, \bibinfo {author}
    {\bibfnamefont {A.}~\bibnamefont {Jones}}, \bibinfo {author} {\bibfnamefont
    {Z.}~\bibnamefont {Fei}}, \bibinfo {author} {\bibfnamefont {W.}~\bibnamefont
    {Zhu}}, \bibinfo {author} {\bibfnamefont {D.}~\bibnamefont {Xiao}}, \bibinfo
    {author} {\bibfnamefont {W.}~\bibnamefont {Yao}}, \bibinfo {author}
    {\bibfnamefont {D.}~\bibnamefont {Cobden}},\ and\ \bibinfo {author}
    {\bibfnamefont {X.}~\bibnamefont {Xu}},\ }\href
    {https://doi.org/10.1038/nphys2524} {\bibfield  {journal} {\bibinfo
    {journal} {Nat. Phys.}\ }\textbf {\bibinfo {volume} {9}},\ \bibinfo {pages}
    {149} (\bibinfo {year} {2013})}\BibitemShut {NoStop}%
  \bibitem [{\citenamefont {Yuan}\ \emph {et~al.}(2013)\citenamefont {Yuan},
    \citenamefont {Bahramy}, \citenamefont {Morimoto}, \citenamefont {Wu},
    \citenamefont {Nomura}, \citenamefont {Yang}, \citenamefont {Shimotani},
    \citenamefont {Suzuki}, \citenamefont {Toh}, \citenamefont {Kloc},
    \citenamefont {Xu}, \citenamefont {Arita}, \citenamefont {Nagaosa},\ and\
    \citenamefont {Iwasa}}]{Yuan2013-oa}%
    \BibitemOpen
    \bibfield  {author} {\bibinfo {author} {\bibfnamefont {H.}~\bibnamefont
    {Yuan}}, \bibinfo {author} {\bibfnamefont {M.~S.}\ \bibnamefont {Bahramy}},
    \bibinfo {author} {\bibfnamefont {K.}~\bibnamefont {Morimoto}}, \bibinfo
    {author} {\bibfnamefont {S.}~\bibnamefont {Wu}}, \bibinfo {author}
    {\bibfnamefont {K.}~\bibnamefont {Nomura}}, \bibinfo {author} {\bibfnamefont
    {B.-J.}\ \bibnamefont {Yang}}, \bibinfo {author} {\bibfnamefont
    {H.}~\bibnamefont {Shimotani}}, \bibinfo {author} {\bibfnamefont
    {R.}~\bibnamefont {Suzuki}}, \bibinfo {author} {\bibfnamefont
    {M.}~\bibnamefont {Toh}}, \bibinfo {author} {\bibfnamefont {C.}~\bibnamefont
    {Kloc}}, \bibinfo {author} {\bibfnamefont {X.}~\bibnamefont {Xu}}, \bibinfo
    {author} {\bibfnamefont {R.}~\bibnamefont {Arita}}, \bibinfo {author}
    {\bibfnamefont {N.}~\bibnamefont {Nagaosa}},\ and\ \bibinfo {author}
    {\bibfnamefont {Y.}~\bibnamefont {Iwasa}},\ }\href
    {https://doi.org/10.1038/nphys2691} {\bibfield  {journal} {\bibinfo
    {journal} {Nat. Phys.}\ }\textbf {\bibinfo {volume} {9}},\ \bibinfo {pages}
    {563} (\bibinfo {year} {2013})}\BibitemShut {NoStop}%
  \bibitem [{\citenamefont {Jones}\ \emph {et~al.}(2013)\citenamefont {Jones},
    \citenamefont {Yu}, \citenamefont {Ghimire}, \citenamefont {Wu},
    \citenamefont {Aivazian}, \citenamefont {Ross}, \citenamefont {Zhao},
    \citenamefont {Yan}, \citenamefont {Mandrus}, \citenamefont {Xiao},
    \citenamefont {Yao},\ and\ \citenamefont {Xu}}]{Jones2013-uy}%
    \BibitemOpen
    \bibfield  {author} {\bibinfo {author} {\bibfnamefont {A.~M.}\ \bibnamefont
    {Jones}}, \bibinfo {author} {\bibfnamefont {H.}~\bibnamefont {Yu}}, \bibinfo
    {author} {\bibfnamefont {N.~J.}\ \bibnamefont {Ghimire}}, \bibinfo {author}
    {\bibfnamefont {S.}~\bibnamefont {Wu}}, \bibinfo {author} {\bibfnamefont
    {G.}~\bibnamefont {Aivazian}}, \bibinfo {author} {\bibfnamefont {J.~S.}\
    \bibnamefont {Ross}}, \bibinfo {author} {\bibfnamefont {B.}~\bibnamefont
    {Zhao}}, \bibinfo {author} {\bibfnamefont {J.}~\bibnamefont {Yan}}, \bibinfo
    {author} {\bibfnamefont {D.~G.}\ \bibnamefont {Mandrus}}, \bibinfo {author}
    {\bibfnamefont {D.}~\bibnamefont {Xiao}}, \bibinfo {author} {\bibfnamefont
    {W.}~\bibnamefont {Yao}},\ and\ \bibinfo {author} {\bibfnamefont
    {X.}~\bibnamefont {Xu}},\ }\href {https://doi.org/10.1038/nnano.2013.151}
    {\bibfield  {journal} {\bibinfo  {journal} {Nat. Nanotechnol.}\ }\textbf
    {\bibinfo {volume} {8}},\ \bibinfo {pages} {634} (\bibinfo {year}
    {2013})}\BibitemShut {NoStop}%
  \bibitem [{\citenamefont {Mak}\ \emph {et~al.}(2014)\citenamefont {Mak},
    \citenamefont {McGill}, \citenamefont {Park},\ and\ \citenamefont
    {McEuen}}]{Mak2014-zs}%
    \BibitemOpen
    \bibfield  {author} {\bibinfo {author} {\bibfnamefont {K.~F.}\ \bibnamefont
    {Mak}}, \bibinfo {author} {\bibfnamefont {K.~L.}\ \bibnamefont {McGill}},
    \bibinfo {author} {\bibfnamefont {J.}~\bibnamefont {Park}},\ and\ \bibinfo
    {author} {\bibfnamefont {P.~L.}\ \bibnamefont {McEuen}},\ }\href
    {https://doi.org/10.1126/science.1250140} {\bibfield  {journal} {\bibinfo
    {journal} {Science}\ }\textbf {\bibinfo {volume} {344}},\ \bibinfo {pages}
    {1489} (\bibinfo {year} {2014})}\BibitemShut {NoStop}%
  \bibitem [{\citenamefont {Tong}\ and\ \citenamefont
    {Duan}(2017)}]{Tong2017-bj}%
    \BibitemOpen
    \bibfield  {author} {\bibinfo {author} {\bibfnamefont {W.-Y.}\ \bibnamefont
    {Tong}}\ and\ \bibinfo {author} {\bibfnamefont {C.-G.}\ \bibnamefont
    {Duan}},\ }\href {https://doi.org/10.1038/s41535-017-0051-6} {\bibfield
    {journal} {\bibinfo  {journal} {npj Quantum Mater.}\ }\textbf {\bibinfo
    {volume} {2}},\ \bibinfo {pages} {1} (\bibinfo {year} {2017})}\BibitemShut
    {NoStop}%
  \bibitem [{\citenamefont {Liu}\ \emph {et~al.}(2019)\citenamefont {Liu},
    \citenamefont {Gao}, \citenamefont {Zhang}, \citenamefont {He}, \citenamefont
    {Yu},\ and\ \citenamefont {Liu}}]{Liu2019-ia}%
    \BibitemOpen
    \bibfield  {author} {\bibinfo {author} {\bibfnamefont {Y.}~\bibnamefont
    {Liu}}, \bibinfo {author} {\bibfnamefont {Y.}~\bibnamefont {Gao}}, \bibinfo
    {author} {\bibfnamefont {S.}~\bibnamefont {Zhang}}, \bibinfo {author}
    {\bibfnamefont {J.}~\bibnamefont {He}}, \bibinfo {author} {\bibfnamefont
    {J.}~\bibnamefont {Yu}},\ and\ \bibinfo {author} {\bibfnamefont
    {Z.}~\bibnamefont {Liu}},\ }\href {https://doi.org/10.1007/s12274-019-2497-2}
    {\bibfield  {journal} {\bibinfo  {journal} {Nano Res.}\ }\textbf {\bibinfo
    {volume} {12}},\ \bibinfo {pages} {2695} (\bibinfo {year}
    {2019})}\BibitemShut {NoStop}%
  \bibitem [{\citenamefont {Benda}(1974)}]{PhysRevB.10.1409}%
    \BibitemOpen
    \bibfield  {author} {\bibinfo {author} {\bibfnamefont {J.~A.}\ \bibnamefont
    {Benda}},\ }\href {https://doi.org/10.1103/PhysRevB.10.1409} {\bibfield
    {journal} {\bibinfo  {journal} {Phys. Rev. B}\ }\textbf {\bibinfo {volume}
    {10}},\ \bibinfo {pages} {1409} (\bibinfo {year} {1974})}\BibitemShut
    {NoStop}%
  \bibitem [{\citenamefont {Thompson}\ \emph {et~al.}(1972)\citenamefont
    {Thompson}, \citenamefont {Pisharody},\ and\ \citenamefont
    {Koehler}}]{PhysRevLett.29.163}%
    \BibitemOpen
    \bibfield  {author} {\bibinfo {author} {\bibfnamefont {A.~H.}\ \bibnamefont
    {Thompson}}, \bibinfo {author} {\bibfnamefont {K.~R.}\ \bibnamefont
    {Pisharody}},\ and\ \bibinfo {author} {\bibfnamefont {R.~F.}\ \bibnamefont
    {Koehler}},\ }\href {https://doi.org/10.1103/PhysRevLett.29.163} {\bibfield
    {journal} {\bibinfo  {journal} {Phys. Rev. Lett.}\ }\textbf {\bibinfo
    {volume} {29}},\ \bibinfo {pages} {163} (\bibinfo {year} {1972})}\BibitemShut
    {NoStop}%
  \bibitem [{\citenamefont {Thompson}(1975)}]{PhysRevLett.35.1786}%
    \BibitemOpen
    \bibfield  {author} {\bibinfo {author} {\bibfnamefont {A.~H.}\ \bibnamefont
    {Thompson}},\ }\href {https://doi.org/10.1103/PhysRevLett.35.1786} {\bibfield
     {journal} {\bibinfo  {journal} {Phys. Rev. Lett.}\ }\textbf {\bibinfo
    {volume} {35}},\ \bibinfo {pages} {1786} (\bibinfo {year}
    {1975})}\BibitemShut {NoStop}%
  \bibitem [{\citenamefont {Klipstein}\ \emph {et~al.}(1981)\citenamefont
    {Klipstein}, \citenamefont {Bagnall}, \citenamefont {Liang}, \citenamefont
    {Marseglia},\ and\ \citenamefont {Friend}}]{Klipstein1981-hy}%
    \BibitemOpen
    \bibfield  {author} {\bibinfo {author} {\bibfnamefont {P.~C.}\ \bibnamefont
    {Klipstein}}, \bibinfo {author} {\bibfnamefont {A.~G.}\ \bibnamefont
    {Bagnall}}, \bibinfo {author} {\bibfnamefont {W.~Y.}\ \bibnamefont {Liang}},
    \bibinfo {author} {\bibfnamefont {E.~A.}\ \bibnamefont {Marseglia}},\ and\
    \bibinfo {author} {\bibfnamefont {R.~H.}\ \bibnamefont {Friend}},\ }\href
    {https://doi.org/10.1088/0022-3719/14/28/009} {\bibfield  {journal} {\bibinfo
     {journal} {J. Phys. C: Solid State Phys.}\ }\textbf {\bibinfo {volume}
    {14}},\ \bibinfo {pages} {4067} (\bibinfo {year} {1981})}\BibitemShut
    {NoStop}%
  \bibitem [{\citenamefont {{\=O}nuki}\ \emph {et~al.}(1982)\citenamefont
    {{\=O}nuki}, \citenamefont {Inada},\ and\ \citenamefont
    {Tanuma}}]{Onuki1982-dz}%
    \BibitemOpen
    \bibfield  {author} {\bibinfo {author} {\bibfnamefont {Y.}~\bibnamefont
    {{\=O}nuki}}, \bibinfo {author} {\bibfnamefont {R.}~\bibnamefont {Inada}},\
    and\ \bibinfo {author} {\bibfnamefont {S.-I.}\ \bibnamefont {Tanuma}},\
    }\href {https://doi.org/10.1143/JPSJ.51.1223} {\bibfield  {journal} {\bibinfo
     {journal} {J. Phys. Soc. Jpn.}\ }\textbf {\bibinfo {volume} {51}},\ \bibinfo
    {pages} {1223} (\bibinfo {year} {1982})}\BibitemShut {NoStop}%
  \bibitem [{\citenamefont {Zheng}\ \emph {et~al.}(1989)\citenamefont {Zheng},
    \citenamefont {Kuriyaki},\ and\ \citenamefont {Hirakawa}}]{Zheng1989-li}%
    \BibitemOpen
    \bibfield  {author} {\bibinfo {author} {\bibfnamefont {X.-G.}\ \bibnamefont
    {Zheng}}, \bibinfo {author} {\bibfnamefont {H.}~\bibnamefont {Kuriyaki}},\
    and\ \bibinfo {author} {\bibfnamefont {K.}~\bibnamefont {Hirakawa}},\ }\href
    {https://doi.org/10.1143/JPSJ.58.622} {\bibfield  {journal} {\bibinfo
    {journal} {J. Phys. Soc. Jpn.}\ }\textbf {\bibinfo {volume} {58}},\ \bibinfo
    {pages} {622} (\bibinfo {year} {1989})}\BibitemShut {NoStop}%
  \bibitem [{\citenamefont {Patel}\ \emph {et~al.}(1998)\citenamefont {Patel},
    \citenamefont {Agarwal}, \citenamefont {Batra},\ and\ \citenamefont
    {Lakshminarayana}}]{Patel1998-zd}%
    \BibitemOpen
    \bibfield  {author} {\bibinfo {author} {\bibfnamefont {S.~G.}\ \bibnamefont
    {Patel}}, \bibinfo {author} {\bibfnamefont {M.~K.}\ \bibnamefont {Agarwal}},
    \bibinfo {author} {\bibfnamefont {N.~M.}\ \bibnamefont {Batra}},\ and\
    \bibinfo {author} {\bibfnamefont {D.}~\bibnamefont {Lakshminarayana}},\
    }\href {https://doi.org/10.1007/bf02744972} {\bibfield  {journal} {\bibinfo
    {journal} {Bull. Mater. Sci.}\ }\textbf {\bibinfo {volume} {21}},\ \bibinfo
    {pages} {213} (\bibinfo {year} {1998})}\BibitemShut {NoStop}%
  \bibitem [{\citenamefont {Zandt}\ \emph {et~al.}(2007)\citenamefont {Zandt},
    \citenamefont {Dwelk}, \citenamefont {Janowitz},\ and\ \citenamefont
    {Manzke}}]{Zandt2007-lm}%
    \BibitemOpen
    \bibfield  {author} {\bibinfo {author} {\bibfnamefont {T.}~\bibnamefont
    {Zandt}}, \bibinfo {author} {\bibfnamefont {H.}~\bibnamefont {Dwelk}},
    \bibinfo {author} {\bibfnamefont {C.}~\bibnamefont {Janowitz}},\ and\
    \bibinfo {author} {\bibfnamefont {R.}~\bibnamefont {Manzke}},\ }\href
    {https://doi.org/10.1016/j.jallcom.2006.09.157} {\bibfield  {journal}
    {\bibinfo  {journal} {J. Alloys Compd.}\ }\textbf {\bibinfo {volume} {442}},\
    \bibinfo {pages} {216} (\bibinfo {year} {2007})}\BibitemShut {NoStop}%
  \bibitem [{\citenamefont {Suri}\ \emph {et~al.}(2017)\citenamefont {Suri},
    \citenamefont {Siva}, \citenamefont {Joshi}, \citenamefont {Senapati},
    \citenamefont {Sahoo}, \citenamefont {Varma},\ and\ \citenamefont
    {Patel}}]{Suri_2017}%
    \BibitemOpen
    \bibfield  {author} {\bibinfo {author} {\bibfnamefont {D.}~\bibnamefont
    {Suri}}, \bibinfo {author} {\bibfnamefont {V.}~\bibnamefont {Siva}}, \bibinfo
    {author} {\bibfnamefont {S.}~\bibnamefont {Joshi}}, \bibinfo {author}
    {\bibfnamefont {K.}~\bibnamefont {Senapati}}, \bibinfo {author}
    {\bibfnamefont {P.~K.}\ \bibnamefont {Sahoo}}, \bibinfo {author}
    {\bibfnamefont {S.}~\bibnamefont {Varma}},\ and\ \bibinfo {author}
    {\bibfnamefont {R.~S.}\ \bibnamefont {Patel}},\ }\href
    {https://doi.org/10.1088/1361-648x/aa90c5} {\bibfield  {journal} {\bibinfo
    {journal} {J. Phys.: Condens. Matter}\ }\textbf {\bibinfo {volume} {29}},\
    \bibinfo {pages} {485708} (\bibinfo {year} {2017})}\BibitemShut {NoStop}%
  \bibitem [{\citenamefont {Herring}(1955)}]{Herring1955-nt}%
    \BibitemOpen
    \bibfield  {author} {\bibinfo {author} {\bibfnamefont {C.}~\bibnamefont
    {Herring}},\ }\href {https://doi.org/10.1002/j.1538-7305.1955.tb01472.x}
    {\bibfield  {journal} {\bibinfo  {journal} {Bell Syst. Tech. J.}\ }\textbf
    {\bibinfo {volume} {34}},\ \bibinfo {pages} {237} (\bibinfo {year}
    {1955})}\BibitemShut {NoStop}%
  \bibitem [{\citenamefont {Popescu}\ and\ \citenamefont
    {Woods}(2012)}]{Popescu2012-xy}%
    \BibitemOpen
    \bibfield  {author} {\bibinfo {author} {\bibfnamefont {A.}~\bibnamefont
    {Popescu}}\ and\ \bibinfo {author} {\bibfnamefont {L.~M.}\ \bibnamefont
    {Woods}},\ }\href {https://doi.org/10.1002/adfm.201200818} {\bibfield
    {journal} {\bibinfo  {journal} {Adv. Funct. Mater.}\ }\textbf {\bibinfo
    {volume} {22}},\ \bibinfo {pages} {3945} (\bibinfo {year}
    {2012})}\BibitemShut {NoStop}%
  \bibitem [{\citenamefont {Pei}\ \emph {et~al.}(2012)\citenamefont {Pei},
    \citenamefont {Wang},\ and\ \citenamefont {Snyder}}]{Pei2012-uj}%
    \BibitemOpen
    \bibfield  {author} {\bibinfo {author} {\bibfnamefont {Y.}~\bibnamefont
    {Pei}}, \bibinfo {author} {\bibfnamefont {H.}~\bibnamefont {Wang}},\ and\
    \bibinfo {author} {\bibfnamefont {G.~J.}\ \bibnamefont {Snyder}},\ }\href
    {https://doi.org/10.1002/adma.201202919} {\bibfield  {journal} {\bibinfo
    {journal} {Adv. Mater.}\ }\textbf {\bibinfo {volume} {24}},\ \bibinfo {pages}
    {6125} (\bibinfo {year} {2012})}\BibitemShut {NoStop}%
  \bibitem [{\citenamefont {Xin}\ \emph {et~al.}(2018)\citenamefont {Xin},
    \citenamefont {Tang}, \citenamefont {Liu}, \citenamefont {Zhao},
    \citenamefont {Pan},\ and\ \citenamefont {Zhu}}]{Xin2018-zw}%
    \BibitemOpen
    \bibfield  {author} {\bibinfo {author} {\bibfnamefont {J.}~\bibnamefont
    {Xin}}, \bibinfo {author} {\bibfnamefont {Y.}~\bibnamefont {Tang}}, \bibinfo
    {author} {\bibfnamefont {Y.}~\bibnamefont {Liu}}, \bibinfo {author}
    {\bibfnamefont {X.}~\bibnamefont {Zhao}}, \bibinfo {author} {\bibfnamefont
    {H.}~\bibnamefont {Pan}},\ and\ \bibinfo {author} {\bibfnamefont
    {T.}~\bibnamefont {Zhu}},\ }\href {https://doi.org/10.1038/s41535-018-0083-6}
    {\bibfield  {journal} {\bibinfo  {journal} {npj Quantum Materials}\ }\textbf
    {\bibinfo {volume} {3}},\ \bibinfo {pages} {1} (\bibinfo {year}
    {2018})}\BibitemShut {NoStop}%
  \bibitem [{\citenamefont {Wilson}(1977)}]{Wilson1977-mt}%
    \BibitemOpen
    \bibfield  {author} {\bibinfo {author} {\bibfnamefont {J.~A.}\ \bibnamefont
    {Wilson}},\ }\href {https://doi.org/10.1016/0038-1098(77)90133-8} {\bibfield
    {journal} {\bibinfo  {journal} {Solid State Commun.}\ }\textbf {\bibinfo
    {volume} {22}},\ \bibinfo {pages} {551} (\bibinfo {year} {1977})}\BibitemShut
    {NoStop}%
  \bibitem [{\citenamefont {Wilson}(1978)}]{Wilson1978-zs}%
    \BibitemOpen
    \bibfield  {author} {\bibinfo {author} {\bibfnamefont {J.~A.}\ \bibnamefont
    {Wilson}},\ }\href {https://doi.org/10.1002/pssb.2220860102} {\bibfield
    {journal} {\bibinfo  {journal} {Phys. Status Solidi B Basic Res.}\ }\textbf
    {\bibinfo {volume} {86}},\ \bibinfo {pages} {11} (\bibinfo {year}
    {1978})}\BibitemShut {NoStop}%
  \bibitem [{\citenamefont {Kukkonen}\ \emph {et~al.}(1981)\citenamefont
    {Kukkonen}, \citenamefont {Kaiser}, \citenamefont {Logothetis}, \citenamefont
    {Blumenstock}, \citenamefont {Schroeder}, \citenamefont {Faile},
    \citenamefont {Colella},\ and\ \citenamefont {Gambold}}]{PhysRevB.24.1691}%
    \BibitemOpen
    \bibfield  {author} {\bibinfo {author} {\bibfnamefont {C.~A.}\ \bibnamefont
    {Kukkonen}}, \bibinfo {author} {\bibfnamefont {W.~J.}\ \bibnamefont
    {Kaiser}}, \bibinfo {author} {\bibfnamefont {E.~M.}\ \bibnamefont
    {Logothetis}}, \bibinfo {author} {\bibfnamefont {B.~J.}\ \bibnamefont
    {Blumenstock}}, \bibinfo {author} {\bibfnamefont {P.~A.}\ \bibnamefont
    {Schroeder}}, \bibinfo {author} {\bibfnamefont {S.~P.}\ \bibnamefont
    {Faile}}, \bibinfo {author} {\bibfnamefont {R.}~\bibnamefont {Colella}},\
    and\ \bibinfo {author} {\bibfnamefont {J.}~\bibnamefont {Gambold}},\ }\href
    {https://doi.org/10.1103/PhysRevB.24.1691} {\bibfield  {journal} {\bibinfo
    {journal} {Phys. Rev. B}\ }\textbf {\bibinfo {volume} {24}},\ \bibinfo
    {pages} {1691} (\bibinfo {year} {1981})}\BibitemShut {NoStop}%
  \bibitem [{\citenamefont {Fivaz}\ and\ \citenamefont
    {Mooser}(1967)}]{PhysRev.163.743}%
    \BibitemOpen
    \bibfield  {author} {\bibinfo {author} {\bibfnamefont {R.}~\bibnamefont
    {Fivaz}}\ and\ \bibinfo {author} {\bibfnamefont {E.}~\bibnamefont {Mooser}},\
    }\href {https://doi.org/10.1103/PhysRev.163.743} {\bibfield  {journal}
    {\bibinfo  {journal} {Phys. Rev.}\ }\textbf {\bibinfo {volume} {163}},\
    \bibinfo {pages} {743} (\bibinfo {year} {1967})}\BibitemShut {NoStop}%
  \bibitem [{\citenamefont {Maldague}\ and\ \citenamefont
    {Kukkonen}(1979)}]{PhysRevB.19.6172}%
    \BibitemOpen
    \bibfield  {author} {\bibinfo {author} {\bibfnamefont {P.~F.}\ \bibnamefont
    {Maldague}}\ and\ \bibinfo {author} {\bibfnamefont {C.~A.}\ \bibnamefont
    {Kukkonen}},\ }\href {https://doi.org/10.1103/PhysRevB.19.6172} {\bibfield
    {journal} {\bibinfo  {journal} {Phys. Rev. B}\ }\textbf {\bibinfo {volume}
    {19}},\ \bibinfo {pages} {6172} (\bibinfo {year} {1979})}\BibitemShut
    {NoStop}%
  \bibitem [{\citenamefont {Kukkonen}\ and\ \citenamefont
    {Maldague}(1976)}]{PhysRevLett.37.782}%
    \BibitemOpen
    \bibfield  {author} {\bibinfo {author} {\bibfnamefont {C.~A.}\ \bibnamefont
    {Kukkonen}}\ and\ \bibinfo {author} {\bibfnamefont {P.~F.}\ \bibnamefont
    {Maldague}},\ }\href {https://doi.org/10.1103/PhysRevLett.37.782} {\bibfield
    {journal} {\bibinfo  {journal} {Phys. Rev. Lett.}\ }\textbf {\bibinfo
    {volume} {37}},\ \bibinfo {pages} {782} (\bibinfo {year} {1976})}\BibitemShut
    {NoStop}%
  \bibitem [{\citenamefont {Kaasbjerg}\ \emph {et~al.}(2012)\citenamefont
    {Kaasbjerg}, \citenamefont {Thygesen},\ and\ \citenamefont
    {Jacobsen}}]{Kaasbjerg2012-za}%
    \BibitemOpen
    \bibfield  {author} {\bibinfo {author} {\bibfnamefont {K.}~\bibnamefont
    {Kaasbjerg}}, \bibinfo {author} {\bibfnamefont {K.~S.}\ \bibnamefont
    {Thygesen}},\ and\ \bibinfo {author} {\bibfnamefont {K.~W.}\ \bibnamefont
    {Jacobsen}},\ }\href {https://doi.org/10.1103/PhysRevB.85.115317} {\bibfield
    {journal} {\bibinfo  {journal} {Phys. Rev. B}\ }\textbf {\bibinfo {volume}
    {85}},\ \bibinfo {pages} {115317} (\bibinfo {year} {2012})}\BibitemShut
    {NoStop}%
  \bibitem [{\citenamefont {Zhao}\ \emph {et~al.}(2018)\citenamefont {Zhao},
    \citenamefont {Dai}, \citenamefont {Zhang}, \citenamefont {Lian},
    \citenamefont {Zeng}, \citenamefont {Li}, \citenamefont {Meng},\ and\
    \citenamefont {Ni}}]{Zhao2018-yq}%
    \BibitemOpen
    \bibfield  {author} {\bibinfo {author} {\bibfnamefont {Y.}~\bibnamefont
    {Zhao}}, \bibinfo {author} {\bibfnamefont {Z.}~\bibnamefont {Dai}}, \bibinfo
    {author} {\bibfnamefont {C.}~\bibnamefont {Zhang}}, \bibinfo {author}
    {\bibfnamefont {C.}~\bibnamefont {Lian}}, \bibinfo {author} {\bibfnamefont
    {S.}~\bibnamefont {Zeng}}, \bibinfo {author} {\bibfnamefont {G.}~\bibnamefont
    {Li}}, \bibinfo {author} {\bibfnamefont {S.}~\bibnamefont {Meng}},\ and\
    \bibinfo {author} {\bibfnamefont {J.}~\bibnamefont {Ni}},\ }\href
    {https://doi.org/10.1088/1367-2630/aab338} {\bibfield  {journal} {\bibinfo
    {journal} {New J. Phys.}\ }\textbf {\bibinfo {volume} {20}},\ \bibinfo
    {pages} {043009} (\bibinfo {year} {2018})}\BibitemShut {NoStop}%
  \bibitem [{\citenamefont {T.~Hung}\ \emph {et~al.}(2019)\citenamefont
    {T.~Hung}, \citenamefont {Nugraha}, \citenamefont {Yang}, \citenamefont
    {Zhang},\ and\ \citenamefont {Saito}}]{doi:10.1063/1.5040752}%
    \BibitemOpen
    \bibfield  {author} {\bibinfo {author} {\bibfnamefont {N.}~\bibnamefont
    {T.~Hung}}, \bibinfo {author} {\bibfnamefont {A.~R.~T.}\ \bibnamefont
    {Nugraha}}, \bibinfo {author} {\bibfnamefont {T.}~\bibnamefont {Yang}},
    \bibinfo {author} {\bibfnamefont {Z.}~\bibnamefont {Zhang}},\ and\ \bibinfo
    {author} {\bibfnamefont {R.}~\bibnamefont {Saito}},\ }\href
    {https://doi.org/10.1063/1.5040752} {\bibfield  {journal} {\bibinfo
    {journal} {J. Appl. Phys.}\ }\textbf {\bibinfo {volume} {125}},\ \bibinfo
    {pages} {082502} (\bibinfo {year} {2019})}\BibitemShut {NoStop}%
  \bibitem [{\citenamefont {Sohier}\ \emph {et~al.}(2019)\citenamefont {Sohier},
    \citenamefont {Ponomarev}, \citenamefont {Gibertini}, \citenamefont {Berger},
    \citenamefont {Marzari}, \citenamefont {Ubrig},\ and\ \citenamefont
    {Morpurgo}}]{PhysRevX.9.031019}%
    \BibitemOpen
    \bibfield  {author} {\bibinfo {author} {\bibfnamefont {T.}~\bibnamefont
    {Sohier}}, \bibinfo {author} {\bibfnamefont {E.}~\bibnamefont {Ponomarev}},
    \bibinfo {author} {\bibfnamefont {M.}~\bibnamefont {Gibertini}}, \bibinfo
    {author} {\bibfnamefont {H.}~\bibnamefont {Berger}}, \bibinfo {author}
    {\bibfnamefont {N.}~\bibnamefont {Marzari}}, \bibinfo {author} {\bibfnamefont
    {N.}~\bibnamefont {Ubrig}},\ and\ \bibinfo {author} {\bibfnamefont {A.~F.}\
    \bibnamefont {Morpurgo}},\ }\href {https://doi.org/10.1103/PhysRevX.9.031019}
    {\bibfield  {journal} {\bibinfo  {journal} {Phys. Rev. X}\ }\textbf {\bibinfo
    {volume} {9}},\ \bibinfo {pages} {031019} (\bibinfo {year}
    {2019})}\BibitemShut {NoStop}%
  \bibitem [{\citenamefont {Wu}\ \emph {et~al.}(2021)\citenamefont {Wu},
    \citenamefont {Hou}, \citenamefont {Ma}, \citenamefont {Cao}, \citenamefont
    {Chen}, \citenamefont {Lu}, \citenamefont {Mei}, \citenamefont {Shao},
    \citenamefont {Xu}, \citenamefont {Zhu}, \citenamefont {Fang}, \citenamefont
    {Zhang},\ and\ \citenamefont {Zhang}}]{Wu2021-db}%
    \BibitemOpen
    \bibfield  {author} {\bibinfo {author} {\bibfnamefont {Y.}~\bibnamefont
    {Wu}}, \bibinfo {author} {\bibfnamefont {B.}~\bibnamefont {Hou}}, \bibinfo
    {author} {\bibfnamefont {C.}~\bibnamefont {Ma}}, \bibinfo {author}
    {\bibfnamefont {J.}~\bibnamefont {Cao}}, \bibinfo {author} {\bibfnamefont
    {Y.}~\bibnamefont {Chen}}, \bibinfo {author} {\bibfnamefont {Z.}~\bibnamefont
    {Lu}}, \bibinfo {author} {\bibfnamefont {H.}~\bibnamefont {Mei}}, \bibinfo
    {author} {\bibfnamefont {H.}~\bibnamefont {Shao}}, \bibinfo {author}
    {\bibfnamefont {Y.}~\bibnamefont {Xu}}, \bibinfo {author} {\bibfnamefont
    {H.}~\bibnamefont {Zhu}}, \bibinfo {author} {\bibfnamefont {Z.}~\bibnamefont
    {Fang}}, \bibinfo {author} {\bibfnamefont {R.}~\bibnamefont {Zhang}},\ and\
    \bibinfo {author} {\bibfnamefont {H.}~\bibnamefont {Zhang}},\ }\href
    {https://doi.org/10.1039/D0MH01802C} {\bibfield  {journal} {\bibinfo
    {journal} {Mater. Horiz.}\ }\textbf {\bibinfo {volume} {8}},\ \bibinfo
    {pages} {1253} (\bibinfo {year} {2021})}\BibitemShut {NoStop}%
  \bibitem [{\citenamefont {Giannozzi}\ \emph {et~al.}(2009)\citenamefont
    {Giannozzi}, \citenamefont {Baroni}, \citenamefont {Bonini}, \citenamefont
    {Calandra}, \citenamefont {Car}, \citenamefont {Cavazzoni}, \citenamefont
    {Ceresoli}, \citenamefont {Chiarotti}, \citenamefont {Cococcioni},
    \citenamefont {Dabo}, \citenamefont {Corso}, \citenamefont {de~Gironcoli},
    \citenamefont {Fabris}, \citenamefont {Fratesi}, \citenamefont {Gebauer},
    \citenamefont {Gerstmann}, \citenamefont {Gougoussis}, \citenamefont
    {Kokalj}, \citenamefont {Lazzeri}, \citenamefont {Martin-Samos},
    \citenamefont {Marzari}, \citenamefont {Mauri}, \citenamefont {Mazzarello},
    \citenamefont {Paolini}, \citenamefont {Pasquarello}, \citenamefont
    {Paulatto}, \citenamefont {Sbraccia}, \citenamefont {Scandolo}, \citenamefont
    {Sclauzero}, \citenamefont {Seitsonen}, \citenamefont {Smogunov},
    \citenamefont {Umari},\ and\ \citenamefont {Wentzcovitch}}]{QE-2009}%
    \BibitemOpen
    \bibfield  {author} {\bibinfo {author} {\bibfnamefont {P.}~\bibnamefont
    {Giannozzi}}, \bibinfo {author} {\bibfnamefont {S.}~\bibnamefont {Baroni}},
    \bibinfo {author} {\bibfnamefont {N.}~\bibnamefont {Bonini}}, \bibinfo
    {author} {\bibfnamefont {M.}~\bibnamefont {Calandra}}, \bibinfo {author}
    {\bibfnamefont {R.}~\bibnamefont {Car}}, \bibinfo {author} {\bibfnamefont
    {C.}~\bibnamefont {Cavazzoni}}, \bibinfo {author} {\bibfnamefont
    {D.}~\bibnamefont {Ceresoli}}, \bibinfo {author} {\bibfnamefont {G.~L.}\
    \bibnamefont {Chiarotti}}, \bibinfo {author} {\bibfnamefont {M.}~\bibnamefont
    {Cococcioni}}, \bibinfo {author} {\bibfnamefont {I.}~\bibnamefont {Dabo}},
    \bibinfo {author} {\bibfnamefont {A.~D.}\ \bibnamefont {Corso}}, \bibinfo
    {author} {\bibfnamefont {S.}~\bibnamefont {de~Gironcoli}}, \bibinfo {author}
    {\bibfnamefont {S.}~\bibnamefont {Fabris}}, \bibinfo {author} {\bibfnamefont
    {G.}~\bibnamefont {Fratesi}}, \bibinfo {author} {\bibfnamefont
    {R.}~\bibnamefont {Gebauer}}, \bibinfo {author} {\bibfnamefont
    {U.}~\bibnamefont {Gerstmann}}, \bibinfo {author} {\bibfnamefont
    {C.}~\bibnamefont {Gougoussis}}, \bibinfo {author} {\bibfnamefont
    {A.}~\bibnamefont {Kokalj}}, \bibinfo {author} {\bibfnamefont
    {M.}~\bibnamefont {Lazzeri}}, \bibinfo {author} {\bibfnamefont
    {L.}~\bibnamefont {Martin-Samos}}, \bibinfo {author} {\bibfnamefont
    {N.}~\bibnamefont {Marzari}}, \bibinfo {author} {\bibfnamefont
    {F.}~\bibnamefont {Mauri}}, \bibinfo {author} {\bibfnamefont
    {R.}~\bibnamefont {Mazzarello}}, \bibinfo {author} {\bibfnamefont
    {S.}~\bibnamefont {Paolini}}, \bibinfo {author} {\bibfnamefont
    {A.}~\bibnamefont {Pasquarello}}, \bibinfo {author} {\bibfnamefont
    {L.}~\bibnamefont {Paulatto}}, \bibinfo {author} {\bibfnamefont
    {C.}~\bibnamefont {Sbraccia}}, \bibinfo {author} {\bibfnamefont
    {S.}~\bibnamefont {Scandolo}}, \bibinfo {author} {\bibfnamefont
    {G.}~\bibnamefont {Sclauzero}}, \bibinfo {author} {\bibfnamefont {A.~P.}\
    \bibnamefont {Seitsonen}}, \bibinfo {author} {\bibfnamefont {A.}~\bibnamefont
    {Smogunov}}, \bibinfo {author} {\bibfnamefont {P.}~\bibnamefont {Umari}},\
    and\ \bibinfo {author} {\bibfnamefont {R.~M.}\ \bibnamefont {Wentzcovitch}},\
    }\href {https://doi.org/10.1088/0953-8984/21/39/395502} {\bibfield  {journal}
    {\bibinfo  {journal} {J. Phys.: Condens. Matter}\ }\textbf {\bibinfo {volume}
    {21}},\ \bibinfo {pages} {395502} (\bibinfo {year} {2009})}\BibitemShut
    {NoStop}%
  \bibitem [{\citenamefont {Giannozzi}\ \emph {et~al.}(2017)\citenamefont
    {Giannozzi}, \citenamefont {Andreussi}, \citenamefont {Brumme}, \citenamefont
    {Bunau}, \citenamefont {Nardelli}, \citenamefont {Calandra}, \citenamefont
    {Car}, \citenamefont {Cavazzoni}, \citenamefont {Ceresoli}, \citenamefont
    {Cococcioni}, \citenamefont {Colonna}, \citenamefont {Carnimeo},
    \citenamefont {Corso}, \citenamefont {de~Gironcoli}, \citenamefont {Delugas},
    \citenamefont {DiStasio}, \citenamefont {Ferretti}, \citenamefont {Floris},
    \citenamefont {Fratesi}, \citenamefont {Fugallo}, \citenamefont {Gebauer},
    \citenamefont {Gerstmann}, \citenamefont {Giustino}, \citenamefont {Gorni},
    \citenamefont {Jia}, \citenamefont {Kawamura}, \citenamefont {Ko},
    \citenamefont {Kokalj}, \citenamefont {K{\"{u}}{\c{c}}{\"{u}}kbenli},
    \citenamefont {Lazzeri}, \citenamefont {Marsili}, \citenamefont {Marzari},
    \citenamefont {Mauri}, \citenamefont {Nguyen}, \citenamefont {Nguyen},
    \citenamefont {de-la Roza}, \citenamefont {Paulatto}, \citenamefont
    {Ponc{\'{e}}}, \citenamefont {Rocca}, \citenamefont {Sabatini}, \citenamefont
    {Santra}, \citenamefont {Schlipf}, \citenamefont {Seitsonen}, \citenamefont
    {Smogunov}, \citenamefont {Timrov}, \citenamefont {Thonhauser}, \citenamefont
    {Umari}, \citenamefont {Vast}, \citenamefont {Wu},\ and\ \citenamefont
    {Baroni}}]{QE-2017}%
    \BibitemOpen
    \bibfield  {author} {\bibinfo {author} {\bibfnamefont {P.}~\bibnamefont
    {Giannozzi}}, \bibinfo {author} {\bibfnamefont {O.}~\bibnamefont
    {Andreussi}}, \bibinfo {author} {\bibfnamefont {T.}~\bibnamefont {Brumme}},
    \bibinfo {author} {\bibfnamefont {O.}~\bibnamefont {Bunau}}, \bibinfo
    {author} {\bibfnamefont {M.~B.}\ \bibnamefont {Nardelli}}, \bibinfo {author}
    {\bibfnamefont {M.}~\bibnamefont {Calandra}}, \bibinfo {author}
    {\bibfnamefont {R.}~\bibnamefont {Car}}, \bibinfo {author} {\bibfnamefont
    {C.}~\bibnamefont {Cavazzoni}}, \bibinfo {author} {\bibfnamefont
    {D.}~\bibnamefont {Ceresoli}}, \bibinfo {author} {\bibfnamefont
    {M.}~\bibnamefont {Cococcioni}}, \bibinfo {author} {\bibfnamefont
    {N.}~\bibnamefont {Colonna}}, \bibinfo {author} {\bibfnamefont
    {I.}~\bibnamefont {Carnimeo}}, \bibinfo {author} {\bibfnamefont {A.~D.}\
    \bibnamefont {Corso}}, \bibinfo {author} {\bibfnamefont {S.}~\bibnamefont
    {de~Gironcoli}}, \bibinfo {author} {\bibfnamefont {P.}~\bibnamefont
    {Delugas}}, \bibinfo {author} {\bibfnamefont {R.~A.}\ \bibnamefont
    {DiStasio}}, \bibinfo {author} {\bibfnamefont {A.}~\bibnamefont {Ferretti}},
    \bibinfo {author} {\bibfnamefont {A.}~\bibnamefont {Floris}}, \bibinfo
    {author} {\bibfnamefont {G.}~\bibnamefont {Fratesi}}, \bibinfo {author}
    {\bibfnamefont {G.}~\bibnamefont {Fugallo}}, \bibinfo {author} {\bibfnamefont
    {R.}~\bibnamefont {Gebauer}}, \bibinfo {author} {\bibfnamefont
    {U.}~\bibnamefont {Gerstmann}}, \bibinfo {author} {\bibfnamefont
    {F.}~\bibnamefont {Giustino}}, \bibinfo {author} {\bibfnamefont
    {T.}~\bibnamefont {Gorni}}, \bibinfo {author} {\bibfnamefont
    {J.}~\bibnamefont {Jia}}, \bibinfo {author} {\bibfnamefont {M.}~\bibnamefont
    {Kawamura}}, \bibinfo {author} {\bibfnamefont {H.-Y.}\ \bibnamefont {Ko}},
    \bibinfo {author} {\bibfnamefont {A.}~\bibnamefont {Kokalj}}, \bibinfo
    {author} {\bibfnamefont {E.}~\bibnamefont {K{\"{u}}{\c{c}}{\"{u}}kbenli}},
    \bibinfo {author} {\bibfnamefont {M.}~\bibnamefont {Lazzeri}}, \bibinfo
    {author} {\bibfnamefont {M.}~\bibnamefont {Marsili}}, \bibinfo {author}
    {\bibfnamefont {N.}~\bibnamefont {Marzari}}, \bibinfo {author} {\bibfnamefont
    {F.}~\bibnamefont {Mauri}}, \bibinfo {author} {\bibfnamefont {N.~L.}\
    \bibnamefont {Nguyen}}, \bibinfo {author} {\bibfnamefont {H.-V.}\
    \bibnamefont {Nguyen}}, \bibinfo {author} {\bibfnamefont {A.~O.}\
    \bibnamefont {de-la Roza}}, \bibinfo {author} {\bibfnamefont
    {L.}~\bibnamefont {Paulatto}}, \bibinfo {author} {\bibfnamefont
    {S.}~\bibnamefont {Ponc{\'{e}}}}, \bibinfo {author} {\bibfnamefont
    {D.}~\bibnamefont {Rocca}}, \bibinfo {author} {\bibfnamefont
    {R.}~\bibnamefont {Sabatini}}, \bibinfo {author} {\bibfnamefont
    {B.}~\bibnamefont {Santra}}, \bibinfo {author} {\bibfnamefont
    {M.}~\bibnamefont {Schlipf}}, \bibinfo {author} {\bibfnamefont {A.~P.}\
    \bibnamefont {Seitsonen}}, \bibinfo {author} {\bibfnamefont {A.}~\bibnamefont
    {Smogunov}}, \bibinfo {author} {\bibfnamefont {I.}~\bibnamefont {Timrov}},
    \bibinfo {author} {\bibfnamefont {T.}~\bibnamefont {Thonhauser}}, \bibinfo
    {author} {\bibfnamefont {P.}~\bibnamefont {Umari}}, \bibinfo {author}
    {\bibfnamefont {N.}~\bibnamefont {Vast}}, \bibinfo {author} {\bibfnamefont
    {X.}~\bibnamefont {Wu}},\ and\ \bibinfo {author} {\bibfnamefont
    {S.}~\bibnamefont {Baroni}},\ }\href
    {https://doi.org/10.1088/1361-648x/aa8f79} {\bibfield  {journal} {\bibinfo
    {journal} {J. Phys.: Condens. Matter}\ }\textbf {\bibinfo {volume} {29}},\
    \bibinfo {pages} {465901} (\bibinfo {year} {2017})}\BibitemShut {NoStop}%
  \bibitem [{\citenamefont {Giannozzi}\ \emph {et~al.}(2020)\citenamefont
    {Giannozzi}, \citenamefont {Baseggio}, \citenamefont {Bonf{\`{a}}},
    \citenamefont {Brunato}, \citenamefont {Car}, \citenamefont {Carnimeo},
    \citenamefont {Cavazzoni}, \citenamefont {de~Gironcoli}, \citenamefont
    {Delugas}, \citenamefont {Ferrari~Ruffino}, \citenamefont {Ferretti},
    \citenamefont {Marzari}, \citenamefont {Timrov}, \citenamefont {Urru},\ and\
    \citenamefont {Baroni}}]{doi:10.1063/5.0005082}%
    \BibitemOpen
    \bibfield  {author} {\bibinfo {author} {\bibfnamefont {P.}~\bibnamefont
    {Giannozzi}}, \bibinfo {author} {\bibfnamefont {O.}~\bibnamefont {Baseggio}},
    \bibinfo {author} {\bibfnamefont {P.}~\bibnamefont {Bonf{\`{a}}}}, \bibinfo
    {author} {\bibfnamefont {D.}~\bibnamefont {Brunato}}, \bibinfo {author}
    {\bibfnamefont {R.}~\bibnamefont {Car}}, \bibinfo {author} {\bibfnamefont
    {I.}~\bibnamefont {Carnimeo}}, \bibinfo {author} {\bibfnamefont
    {C.}~\bibnamefont {Cavazzoni}}, \bibinfo {author} {\bibfnamefont
    {S.}~\bibnamefont {de~Gironcoli}}, \bibinfo {author} {\bibfnamefont
    {P.}~\bibnamefont {Delugas}}, \bibinfo {author} {\bibfnamefont
    {F.}~\bibnamefont {Ferrari~Ruffino}}, \bibinfo {author} {\bibfnamefont
    {A.}~\bibnamefont {Ferretti}}, \bibinfo {author} {\bibfnamefont
    {N.}~\bibnamefont {Marzari}}, \bibinfo {author} {\bibfnamefont
    {I.}~\bibnamefont {Timrov}}, \bibinfo {author} {\bibfnamefont
    {A.}~\bibnamefont {Urru}},\ and\ \bibinfo {author} {\bibfnamefont
    {S.}~\bibnamefont {Baroni}},\ }\href {https://doi.org/10.1063/5.0005082}
    {\bibfield  {journal} {\bibinfo  {journal} {J. Chem. Phys.}\ }\textbf
    {\bibinfo {volume} {152}},\ \bibinfo {pages} {154105} (\bibinfo {year}
    {2020})}\BibitemShut {NoStop}%
  \bibitem [{\citenamefont {Constantin}\ \emph {et~al.}(2009)\citenamefont
    {Constantin}, \citenamefont {Perdew},\ and\ \citenamefont
    {Pitarke}}]{PhysRevB.79.075126}%
    \BibitemOpen
    \bibfield  {author} {\bibinfo {author} {\bibfnamefont {L.~A.}\ \bibnamefont
    {Constantin}}, \bibinfo {author} {\bibfnamefont {J.~P.}\ \bibnamefont
    {Perdew}},\ and\ \bibinfo {author} {\bibfnamefont {J.~M.}\ \bibnamefont
    {Pitarke}},\ }\href {https://doi.org/10.1103/PhysRevB.79.075126} {\bibfield
    {journal} {\bibinfo  {journal} {Phys. Rev. B}\ }\textbf {\bibinfo {volume}
    {79}},\ \bibinfo {pages} {075126} (\bibinfo {year} {2009})}\BibitemShut
    {NoStop}%
  \bibitem [{\citenamefont {Perdew}\ \emph {et~al.}(2008)\citenamefont {Perdew},
    \citenamefont {Ruzsinszky}, \citenamefont {Csonka}, \citenamefont {Vydrov},
    \citenamefont {Scuseria}, \citenamefont {Constantin}, \citenamefont {Zhou},\
    and\ \citenamefont {Burke}}]{PhysRevLett.100.136406}%
    \BibitemOpen
    \bibfield  {author} {\bibinfo {author} {\bibfnamefont {J.~P.}\ \bibnamefont
    {Perdew}}, \bibinfo {author} {\bibfnamefont {A.}~\bibnamefont {Ruzsinszky}},
    \bibinfo {author} {\bibfnamefont {G.~I.}\ \bibnamefont {Csonka}}, \bibinfo
    {author} {\bibfnamefont {O.~A.}\ \bibnamefont {Vydrov}}, \bibinfo {author}
    {\bibfnamefont {G.~E.}\ \bibnamefont {Scuseria}}, \bibinfo {author}
    {\bibfnamefont {L.~A.}\ \bibnamefont {Constantin}}, \bibinfo {author}
    {\bibfnamefont {X.}~\bibnamefont {Zhou}},\ and\ \bibinfo {author}
    {\bibfnamefont {K.}~\bibnamefont {Burke}},\ }\href
    {https://doi.org/10.1103/PhysRevLett.100.136406} {\bibfield  {journal}
    {\bibinfo  {journal} {Phys. Rev. Lett.}\ }\textbf {\bibinfo {volume} {100}},\
    \bibinfo {pages} {136406} (\bibinfo {year} {2008})}\BibitemShut {NoStop}%
  \bibitem [{\citenamefont {Hamann}(2013)}]{PhysRevB.88.085117}%
    \BibitemOpen
    \bibfield  {author} {\bibinfo {author} {\bibfnamefont {D.~R.}\ \bibnamefont
    {Hamann}},\ }\href {https://doi.org/10.1103/PhysRevB.88.085117} {\bibfield
    {journal} {\bibinfo  {journal} {Phys. Rev. B}\ }\textbf {\bibinfo {volume}
    {88}},\ \bibinfo {pages} {085117} (\bibinfo {year} {2013})}\BibitemShut
    {NoStop}%
  \bibitem [{\citenamefont {{van Setten}}\ \emph {et~al.}(2018)\citenamefont
    {{van Setten}}, \citenamefont {Giantomassi}, \citenamefont {Bousquet},
    \citenamefont {Verstraete}, \citenamefont {Hamann}, \citenamefont {Gonze},\
    and\ \citenamefont {Rignanese}}]{VANSETTEN201839}%
    \BibitemOpen
    \bibfield  {author} {\bibinfo {author} {\bibfnamefont {M.}~\bibnamefont {{van
    Setten}}}, \bibinfo {author} {\bibfnamefont {M.}~\bibnamefont {Giantomassi}},
    \bibinfo {author} {\bibfnamefont {E.}~\bibnamefont {Bousquet}}, \bibinfo
    {author} {\bibfnamefont {M.}~\bibnamefont {Verstraete}}, \bibinfo {author}
    {\bibfnamefont {D.}~\bibnamefont {Hamann}}, \bibinfo {author} {\bibfnamefont
    {X.}~\bibnamefont {Gonze}},\ and\ \bibinfo {author} {\bibfnamefont {G.-M.}\
    \bibnamefont {Rignanese}},\ }\href
    {https://doi.org/doi.org/10.1016/j.cpc.2018.01.012} {\bibfield  {journal}
    {\bibinfo  {journal} {Comput. Phys. Commun.}\ }\textbf {\bibinfo {volume}
    {226}},\ \bibinfo {pages} {39 } (\bibinfo {year} {2018})}\BibitemShut
    {NoStop}%
  \bibitem [{\citenamefont {Marzari}\ and\ \citenamefont
    {Vanderbilt}(1997)}]{wannier1}%
    \BibitemOpen
    \bibfield  {author} {\bibinfo {author} {\bibfnamefont {N.}~\bibnamefont
    {Marzari}}\ and\ \bibinfo {author} {\bibfnamefont {D.}~\bibnamefont
    {Vanderbilt}},\ }\href {https://doi.org/10.1103/PhysRevB.56.12847} {\bibfield
     {journal} {\bibinfo  {journal} {Phys. Rev. B}\ }\textbf {\bibinfo {volume}
    {56}},\ \bibinfo {pages} {12847} (\bibinfo {year} {1997})}\BibitemShut
    {NoStop}%
  \bibitem [{\citenamefont {Giustino}\ \emph {et~al.}(2007)\citenamefont
    {Giustino}, \citenamefont {Cohen},\ and\ \citenamefont
    {Louie}}]{PhysRevB.76.165108}%
    \BibitemOpen
    \bibfield  {author} {\bibinfo {author} {\bibfnamefont {F.}~\bibnamefont
    {Giustino}}, \bibinfo {author} {\bibfnamefont {M.~L.}\ \bibnamefont
    {Cohen}},\ and\ \bibinfo {author} {\bibfnamefont {S.~G.}\ \bibnamefont
    {Louie}},\ }\href {https://doi.org/10.1103/PhysRevB.76.165108} {\bibfield
    {journal} {\bibinfo  {journal} {Phys. Rev. B}\ }\textbf {\bibinfo {volume}
    {76}},\ \bibinfo {pages} {165108} (\bibinfo {year} {2007})}\BibitemShut
    {NoStop}%
  \bibitem [{\citenamefont {Noffsinger}\ \emph {et~al.}(2010)\citenamefont
    {Noffsinger}, \citenamefont {Giustino}, \citenamefont {Malone}, \citenamefont
    {Park}, \citenamefont {Louie},\ and\ \citenamefont
    {Cohen}}]{NOFFSINGER20102140}%
    \BibitemOpen
    \bibfield  {author} {\bibinfo {author} {\bibfnamefont {J.}~\bibnamefont
    {Noffsinger}}, \bibinfo {author} {\bibfnamefont {F.}~\bibnamefont
    {Giustino}}, \bibinfo {author} {\bibfnamefont {B.~D.}\ \bibnamefont
    {Malone}}, \bibinfo {author} {\bibfnamefont {C.-H.}\ \bibnamefont {Park}},
    \bibinfo {author} {\bibfnamefont {S.~G.}\ \bibnamefont {Louie}},\ and\
    \bibinfo {author} {\bibfnamefont {M.~L.}\ \bibnamefont {Cohen}},\ }\href
    {https://doi.org/10.1016/j.cpc.2010.08.027} {\bibfield  {journal} {\bibinfo
    {journal} {Comput. Phys. Commun.}\ }\textbf {\bibinfo {volume} {181}},\
    \bibinfo {pages} {2140 } (\bibinfo {year} {2010})}\BibitemShut {NoStop}%
  \bibitem [{\citenamefont {Ponc{\'{e}}}\ \emph {et~al.}(2016)\citenamefont
    {Ponc{\'{e}}}, \citenamefont {Margine}, \citenamefont {Verdi},\ and\
    \citenamefont {Giustino}}]{PONCE2016116}%
    \BibitemOpen
    \bibfield  {author} {\bibinfo {author} {\bibfnamefont {S.}~\bibnamefont
    {Ponc{\'{e}}}}, \bibinfo {author} {\bibfnamefont {E.}~\bibnamefont
    {Margine}}, \bibinfo {author} {\bibfnamefont {C.}~\bibnamefont {Verdi}},\
    and\ \bibinfo {author} {\bibfnamefont {F.}~\bibnamefont {Giustino}},\ }\href
    {https://doi.org/10.1016/j.cpc.2016.07.028} {\bibfield  {journal} {\bibinfo
    {journal} {Comput. Phys. Commun.}\ }\textbf {\bibinfo {volume} {209}},\
    \bibinfo {pages} {116 } (\bibinfo {year} {2016})}\BibitemShut {NoStop}%
  \bibitem [{\citenamefont {Ponc\'e}\ \emph {et~al.}(2018)\citenamefont
    {Ponc\'e}, \citenamefont {Margine},\ and\ \citenamefont
    {Giustino}}]{PhysRevB.97.121201}%
    \BibitemOpen
    \bibfield  {author} {\bibinfo {author} {\bibfnamefont {S.}~\bibnamefont
    {Ponc\'e}}, \bibinfo {author} {\bibfnamefont {E.~R.}\ \bibnamefont
    {Margine}},\ and\ \bibinfo {author} {\bibfnamefont {F.}~\bibnamefont
    {Giustino}},\ }\href {https://doi.org/10.1103/PhysRevB.97.121201} {\bibfield
    {journal} {\bibinfo  {journal} {Phys. Rev. B}\ }\textbf {\bibinfo {volume}
    {97}},\ \bibinfo {pages} {121201(R)} (\bibinfo {year} {2018})}\BibitemShut
    {NoStop}%
  \bibitem [{\citenamefont {Souza}\ \emph {et~al.}(2001)\citenamefont {Souza},
    \citenamefont {Marzari},\ and\ \citenamefont {Vanderbilt}}]{wannier2}%
    \BibitemOpen
    \bibfield  {author} {\bibinfo {author} {\bibfnamefont {I.}~\bibnamefont
    {Souza}}, \bibinfo {author} {\bibfnamefont {N.}~\bibnamefont {Marzari}},\
    and\ \bibinfo {author} {\bibfnamefont {D.}~\bibnamefont {Vanderbilt}},\
    }\href {https://doi.org/10.1103/PhysRevB.65.035109} {\bibfield  {journal}
    {\bibinfo  {journal} {Phys. Rev. B}\ }\textbf {\bibinfo {volume} {65}},\
    \bibinfo {pages} {035109} (\bibinfo {year} {2001})}\BibitemShut {NoStop}%
  \bibitem [{\citenamefont {Mostofi}\ \emph {et~al.}(2008)\citenamefont
    {Mostofi}, \citenamefont {Yates}, \citenamefont {Lee}, \citenamefont {Souza},
    \citenamefont {Vanderbilt},\ and\ \citenamefont {Marzari}}]{wannier3}%
    \BibitemOpen
    \bibfield  {author} {\bibinfo {author} {\bibfnamefont {A.~A.}\ \bibnamefont
    {Mostofi}}, \bibinfo {author} {\bibfnamefont {J.~R.}\ \bibnamefont {Yates}},
    \bibinfo {author} {\bibfnamefont {Y.-S.}\ \bibnamefont {Lee}}, \bibinfo
    {author} {\bibfnamefont {I.}~\bibnamefont {Souza}}, \bibinfo {author}
    {\bibfnamefont {D.}~\bibnamefont {Vanderbilt}},\ and\ \bibinfo {author}
    {\bibfnamefont {N.}~\bibnamefont {Marzari}},\ }\href
    {https://doi.org/10.1016/j.cpc.2007.11.016} {\bibfield  {journal} {\bibinfo
    {journal} {Comput. Phys. Commun.}\ }\textbf {\bibinfo {volume} {178}},\
    \bibinfo {pages} {685 } (\bibinfo {year} {2008})}\BibitemShut {NoStop}%
  \bibitem [{\citenamefont {Mostofi}\ \emph {et~al.}(2014)\citenamefont
    {Mostofi}, \citenamefont {Yates}, \citenamefont {Pizzi}, \citenamefont {Lee},
    \citenamefont {Souza}, \citenamefont {Vanderbilt},\ and\ \citenamefont
    {Marzari}}]{wannier4}%
    \BibitemOpen
    \bibfield  {author} {\bibinfo {author} {\bibfnamefont {A.~A.}\ \bibnamefont
    {Mostofi}}, \bibinfo {author} {\bibfnamefont {J.~R.}\ \bibnamefont {Yates}},
    \bibinfo {author} {\bibfnamefont {G.}~\bibnamefont {Pizzi}}, \bibinfo
    {author} {\bibfnamefont {Y.-S.}\ \bibnamefont {Lee}}, \bibinfo {author}
    {\bibfnamefont {I.}~\bibnamefont {Souza}}, \bibinfo {author} {\bibfnamefont
    {D.}~\bibnamefont {Vanderbilt}},\ and\ \bibinfo {author} {\bibfnamefont
    {N.}~\bibnamefont {Marzari}},\ }\href
    {https://doi.org/10.1016/j.cpc.2007.11.016} {\bibfield  {journal} {\bibinfo
    {journal} {Comput. Phys. Commun.}\ }\textbf {\bibinfo {volume} {185}},\
    \bibinfo {pages} {2309} (\bibinfo {year} {2014})}\BibitemShut {NoStop}%
  \bibitem [{\citenamefont {Pizzi}\ \emph {et~al.}(2020)\citenamefont {Pizzi},
    \citenamefont {Vitale}, \citenamefont {Arita}, \citenamefont {Bl{\"{u}}gel},
    \citenamefont {Freimuth}, \citenamefont {G{\'{e}}ranton}, \citenamefont
    {Gibertini}, \citenamefont {Gresch}, \citenamefont {Johnson}, \citenamefont
    {Koretsune}, \citenamefont {Iba{\~{n}}ez-Azpiroz}, \citenamefont {Lee},
    \citenamefont {Lihm}, \citenamefont {Marchand}, \citenamefont {Marrazzo},
    \citenamefont {Mokrousov}, \citenamefont {Mustafa}, \citenamefont {Nohara},
    \citenamefont {Nomura}, \citenamefont {Paulatto}, \citenamefont
    {Ponc{\'{e}}}, \citenamefont {Ponweiser}, \citenamefont {Qiao}, \citenamefont
    {Th{\"{o}}le}, \citenamefont {Tsirkin}, \citenamefont {Wierzbowska},
    \citenamefont {Marzari}, \citenamefont {Vanderbilt}, \citenamefont {Souza},
    \citenamefont {Mostofi},\ and\ \citenamefont {Yates}}]{wannier5}%
    \BibitemOpen
    \bibfield  {author} {\bibinfo {author} {\bibfnamefont {G.}~\bibnamefont
    {Pizzi}}, \bibinfo {author} {\bibfnamefont {V.}~\bibnamefont {Vitale}},
    \bibinfo {author} {\bibfnamefont {R.}~\bibnamefont {Arita}}, \bibinfo
    {author} {\bibfnamefont {S.}~\bibnamefont {Bl{\"{u}}gel}}, \bibinfo {author}
    {\bibfnamefont {F.}~\bibnamefont {Freimuth}}, \bibinfo {author}
    {\bibfnamefont {G.}~\bibnamefont {G{\'{e}}ranton}}, \bibinfo {author}
    {\bibfnamefont {M.}~\bibnamefont {Gibertini}}, \bibinfo {author}
    {\bibfnamefont {D.}~\bibnamefont {Gresch}}, \bibinfo {author} {\bibfnamefont
    {C.}~\bibnamefont {Johnson}}, \bibinfo {author} {\bibfnamefont
    {T.}~\bibnamefont {Koretsune}}, \bibinfo {author} {\bibfnamefont
    {J.}~\bibnamefont {Iba{\~{n}}ez-Azpiroz}}, \bibinfo {author} {\bibfnamefont
    {H.}~\bibnamefont {Lee}}, \bibinfo {author} {\bibfnamefont {J.-M.}\
    \bibnamefont {Lihm}}, \bibinfo {author} {\bibfnamefont {D.}~\bibnamefont
    {Marchand}}, \bibinfo {author} {\bibfnamefont {A.}~\bibnamefont {Marrazzo}},
    \bibinfo {author} {\bibfnamefont {Y.}~\bibnamefont {Mokrousov}}, \bibinfo
    {author} {\bibfnamefont {J.~I.}\ \bibnamefont {Mustafa}}, \bibinfo {author}
    {\bibfnamefont {Y.}~\bibnamefont {Nohara}}, \bibinfo {author} {\bibfnamefont
    {Y.}~\bibnamefont {Nomura}}, \bibinfo {author} {\bibfnamefont
    {L.}~\bibnamefont {Paulatto}}, \bibinfo {author} {\bibfnamefont
    {S.}~\bibnamefont {Ponc{\'{e}}}}, \bibinfo {author} {\bibfnamefont
    {T.}~\bibnamefont {Ponweiser}}, \bibinfo {author} {\bibfnamefont
    {J.}~\bibnamefont {Qiao}}, \bibinfo {author} {\bibfnamefont {F.}~\bibnamefont
    {Th{\"{o}}le}}, \bibinfo {author} {\bibfnamefont {S.~S.}\ \bibnamefont
    {Tsirkin}}, \bibinfo {author} {\bibfnamefont {M.}~\bibnamefont
    {Wierzbowska}}, \bibinfo {author} {\bibfnamefont {N.}~\bibnamefont
    {Marzari}}, \bibinfo {author} {\bibfnamefont {D.}~\bibnamefont {Vanderbilt}},
    \bibinfo {author} {\bibfnamefont {I.}~\bibnamefont {Souza}}, \bibinfo
    {author} {\bibfnamefont {A.~A.}\ \bibnamefont {Mostofi}},\ and\ \bibinfo
    {author} {\bibfnamefont {J.~R.}\ \bibnamefont {Yates}},\ }\href
    {https://doi.org/10.1088/1361-648x/ab51ff} {\bibfield  {journal} {\bibinfo
    {journal} {J. Phys.: Condens. Matter}\ }\textbf {\bibinfo {volume} {32}},\
    \bibinfo {pages} {165902} (\bibinfo {year} {2020})}\BibitemShut {NoStop}%
  \bibitem [{\citenamefont {Albers}\ \emph {et~al.}(1976)\citenamefont {Albers},
    \citenamefont {Bohlin}, \citenamefont {Roy},\ and\ \citenamefont
    {Wilkins}}]{PhysRevB.13.768}%
    \BibitemOpen
    \bibfield  {author} {\bibinfo {author} {\bibfnamefont {R.~C.}\ \bibnamefont
    {Albers}}, \bibinfo {author} {\bibfnamefont {L.}~\bibnamefont {Bohlin}},
    \bibinfo {author} {\bibfnamefont {M.}~\bibnamefont {Roy}},\ and\ \bibinfo
    {author} {\bibfnamefont {J.~W.}\ \bibnamefont {Wilkins}},\ }\href
    {https://doi.org/10.1103/PhysRevB.13.768} {\bibfield  {journal} {\bibinfo
    {journal} {Phys. Rev. B}\ }\textbf {\bibinfo {volume} {13}},\ \bibinfo
    {pages} {768} (\bibinfo {year} {1976})}\BibitemShut {NoStop}%
  \bibitem [{\citenamefont {Wagner}\ and\ \citenamefont
    {Bowers}(1978)}]{wagner1978radio}%
    \BibitemOpen
    \bibfield  {author} {\bibinfo {author} {\bibfnamefont {D.}~\bibnamefont
    {Wagner}}\ and\ \bibinfo {author} {\bibfnamefont {R.}~\bibnamefont
    {Bowers}},\ }\href {https://doi.org/10.1080/00018737800101464} {\bibfield
    {journal} {\bibinfo  {journal} {Adv. Phys.}\ }\textbf {\bibinfo {volume}
    {27}},\ \bibinfo {pages} {651} (\bibinfo {year} {1978})}\BibitemShut
    {NoStop}%
  \bibitem [{\citenamefont {Grimvall}(1981)}]{grimvall1981electron}%
    \BibitemOpen
    \bibfield  {author} {\bibinfo {author} {\bibfnamefont {G.}~\bibnamefont
    {Grimvall}},\ }\href@noop {} {\emph {\bibinfo {title} {The Electron-phonon
    Interaction in Metals}}}\ (\bibinfo  {publisher} {North-Holland Publishing
    Company},\ \bibinfo {year} {1981})\BibitemShut {NoStop}%
  \bibitem [{\citenamefont {Mulliken}(1955)}]{Mulliken1955-rx}%
    \BibitemOpen
    \bibfield  {author} {\bibinfo {author} {\bibfnamefont {R.~S.}\ \bibnamefont
    {Mulliken}},\ }\href {https://doi.org/10.1063/1.1740655} {\bibfield
    {journal} {\bibinfo  {journal} {J. Chem. Phys.}\ }\textbf {\bibinfo {volume}
    {23}},\ \bibinfo {pages} {1997} (\bibinfo {year} {1955})}\BibitemShut
    {NoStop}%
  \bibitem [{\citenamefont {Kokalj}(1999)}]{Kokalj1999-al}%
    \BibitemOpen
    \bibfield  {author} {\bibinfo {author} {\bibfnamefont {A.}~\bibnamefont
    {Kokalj}},\ }\href {https://doi.org/10.1016/S1093-3263(99)00028-5} {\bibfield
     {journal} {\bibinfo  {journal} {J. Mol. Graph. Model.}\ }\textbf {\bibinfo
    {volume} {17}},\ \bibinfo {pages} {176} (\bibinfo {year} {1999})}\BibitemShut
    {NoStop}%
  \bibitem [{Note1()}]{Note1}%
    \BibitemOpen
    \bibinfo {note} {In actual calculations, we replaced the step function
    $\theta (x)$ by $[1+\protect \text {erf}(x)]/2$, where $\protect \text
    {erf}(x)$ is the Gauss error function.}\BibitemShut {Stop}%
  \bibitem [{\citenamefont {Fr{\"o}hlich}(1954)}]{Frohlich1954-zf}%
    \BibitemOpen
    \bibfield  {author} {\bibinfo {author} {\bibfnamefont {H.}~\bibnamefont
    {Fr{\"o}hlich}},\ }\href {https://doi.org/10.1080/00018735400101213}
    {\bibfield  {journal} {\bibinfo  {journal} {Adv. Phys.}\ }\textbf {\bibinfo
    {volume} {3}},\ \bibinfo {pages} {325} (\bibinfo {year} {1954})}\BibitemShut
    {NoStop}%
  \bibitem [{\citenamefont {Zhou}\ and\ \citenamefont
    {Bernardi}(2016)}]{PhysRevB.94.201201}%
    \BibitemOpen
    \bibfield  {author} {\bibinfo {author} {\bibfnamefont {J.-J.}\ \bibnamefont
    {Zhou}}\ and\ \bibinfo {author} {\bibfnamefont {M.}~\bibnamefont
    {Bernardi}},\ }\href {https://doi.org/10.1103/PhysRevB.94.201201} {\bibfield
    {journal} {\bibinfo  {journal} {Phys. Rev. B}\ }\textbf {\bibinfo {volume}
    {94}},\ \bibinfo {pages} {201201(R)} (\bibinfo {year} {2016})}\BibitemShut
    {NoStop}%
  \bibitem [{\citenamefont {Liu}\ \emph {et~al.}(2017)\citenamefont {Liu},
    \citenamefont {Zhou}, \citenamefont {Liao}, \citenamefont {Singh},\ and\
    \citenamefont {Chen}}]{PhysRevB.95.075206}%
    \BibitemOpen
    \bibfield  {author} {\bibinfo {author} {\bibfnamefont {T.-H.}\ \bibnamefont
    {Liu}}, \bibinfo {author} {\bibfnamefont {J.}~\bibnamefont {Zhou}}, \bibinfo
    {author} {\bibfnamefont {B.}~\bibnamefont {Liao}}, \bibinfo {author}
    {\bibfnamefont {D.~J.}\ \bibnamefont {Singh}},\ and\ \bibinfo {author}
    {\bibfnamefont {G.}~\bibnamefont {Chen}},\ }\href
    {https://doi.org/10.1103/PhysRevB.95.075206} {\bibfield  {journal} {\bibinfo
    {journal} {Phys. Rev. B}\ }\textbf {\bibinfo {volume} {95}},\ \bibinfo
    {pages} {075206} (\bibinfo {year} {2017})}\BibitemShut {NoStop}%
  \bibitem [{\citenamefont {Ma}\ \emph {et~al.}(2018)\citenamefont {Ma},
    \citenamefont {Nissimagoudar},\ and\ \citenamefont
    {Li}}]{PhysRevB.97.045201}%
    \BibitemOpen
    \bibfield  {author} {\bibinfo {author} {\bibfnamefont {J.}~\bibnamefont
    {Ma}}, \bibinfo {author} {\bibfnamefont {A.~S.}\ \bibnamefont
    {Nissimagoudar}},\ and\ \bibinfo {author} {\bibfnamefont {W.}~\bibnamefont
    {Li}},\ }\href {https://doi.org/10.1103/PhysRevB.97.045201} {\bibfield
    {journal} {\bibinfo  {journal} {Phys. Rev. B}\ }\textbf {\bibinfo {volume}
    {97}},\ \bibinfo {pages} {045201} (\bibinfo {year} {2018})}\BibitemShut
    {NoStop}%
  \bibitem [{\citenamefont {Chang}\ \emph {et~al.}(2019)\citenamefont {Chang},
    \citenamefont {Liu}, \citenamefont {Liu},\ and\ \citenamefont
    {Du}}]{Chang2019-ii}%
    \BibitemOpen
    \bibfield  {author} {\bibinfo {author} {\bibfnamefont {P.}~\bibnamefont
    {Chang}}, \bibinfo {author} {\bibfnamefont {X.}~\bibnamefont {Liu}}, \bibinfo
    {author} {\bibfnamefont {F.}~\bibnamefont {Liu}},\ and\ \bibinfo {author}
    {\bibfnamefont {G.}~\bibnamefont {Du}},\ }\href
    {https://doi.org/10.1109/LED.2018.2886842} {\bibfield  {journal} {\bibinfo
    {journal} {IEEE Electron Device Lett.}\ }\textbf {\bibinfo {volume} {40}},\
    \bibinfo {pages} {333} (\bibinfo {year} {2019})}\BibitemShut {NoStop}%
  \bibitem [{\citenamefont {Shi}\ \emph {et~al.}(2020)\citenamefont {Shi},
    \citenamefont {Cao}, \citenamefont {Yang}, \citenamefont {You}, \citenamefont
    {Zhang}, \citenamefont {Bao}, \citenamefont {Zhang}, \citenamefont {Niu},\
    and\ \citenamefont {Qian}}]{Shi2020-sf}%
    \BibitemOpen
    \bibfield  {author} {\bibinfo {author} {\bibfnamefont {L.-B.}\ \bibnamefont
    {Shi}}, \bibinfo {author} {\bibfnamefont {S.}~\bibnamefont {Cao}}, \bibinfo
    {author} {\bibfnamefont {M.}~\bibnamefont {Yang}}, \bibinfo {author}
    {\bibfnamefont {Q.}~\bibnamefont {You}}, \bibinfo {author} {\bibfnamefont
    {K.-C.}\ \bibnamefont {Zhang}}, \bibinfo {author} {\bibfnamefont
    {Y.}~\bibnamefont {Bao}}, \bibinfo {author} {\bibfnamefont {Y.-J.}\
    \bibnamefont {Zhang}}, \bibinfo {author} {\bibfnamefont {Y.-Y.}\ \bibnamefont
    {Niu}},\ and\ \bibinfo {author} {\bibfnamefont {P.}~\bibnamefont {Qian}},\
    }\href {https://doi.org/10.1088/1361-648X/ab534f} {\bibfield  {journal}
    {\bibinfo  {journal} {J. Phys. Condens. Matter}\ }\textbf {\bibinfo {volume}
    {32}},\ \bibinfo {pages} {065306} (\bibinfo {year} {2020})}\BibitemShut
    {NoStop}%
  \bibitem [{\citenamefont {Momma}\ and\ \citenamefont
    {Izumi}(2011)}]{Momma2011-fj}%
    \BibitemOpen
    \bibfield  {author} {\bibinfo {author} {\bibfnamefont {K.}~\bibnamefont
    {Momma}}\ and\ \bibinfo {author} {\bibfnamefont {F.}~\bibnamefont {Izumi}},\
    }\href {https://doi.org/10.1107/s0021889811038970} {\bibfield  {journal}
    {\bibinfo  {journal} {J. Appl. Crystallogr.}\ }\textbf {\bibinfo {volume}
    {44}},\ \bibinfo {pages} {1272} (\bibinfo {year} {2011})}\BibitemShut
    {NoStop}%
\end{thebibliography}
\end{document}